\documentclass{aa}  

\usepackage[multidot]{grffile}
\usepackage{graphicx}
\usepackage{txfonts}
\usepackage[caption=false]{subfig}

\usepackage[colorlinks=true, allcolors=blue]{hyperref} 
\usepackage[colorinlistoftodos]{todonotes}

\newcommand{\Msun}{\mbox{$M_\odot$}}
\newcommand{\msun}{\mbox{$M_\odot$}}

\def\arcsec{\hbox{$^{\prime\prime}$}}

\begin{document}

   \title{Strong dependence of the physical properties of cores\\ on spatial resolution in observations and simulations}


   \author{F.~Louvet      \inst{1,2} \and
           P.~Hennebelle    \inst{1,2} \and
           A.~Men'shchikov\inst{1}    \and
           P.~Didelon          \inst{1}    \and
           E.~Ntormousi      \inst{3}    \and
           F. Motte              \inst{4}
          }

   \institute{AIM, CEA, CNRS, Universit{\'e} Paris-Saclay, Universit{\'e} Paris Diderot, Sorbonne Paris Cit{\'e}, F-91191 
              Gif-sur-Yvette, France\\ \email{fabien.louvet@cea.fr}
         \and
        LERMA (UMR CNRS 8112), \'Ecole Normale Sup\'erieure, 75231 Paris Cedex, France
         \and
         Scuola Normale Superiore di Pisa, Piazza dei Cavalieri, 7, 56126 Pisa, Italy
         \and
            Universit\'{e} Grenoble Alpe, CNRS, IPAG, F-38000 Grenoble, France\\
             }

   \date{Received ; Accepted}

\offprints{Fabien Louvet}
\titlerunning{Derived properties of putative prestellar cores and angular resolution}
\authorrunning{F.~Louvet et al.}

  \abstract
   { 
 The angular resolution of a telescope is the primary observational parameter, along with the detector sensitivity in defining the quality of the observed images and of the subsequent scientific exploitation of the data. During the last decade in star formation research, many studies have targeted low- and high-mass star formation regions located at different distances, with different telescopes having specific angular resolution capabilities. However, no dedicated studies of the spatial resolution effects on the derived sizes and masses of the sources extracted from the observed images have been published. 
   
   We present a systematic investigation of the angular resolution effects, with special attention being paid to the derived masses of sources as well as the shape of the resulting source mass functions (SMFs) and to their comparison with the initial stellar mass function (IMF). For our study, we chose two star-forming regions observed with \emph{Herschel}, \object{NGC\,6334} and \object{Aquila} distant of 1750 and 460\,pc respectively, and three (magneto)-hydrodynamical simulations, virtually positioned at the same distances as the observed regions. We built surface density maps with different angular resolutions by convolving the surface density images of the five regions to a set of four resolutions differing by a factor of two (9, 18, 36, and 72{\arcsec}), which allowed us to cover spatial resolutions from 0.6 down to 0.02\,pc. 
 Then we detected and measured sources in each of the images at each resolution using \textsl{getsf} and we analysed the derived masses and sizes of the extracted sources.
  
We find that the number of sources does not converge from 0.6 to $\gtrsim$0.05\,pc. It increases by about two when the angular resolution increases with a similar factor, which confirms that these large sources are cluster-forming clumps. Below 0.05\,pc, the number of source still increases by about 1.3 when the angular resolution increases by two, suggesting that we are close to, but not yet at, convergence. In this regime of physical scales, we find that the measured sizes and masses of sources linearly depend on the angular resolution with no sign of convergence to a resolution-independent value, implying that these sources cannot be assimilated to isolated prestellar cores. 
The corresponding SMF peak also shifts with angular resolution, while the slope of the high-mass tail of the SMFs remains almost invariant. We propose that these angular resolution effects could be caused by the underestimated background of the unresolved sources observed against the sloping, hill-like backgrounds of the molecular clouds.
   
If prestellar cores physically distinct from their background exist in cluster-forming molecular clouds, we conclude that their mass must be lower than reported so far in the literature. We discuss various implications for the studies of star formation: the problem of determining the mass reservoirs involved in the star-formation process; the inapplicability of the Gaussian beam deconvolution to infer source sizes; and the impossibility to determine the efficiency of the mass conversion from the cores to the stars. Our approach constitutes a simple convergence test to determine whether an observation is affected by angular resolution.

   } 
   \maketitle


\section{Introduction}  

One of the challenges of modern astrophysics is to understand what controls the mass distribution of stars at their birth, the so-called initial mass function (IMF). Most of the observational studies since the work of \cite{salpeter55} have found the shape of the IMF to be universal \citep[e.g.][]{kroupa02,LL03,bastian10}, except possibly in young massive clusters \citep[e.g.][]{lu13,schneider18}. Understanding its origin is crucial for both star formation and galactic evolution. The IMF exhibits a peak at mass $M{\,\approx\,}$0.3\,$M_\sun$ and a power-law high-mass end ${\rm d}N/{\rm
d}\log(M){\,\propto\,}M^{-1.35}$. Similarities between the slope of the IMF and the slope of the core mass function (CMF), derived
from observations of molecular clouds, led to the suggestion that the IMF results from the fragmentation of molecular clouds
\citep[e.g.][]{motte98, TS98,alves07,konyves15,konyves20}. Such similarities have also been reported in numerical studies
\citep[e.g.][]{klessen01,TP04,ntormousi19}.

With the possibility of a relationship between the IMF and the CMF, studies on the origin of the IMF proceeded in two directions: to understand (1) the origin of the CMF and its link with the fragmentation processes of molecular clouds \cite[e.g.][]{padoan07,schmidt10} and (2) the connection between the prestellar cores and stars \cite[e.g.][]{smith09, lomax14, pelkonen21}, with a commonly used broad assumption of a one-to-one correspondence between the two types of objects. The assumption means that the gravitational collapse of a prestellar core forms a single star or a close binary, with a certain mass transfer efficiency from the core to the stars \citep[e.g.][]{padoan97,HC08,hopkins12}. In this picture of star formation, the cores must remain not fragmented during the collapse. It is believed that prestellar cores are the result of gravo-thermal fragmentation in molecular clouds, resulting in an average core mass of ${\sim\,}$2\,$M_\sun$ and size of ${\sim\,}$0.2\,pc that correspond to the Jeans mass and length for gas with a temperature and mean density of $T{\,\simeq\,}$10\,K and $n_\mathrm{H_2}{\,\simeq\,}$10$^4$\,cm$^{-3}$, typical of dense regions in molecular clouds.

Modern high-resolution and sensitive imaging in the far-infrared and sub-millimetre wavelengths lead to a standard approach to their analysis. In that approach, the single or multi-wavelength images of star-forming regions are analysed using source extraction methods. The extraction tools detect, measure, and catalogue all sources of emissions in the images, that is, all strong intensity peaks that stand out against the local background and noise fluctuations. 

Despite the importance of the angular resolutions for observational studies of star formation, no detailed systematic study has been done to clarify their effects on the derived properties of the physical objects. We investigate how the sizes and masses of the same sources behave when analysed at different spatial resolutions. We extracted and examined sources at different angular
resolutions in the \object{NGC\,6334} high-mass star-forming region and in the \object{Aquila} low-mass star-forming region. Because they are physically different and present very different resolutions, these two regions are highly complementary for our analysis. In particular the sources in \object{NGC\,6334} are likely to be small clumps  while the sources of \object{Aquila} have been described in the literature as candidate pre- and protostellar dense cores. These two sets of observations complement the analysis with a similar approach using numerical simulations that allowed us to reach higher angular resolutions and to assess the impact of the projection effects. In Sect.~\ref{s:obs-sim}, we present the different observational and numerical data sets that we used. In Sect.~\ref{s:analyse}, we show that angular resolution affects the properties of extracted sources, even at scales lower than 0.05\,pc where we expect to retrieve candidate prestellar cores. In Sect.~\ref{s:disc}, we discuss the similarities and differences between observations and simulations and put our study in perspective with recent investigations on the link between the CMF and the IMF. In Sect.~\ref{s:concl}, we summarise our conclusions.


\section{Observations and numerical simulations}
\label{s:obs-sim}

This paper aims to investigate the effect of angular resolution on the properties of sources and resulting source mass function (SMF), from large scale (0.6\,pc) where we expect to observe cluster-forming clumps down to small-scales ($\lesssim$0.05\,pc) where we expect to see individual pre-stellar cores. To that end, we use \emph{Herschel} observations and numerical simulations of both low- and high-mass star-forming regions to understand the effects of angular resolution on the derivation of the physical parameters of the gravitationally bound sources, in the framework of the standard observational approach of source extractions.


\subsection{A high-mass star-forming region: NGC\,6334}
\label{ss:ngc}

\begin{figure*}                  
\centering                       
\centerline{\resizebox{\hsize}{!}{\includegraphics{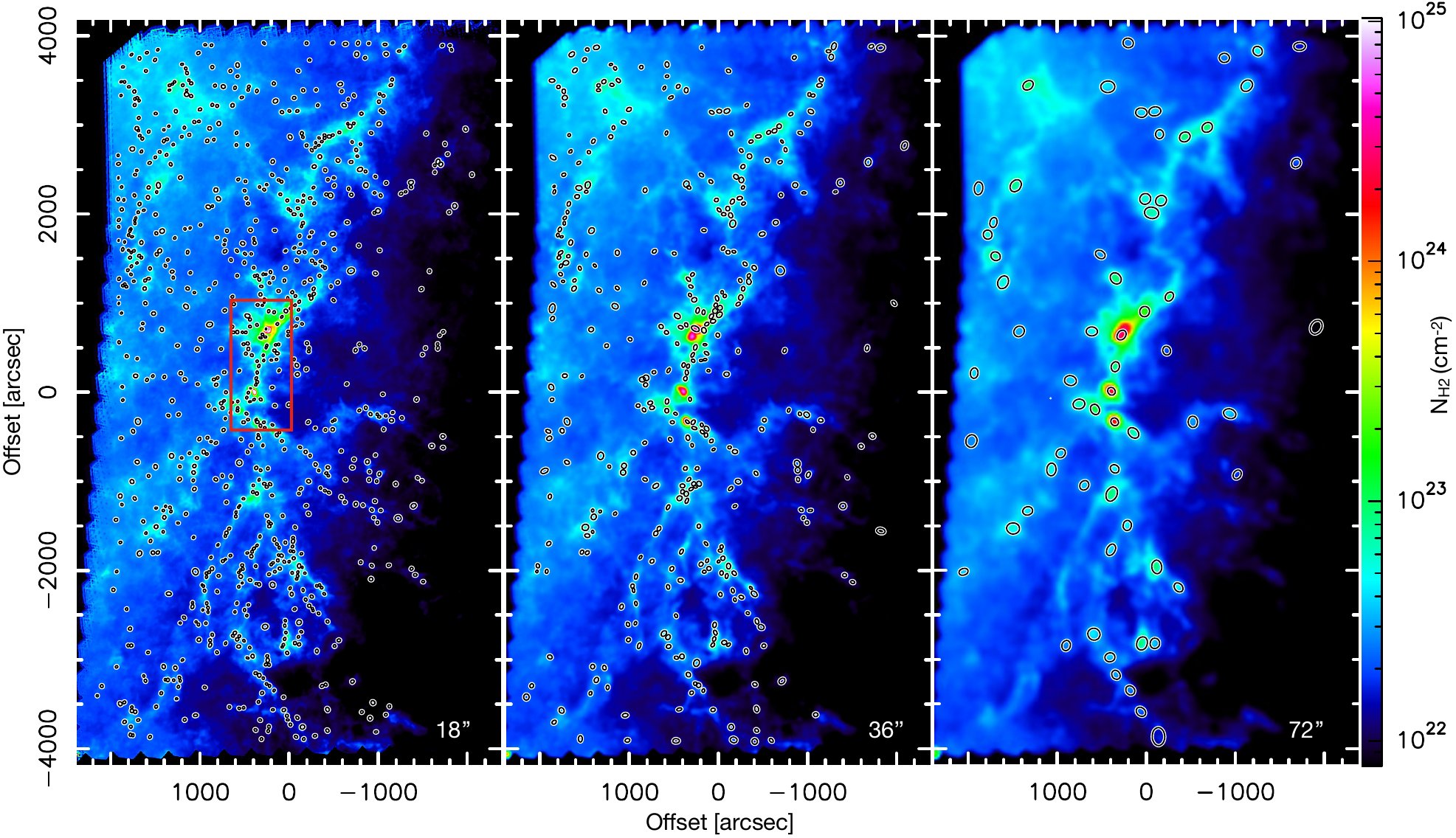}}}
\caption
{ 
Column densities of the high-mass star-forming region \object{NGC\,6334} ($D{\,\simeq\,}1750$\,pc). The images are overlaid with the half-maximum ellipses of the extracted bound sources (Sect.~\ref{ss:getsf}), with the angular resolutions indicated in each panel. The red rectangle in the left panel indicates the central region of \object{NGC\,6334} discussed in Sect.~\ref{sss:compa}.
} 
\label{f:ngc}
\end{figure*}

\object{NGC\,6334} is a complex of giant molecular clouds in the Sagit\-tarius-Carina arm of the Milky Way, with an estimated mass $M{\,\simeq\,}$7${\,\times\,}10^5$\,$M_\sun$ \citep[e.g.][]{russeil12}, located at a distance $D{\,\simeq\,}1.75$\,kpc from the Sun \citep{matthews08}. This very active high-mass star-forming region hosts more than 2000 young stellar objects (YSOs) identified
with \textit{Spitzer} \citep{willis13}, numerous \ion{H}{ii} regions, maser sources, and molecular outflows
\citep[see][]{loughran86,carral02,PT08,louvet19}. In this paper, we use a column density image of \object{NGC\,6334}, derived from the \textit{Herschel} images at 160, 250, 350, and 500\,$\mu$m with an effective angular resolution of 18{\arcsec} from the \textsc{HOBYS} key programme \citep{motte10}\footnote{See http://hobys-herschel.cea.fr}. We extracted and analysed sources detected in the surface density map at the initial 18{\arcsec} resolution and after degrading the image to the 36 and
72{\arcsec} resolutions (Fig.~\ref{f:ngc}, Sect.~\ref{sss:obs}).
                            
\subsection{A low-mass star-forming region: Aquila}
\label{ss:aquila}

\begin{figure*}
\centering
\centerline{\resizebox{\hsize}{!}{\includegraphics{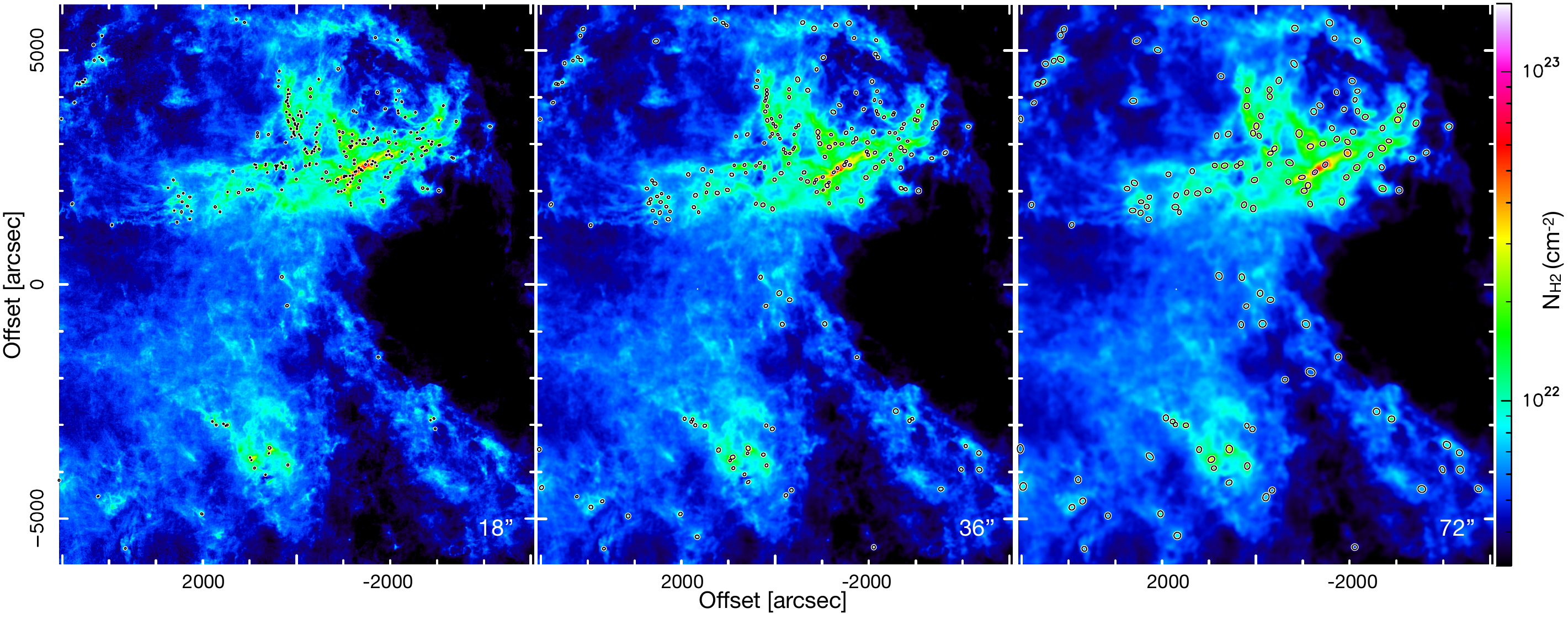}}}
\caption
{ 
Column densities of the low-mass star-forming region \object{Aquila} ($D{\,\simeq\,}460$\,pc). The images are overlaid with the half-maximum ellipses of the extracted bound sources (Sect.~\ref{ss:getsf}), with the angular resolutions indicated in each panel.
} 
\label{f:aquila}
\end{figure*}

\object{Aquila} is a complex of molecular clouds at the end of the \object{Aquila} rift, located at a distance of 435 to 490\,pc \citep[see ][respectively]{ortiz-leon07,zucker20}. In this study we adopt a distance of 460\,pc. Using the 2MASS extinction maps, \cite{bontemps10} derived its total mass $M{\,\simeq\,}$9${\,\times\,}$10$^4$\,$M_\sun$\footnote{We adjusted the mass for a distance of 460\,pc, while \cite{bontemps10} reported $\simeq$3$\times 10^4$\,\msun~for a source distant of 260\,pc. } and, using the \textit{Herschel} 70 and 160\,$\mu$m images, they identified about 200 YSOs in \object{Aquila}. In this work, we use the high-resolution column density map presented in \cite{konyves15}. This map has a native angular resolution of 18{\arcsec}. It was built from the \textit{Herschel} images at 160, 250, 350, and 500\,$\mu$m taken during the \textit{Herschel} Gould Belt Survey \citep[HGBS, see][]{andre10}. We extracted and analysed sources in the surface density image at the initial 18{\arcsec} resolution and after degrading the image to the 36 and 72{\arcsec} resolutions (Fig.~\ref{f:aquila}, Sect.~\ref{sss:obs}).

\subsection{Simulations of star-forming regions}
\label{ss:simus}

To overcome the limited angular resolution provided by currently available telescopes, we employed several
numerical simulations of a 3D turbulent cloud designed to represent an \object{Orion}-sized molecular cloud \citep[][hereafter
NH19]{ntormousi19}. They were produced with the RAMSES code (\citealt{teyssier02, fromang06}) that solves the
magneto-hydrodynamical (MHD) equations on a Cartesian grid with adaptive mesh refinement (AMR). 

The initial conditions of the simulations resembled a molecular cloud, that was approximated by an ellipsoid with a full extent of
{33\,pc}{$\,\times\,$}13.2\,pc{$\,\times\,$}13.2\,pc and a mass of $10^5$\,$M_\sun$, with a volume density profile defined by
\begin{equation} 
\rho(r) = \rho_{0}\left(1+\frac{(x^2+y^2)}{r_0^2}+\frac{z^2}{z_0^2}\right)^{-1},
\label{mcloud}
\end{equation} 
where $r{\,=\,}(x^2{\,+\,}y^2{\,+\,}z^2)^{1/2}$ is the radial coordinate and ($z_0, r_0$) is the point at which the density
profile flattens to $\rho_{0}{\,=\,}1500$\,cm$^{-3}$, the initial peak density in the cloud. The simulations were computed on an
adaptive mesh, such that they always resolve the Jeans length with at least ten cells. The coarsest 512$^{3}$ grid occupied a volume of $66^{3}$\,pc$^3$.

\begin{figure*}
\centering
\centerline{\resizebox{\hsize}{!}{\includegraphics{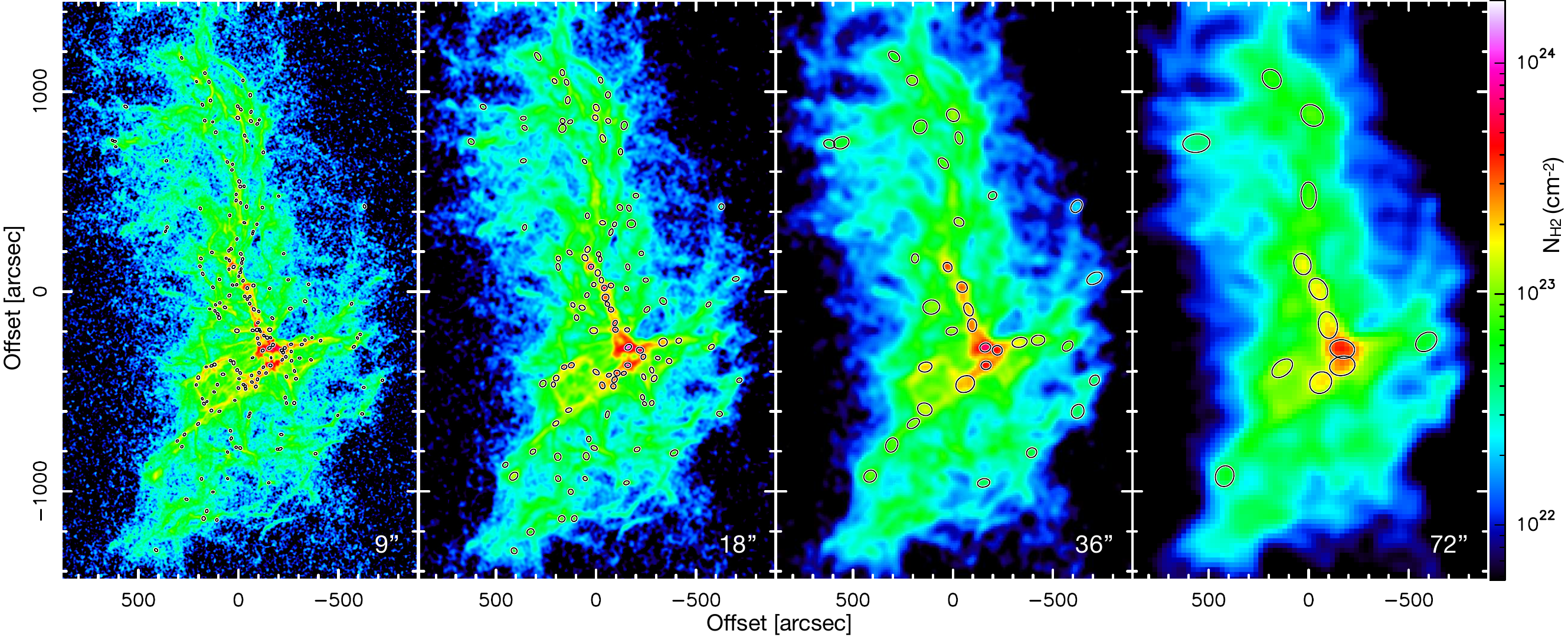}}}
\caption
{ 
Column densities obtained by projecting the HD simulation along the \emph{x} axis, overlaid with the half-maximum ellipses of the extracted bound sources (Sect.~\ref{ss:getsf}). The angular resolutions, indicated in the panels, correspond to the linear scales of 0.07, 0.15, 0.3, and 0.6\,pc.
} 
\label{f:cores-projx}
\end{figure*}

\begin{figure*}
\centering
\centerline{\resizebox{\hsize}{!}{\includegraphics{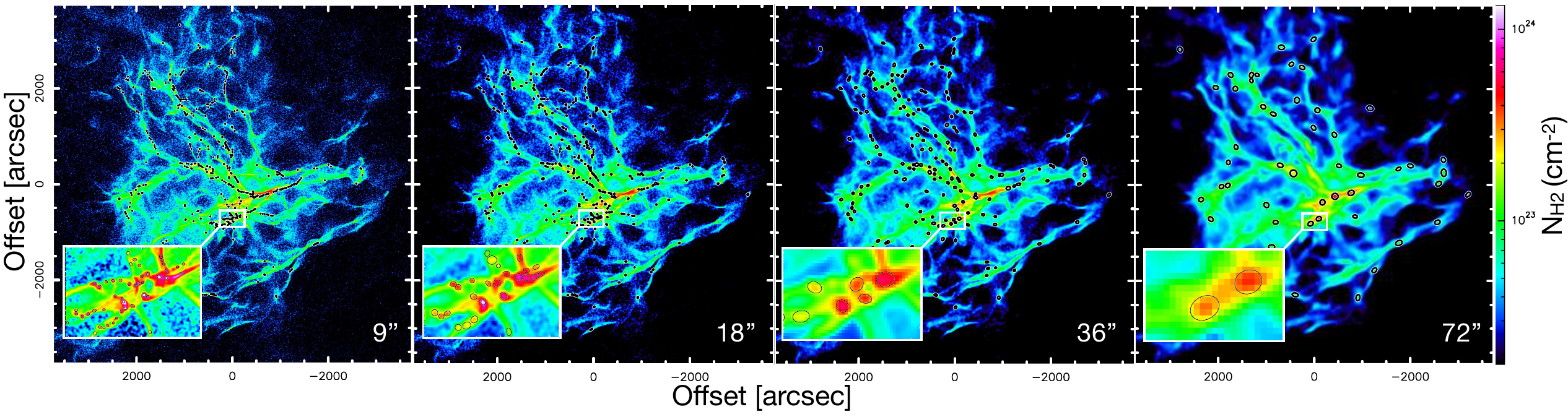}}}
\caption
{ 
Surface densities obtained by projecting the HD$^\mathrm{h}$ simulation along the \emph{x} axis, overlaid with the half-maximum ellipses of the extracted bound sources (Sect.~\ref{ss:getsf}). The angular resolutions, indicated in the panels, correspond to linear scales of 0.02, 0.04, 0.08, and 0.16\,pc.
} 
\label{f:cores-projx-hr}
\end{figure*}

\begin{table} 
\caption{Parameters of the numerical simulations.} 
\label{t:model-parameters}      
\begin{tabular}{lcccc} 
\hline\hline                        
\noalign{\smallskip}
Model&\,\,\,\,\,\,$M$$^{\,(1)}$&\,\,\,\,\,\,$\Delta$$^{\,(2)}$&\,\,\,\,\,\,$\langle{B}\rangle$$^{\,(3)}$&\,\,\,\,\,\,$D$$^{\,(4)}$\\
     & (\,$M_\sun$) & (au) & ($\mu$G) & (pc)\\
\noalign{\smallskip}
\hline
\noalign{\smallskip}
HD              & 10$^5$ & ${\sim\,}$400 & 0 &     1750 \\
MHD             & 10$^5$ & ${\sim\,}$400 & 5 &     1750 \\
HD$^\mathrm{h}$ & 10$^5$ & ${\sim\,}$200 & 0 &\,\,\,460 \\
\noalign{\smallskip}
\hline
\end{tabular}
\tablefoot
{
(1) Mass of the simulated molecular cloud.
(2) Smallest cell size of the refined mesh.
(3) Strength of the magnetic field.
(4) Distance to the simulated region.
}
\end{table} 

We performed two hydrodynamical simulations (HD, HD$^\mathrm{h}$) and one magneto-hydrodynamical simulation (MHD) with a
magnetic field that was initially oriented along the \emph{x}-axis. The HD and MHD runs had 7 additional levels of mesh refinement and the
high-resolution HD$^\mathrm{h}$ run used 8 additional refinement levels, corresponding to the maximum resolutions of ${\sim\,}$400 and
200\,au, respectively. The models included sink particles that approximate the evolution of unresolved small scales by an immediate
collapse onto a point mass \citep{bleuler-teyssier14}. These sink particles are disconnected from the hydrodynamical evolution and
interact with the remaining gas through gravity and accretion only. New sink particles form when the volume density exceeds
10$^8$\,cm$^{-3}$ (HD, MHD) or 10$^9$\,cm$^{-3}$ (HD$^\mathrm{h}$) and when the gas inside a small volume around the density peak
undergoes gravitational collapse. With these formation criteria, the sink particles may be considered as individual protostars.
However, the stellar feedback is not included and the models used an isothermal equation of state. The parameters of the
simulations are listed in Table~\ref{t:model-parameters} and a more detailed description of the models is presented in NH19.

For a meaningful comparison of the numerical simulations with real observations, it is necessary to convert their numerical output
into a form that resembles the observations. We isolated the inner parts of the numerical boxes, limited by ${\pm\,}0.5 x_{\rm
max}$, ${\pm\,}0.5 y_{\rm max}$, and ${\pm\,}0.5 z_{\rm max}$, containing most of the dense gas. These boxes were projected onto
the grids of 4096\,$\times$\,4096 pixels for the HD and MHD simulations and 8192\,$\times$\,8192 pixels for HD$^\mathrm{h}$ along the \emph{x}, \emph{y}, and
\emph{z} axes, allowing us to investigate the projection effects (Sect.~\ref{ss:vieweffect}). The pixel units of the resulting
surface density maps were converted from pc to arcsec, using the distances of 1750\,pc for the H and MHD models and 460\,pc for
HD$^\mathrm{h}$, corresponding to \object{NGC\,6334} and \object{Aquila}, respectively. A uniform Gaussian noise was added to the resulting
surface densities, at the levels corresponding to the first quartile of the image pixel distributions
(${\sim\,}$10$^{21}$\,cm$^{-2}$). The maps were smoothed to the 9, 18, 36, and 72{\arcsec} resolutions for the HD, MHD, and HD$^\mathrm{h}$ simulations, which cover the angular resolutions of the observed regions. In total, we obtained 36 synthetic surface density images for the H, MHD, and HD$^\mathrm{h}$ simulations.
The surface density maps projected along the \emph{x} axis for HD and HD$^\mathrm{h}$ are shown in Figs.~\ref{f:cores-projx} and
\ref{f:cores-projx-hr}, whereas the maps for the other projections and for MHD are presented in Appendix~\ref{a:plots}.


\section{Extraction and analysis of sources}
\label{s:analyse}

The column density images of the observed and simulated star-forming regions were given as an input to a source extraction algorithm that produced catalogues of sources, from which we selected self-gravitating sources. In the following, we describe the angular resolution effects on the measured sizes and masses of the selected sources and the resulting source mass function (SMF). We present the results obtained for the observed fields, \object{NGC\,6334} and \object{Aquila}, and we analyse the results for the HD, MHD, and HD$^\mathrm{h}$ simulations to follow the resolution effects beyond the angular resolutions accessible in present-day observations.

\begin{table*} 
\caption{Source extractions in the observed and simulated regions at different angular resolutions.}
\label{t:core-extraction}      
\begin{tabular}{lcccccccc} 
\hline\hline
\noalign{\smallskip}
Extraction&\,\,\,\,\,\,$D$$^{\,(2)}$&\,\,\,\,\,$O$$^{\,(3)}$&\,\,\,\,\,\,$S$$^{\,(4)}$&\,\,\,\,\,\,\,\,$N$$^{\,(5)}$
&\,\,\,\,\,$N_{\rm B}$$^{(6)}$&\,\,\,\,\,$\tilde{M}$$^{\,(7)}$ & \,\,\,\,\,$\bar{M}$$^{\,(8)}$ & \,\,\,\,\,$H$$^{\,(9)}$\\
 & (pc) & ($\arcsec$) & (pc) & & & ($M_\sun$) & ($M_\sun$) & (pc) \\ 
\noalign{\smallskip}
\hline
\noalign{\smallskip}
NGC6334                & 1750         &     \,\,\,18 &          0.15 &                   1100 &           832 &          \,\,\,\,\,\,15 &                    \,\,\,\,\,103 &          0.27 \\
                       &              &     \,\,\,36 &          0.31 &              \,\,\,416 &           375 &          \,\,\,\,\,\,61 &                    \,\,\,\,\,293 &          0.38 \\
                       &              &     \,\,\,72 &          0.62 &         \,\,\,\,\,\,75 &      \,\,\,64 &      \,\,\,423 &     \,\,\,\,\,\,\,\,\,\,\,2352\,$^{(10)}$ &          0.95 \\
\noalign{\smallskip}
\hline
\noalign{\smallskip}
HD\,$^{(1)}$           & 1750         &\,\,\,\,\,\,9 &          0.07 &              \,\,\,178 &                    174 &          \,\,\,\,\,\,37 &                        \,\,\,\,\,\,\,\,54 &          0.11 \\
                       &              &     \,\,\,18 &          0.15 &         \,\,\,\,\,\,95 &               \,\,\,90 &          \,\,\,\,\,\,95 &                             \,\,\,\,\,140 &          0.23 \\
                       &              &     \,\,\,36 &          0.31 &         \,\,\,\,\,\,31 &               \,\,\,29 &               \,\,\,297 &                             \,\,\,\,\,434 & 0.46 \\
                       &              &     \,\,\,72 &          0.62 &         \,\,\,\,\,\,20 &               \,\,\,14 &                    1750 &                                  \,\,2286 &          0.98 \\
\noalign{\smallskip}         
\hline
\noalign{\smallskip}
MHD\,$^{(1)}$          & 1750         &\,\,\,\,\,\,9 &          0.07 &         \,\,\,\,\,\,75 &               \,\,\,69 &          \,\,\,\,\,\,51 &                        \,\,\,\,\,\,\,\,75 &          0.12 \\
                       &              &     \,\,\,18 &          0.15 &         \,\,\,\,\,\,54 &               \,\,\,50 &               \,\,\,112 &                             \,\,\,\,\,179 &          0.24 \\
                       &              &     \,\,\,36 &          0.31 &         \,\,\,\,\,\,30 &               \,\,\,26 &               \,\,\,303 &                             \,\,\,\,\,484 & 0.47 \\
                       &              &     \,\,\,72 &          0.62 &         \,\,\,\,\,\,13 &          \,\,\,\,\,\,9 &                    2516 &                                  \,\,2424 &          1.04 \\
\noalign{\smallskip}
\hline
\noalign{\smallskip}
Aquila                 & 460 &     \,\,\,18 & 0.04 &              \,\,\,801 &           244 &   \,\,\,\,1.2 &                   \,\,\,\,\,\,1.7 & 0.08 \\
                       &              &     \,\,\,36 & 0.08 &              \,\,\,403 &           207 &   \,\,\,\,2.7 &                   \,\,\,\,\,\,4.0 & 0.15 \\
                       &              &     \,\,\,72 & 0.16 &              \,\,\,166 &           142 &   \,\,\,\,5.4 &                   \,\,\,\,\,\,8.7 & 0.27 \\
\noalign{\smallskip}
\hline
\noalign{\smallskip}
HD$^{\rm h}$\,$^{(1)}$ & 460 &\,\,\,\,\,\,9 & 0.02 &     \,\,\,454 &           448 &   \,\,\,\,5.2 &                   \,\,\,\,\,\,6.9 & 0.03 \\
                       &              &     \,\,\,18 & 0.04 &     \,\,\,355 &           346 &        \,11.3 &                        \,\,\,14.8 & 0.06 \\
                       &              &     \,\,\,36 & 0.08 &     \,\,\,186 &           180 &        \,27.5 &                        \,\,\,35.4 & 0.12 \\
                       &              &     \,\,\,72 & 0.16 &\,\,\,\,\,\,57 &      \,\,\,55 &        \,92.9 &                             121.0 & 0.24 \\
\noalign{\smallskip}
\hline
\end{tabular}
\tablefoot
{
(1) Quantities are averaged over the \emph{x}, \emph{y}, and \emph{z} projections. 
(2) Distance to the region.
(3) Angular resolution.
(4) Linear scale corresponding to the angular resolution.
(5) Total number of sources in the extraction catalogues.
(6) Number of selected bound sources.
(7) Median mass of bound sources.
(8) Mean mass of bound sources.
(9) Equivalent half-maximum diameter $(A+B)/2$.
(10) The mass drops to ${\simeq}$\,1585\,$M_\sun$, if the most massive source is ignored.
}
\end{table*} 


\subsection{Identification and selection of sources}
\label{ss:getsf}

To extract sources in the observed and simulated images, we used the new multi-scale source and filament extraction method \textsl{getsf} \citep{sasha21}. The method is the successor of \textsl{getsources}, \textsl{getfilaments}, and \textsl{getimages} \citep[][]{sasha12,sasha13,sasha17} that have been widely used, primarily in the studies of low- and high-mass star formation with \textit{Herschel}. This method decomposes spatially the image(s) to effectively isolate structures of different widths and shapes. It separates the structural components of sources, filaments, and their backgrounds. Having separated the three components, \textsl{getsf} flattens the sources and filaments images to produce the flat detection images with uniform fluctuations over the images. Then it applies thresholding to remove insignificant fluctuations and uses only significant peaks to detect sources. Finally, it measures detected sources in the original images after subtracting their background and after deblending the sources. The \textsl{getsf} method only takes one input parameter: the maximum size of sources of interest, that the user determines from the images. The value of this parameter affects marginally the extraction results. The method is publicly available on its website\footnote{\url{http://irfu.cea.fr/Pisp/alexander.menshchikov/}} and fully described in \cite{sasha21}.

The column density integrated over the source  area at the adopted distance $D$ gives the source mass $M_{\rm S}$. We use the half-maximum sizes $A$ and $B$ (FWHM) to obtain the equivalent spherical radius $R_{\rm S}$ of the source with the volume $\pi A B^2/6$. To determine whether a source may be considered as gravitationally bound (self-gravitating) we followed \cite{konyves15} and computed the ratio $\alpha_{\rm BE}=\frac{M_{\rm BE, cr}}{M_{\rm  S}}$ between the source mass and the critical Bonnor-Ebert mass:
\begin{equation} 
M_{\rm BE, cr} \simeq \frac{2.4 R_{\rm S} c_{\rm s}^2}{G}, 
\label{e:BE}
\end{equation} 
where $R_{\rm S}$ is the source radius, $c_{\rm s}$ the sound speed at $T$=20\,K, and $G$ the gravitational constant. The sources with $\alpha_{\rm
BE}{\,<\,}2$ were considered as self-gravitating, hereafter called \emph{bound} sources for brevity. In the analysis below we consider only the bound sources. In all column density maps (Figs.~\ref{f:ngc}--\ref{f:cores-projx-hr} and \ref{f:cores-projy-projz}--\ref{f:cores-mhd}), bound sources are shown by their ellipses, whose major and minor axes are equal to the source sizes $A$ and $B$ that are estimated by
\textsl{getsf} at half-maximum intensity after subtraction of the sources background. The extraction results for all regions (NGC\,6334, Aquila, simulations HD, MHD, and HD$^\mathrm{h}$) and angular resolutions are summarised in Table~\ref{t:core-extraction}.


\subsection{Angular resolution effects in observations}
\label{sss:obs}

\begin{figure*} 
    \centering
\subfloat{\includegraphics[trim=2cm 3cm 0cm 0cm, width=0.45\linewidth]{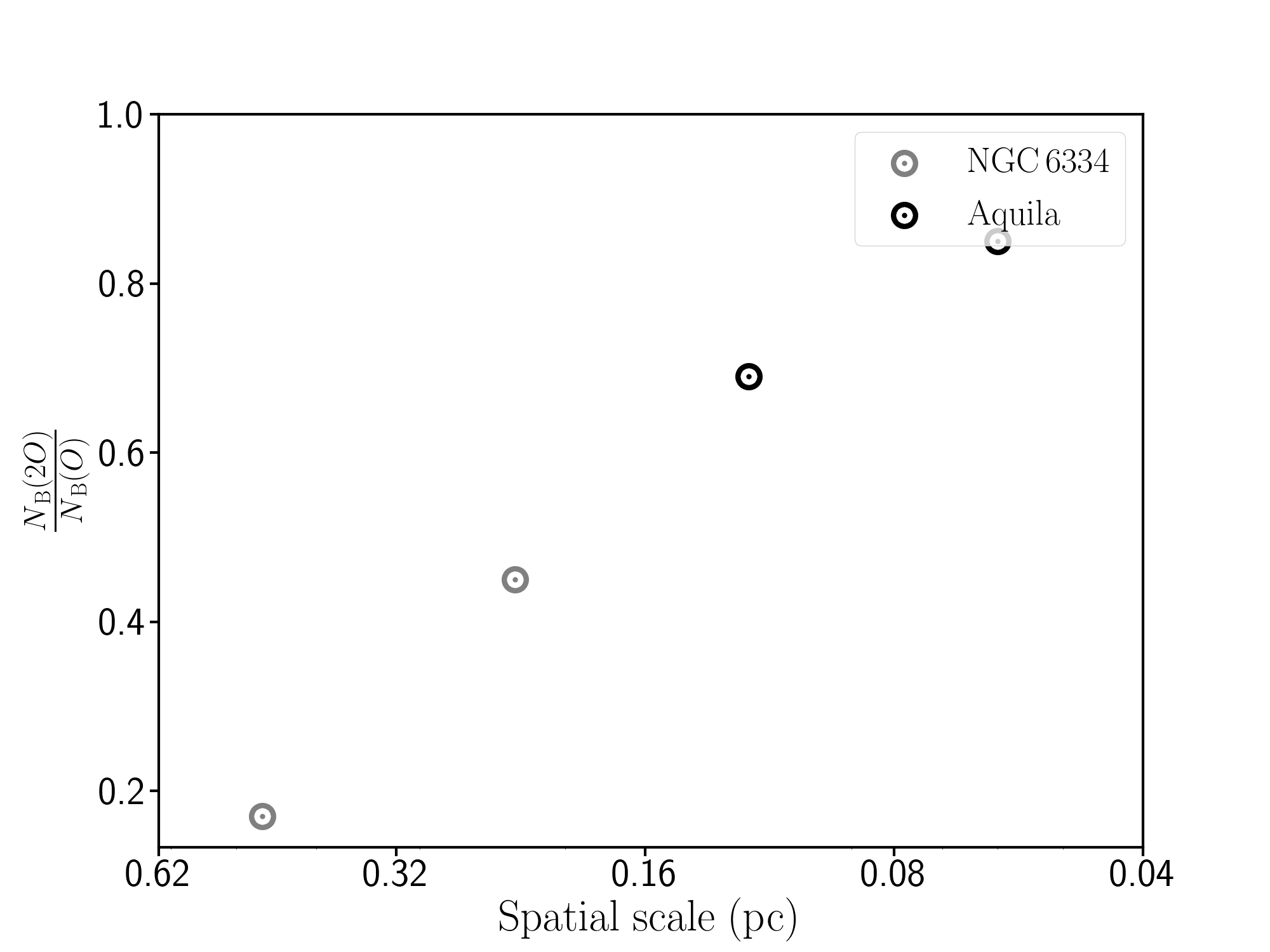}}
\subfloat{\includegraphics[trim=2cm 3cm 0cm 0cm, width=0.45\linewidth]{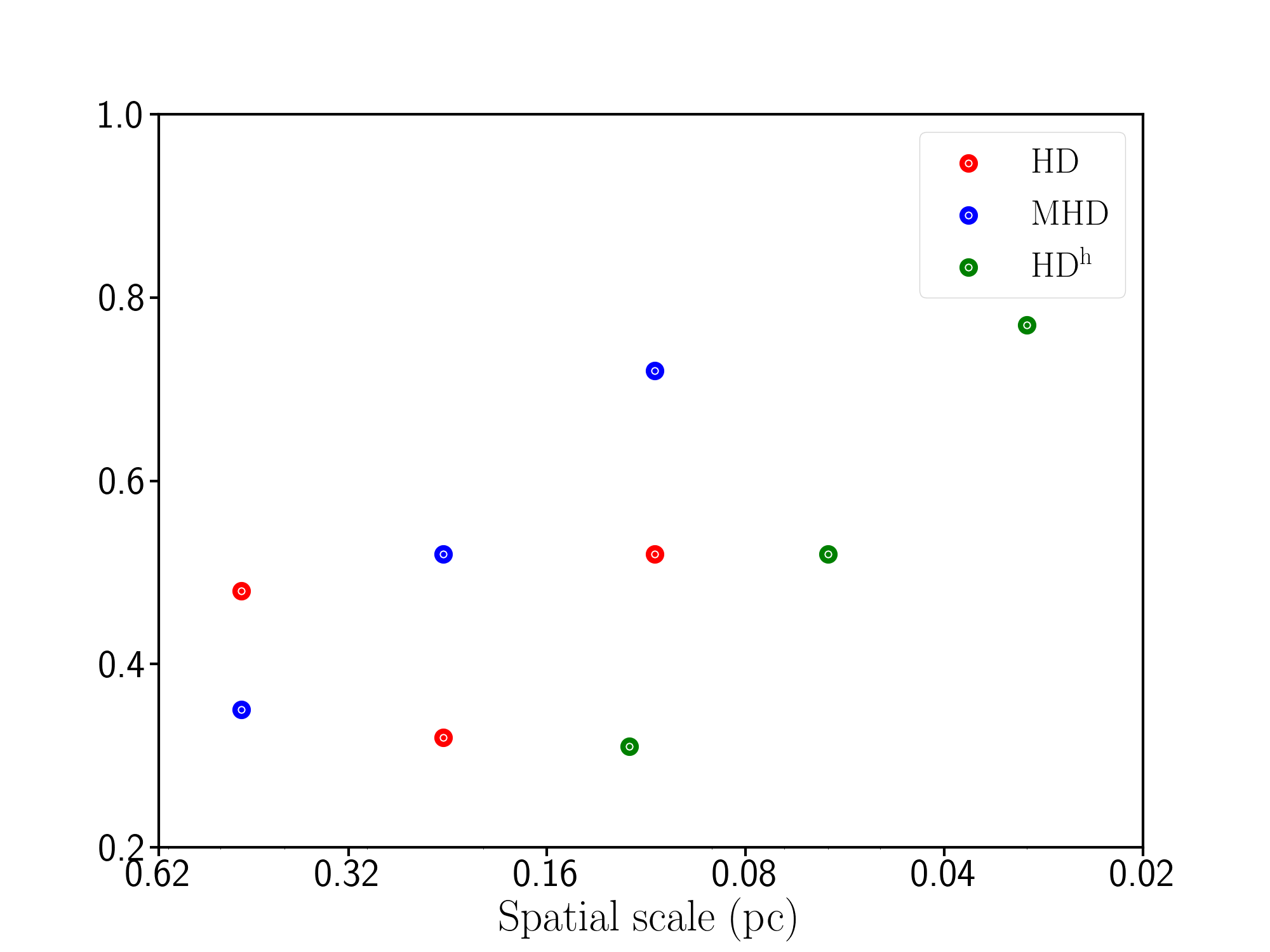}}\\
\subfloat{\includegraphics[trim=2cm 1cm 0cm 0cm, width=0.45\linewidth]{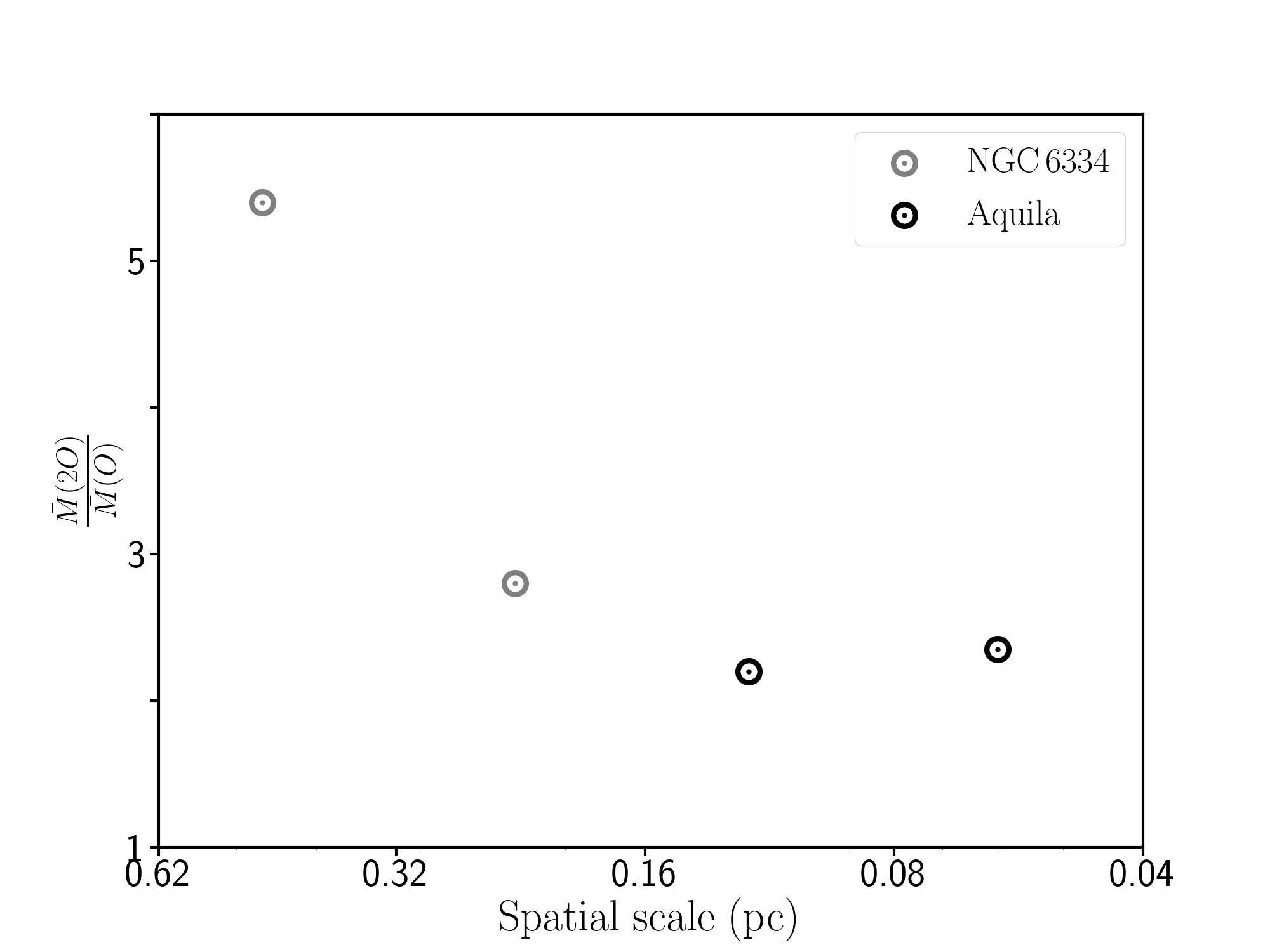}}
\subfloat{\includegraphics[trim=2cm 1cm 0cm 0cm, width=0.45\linewidth]{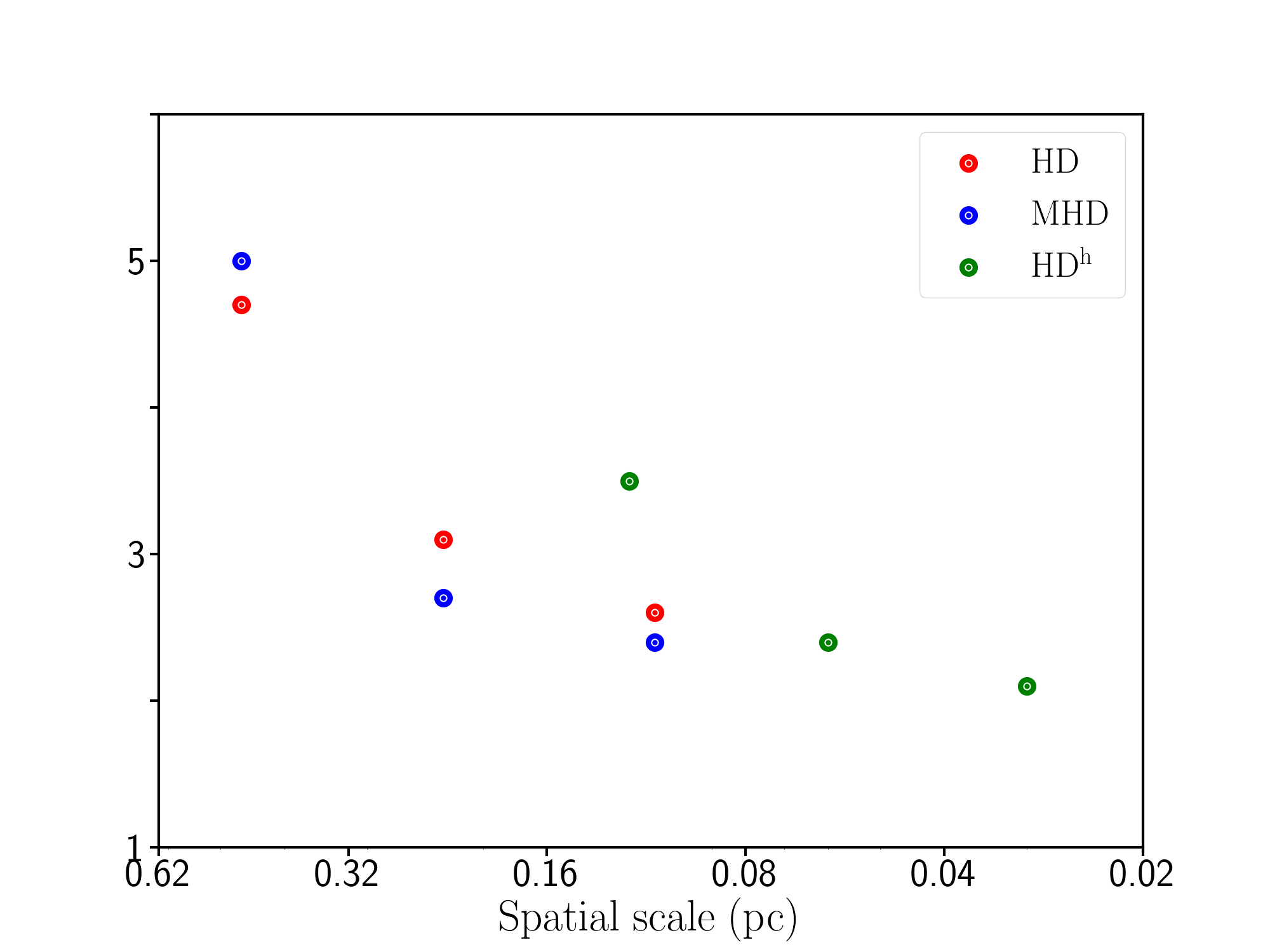}}\\
\caption{
Each data point represents the ratio, $R=F(2O)/F(O)$, at the spatial resolutions $O$ and $2O$ of either the number (top, $F=N_B$) or  mean mass (bottom, $F=\bar{M}$) of sources, and plotted at the mean  spatial resolution  $x=3O/2$. The left panels correspond to the observations of \object{Aquila} and \object{NGC\,6334}. The right panels correspond to the numerical models.
} 
\label{af:ratio}
\end{figure*}

Figures~\ref{f:ngc} and~\ref{f:aquila} show the surface density maps of \object{NGC\,6334} and \object{Aquila} at three angular resolutions, differing by a factor of two. The maps are overlaid with the half-maximum ellipses of the bound sources. In both cases, the coarser the angular resolution, the fewer bound sources are found. Figure~\ref{af:ratio} presents the ratio of bound sources from one scale to the next: $N_B(2O)/N_B(O)$. The fragmentation cascade decreases with increasing spatial resolution for both \object{NGC\,6334} and \object{Aquila}. The effect is the strongest in \object{NGC\,6334}. We argue it is because the physical scales we probe in \object{NGC\,6334} are larger than those we probe in \object{Aquila}.

\begin{figure*} 
    \centering
\includegraphics[trim=0cm 12cm 5cm 0cm, width=1.0\linewidth]{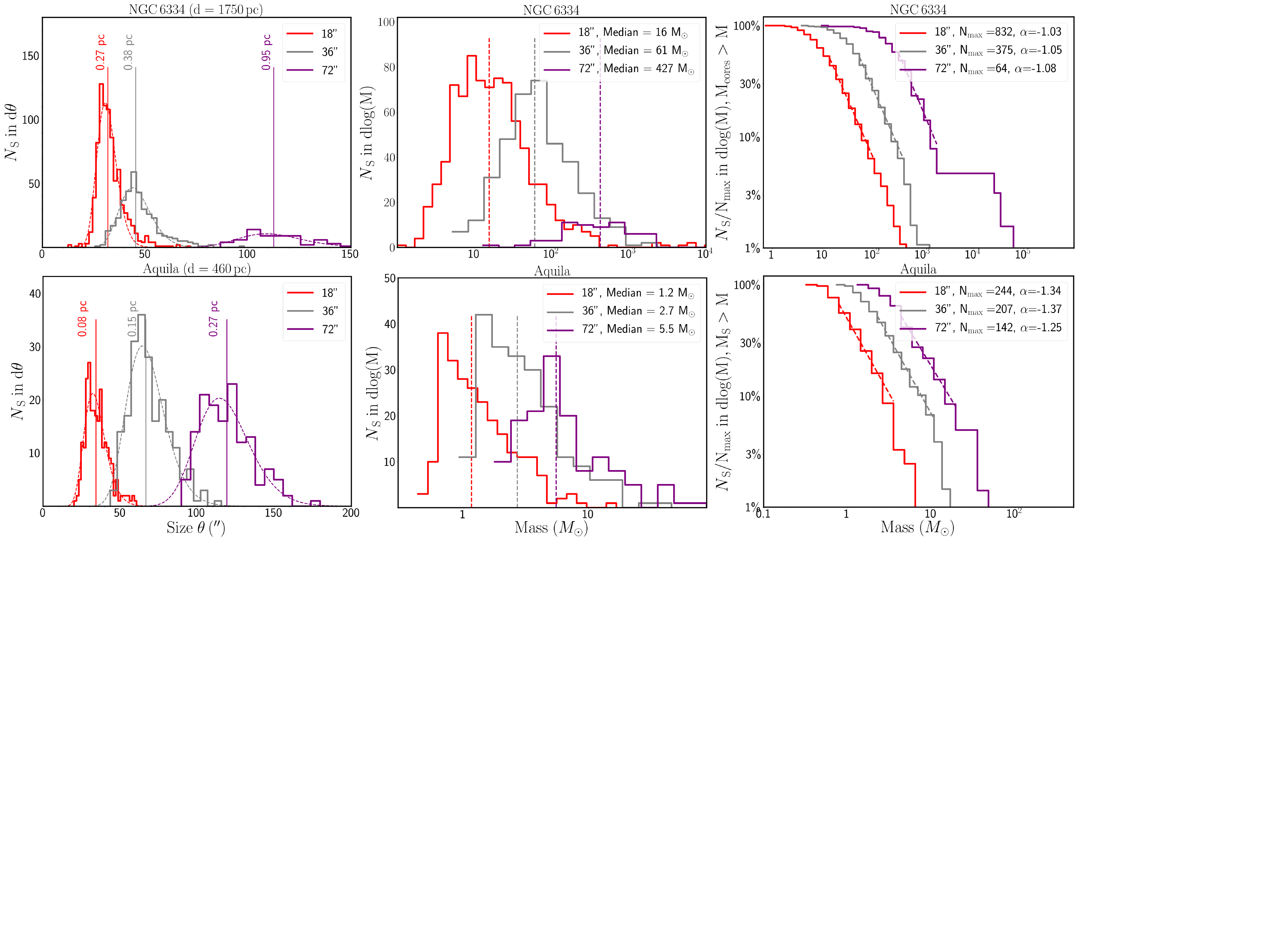}
\caption
{ 
Effects of different angular resolutions on the derived sizes and masses of bound sources in the star-forming regions \object{NGC\,6334} (\emph{top}) and \object{Aquila} (\emph{bottom}). We show the source size function (\emph{left}), the source mass function (\emph{middle}), and the cumulative mass function (\emph{right}). The dashed curves in the left panels are the lognormal fits to the source size distributions and the vertical lines indicate the median values for each distribution. The dashed
lines in the right panels are the fits to the cumulative SMFs within the range 5${-}$50{\%}. For \object{NGC\,6334} (resp. \object{Aquila}), these intervals correspond to 14-107\,\Msun~(resp. 1-4) at 18\arcsec, 54-418 \Msun~(resp. 2-11) at 36\arcsec, and 323-1870 \Msun~(resp. 4-20) at 72\arcsec.
} 
\label{f:reseffect-obs}
\end{figure*}

Figure~\ref{f:reseffect-obs} demonstrates that degrading angular resolution affects the measured sizes of the sources. As a result of the angular resolution being lower by a factor of two, the average size of sources becomes twice larger. The mean size of the sources is of approximately 1.5 times the beam size for the three angular resolutions investigated. The size
distribution is well represented by a lognormal distribution with a standard deviation of ${\sim}$15{\%}
(Fig.~\ref{f:reseffect-obs}), attributable to statistical fluctuations.

The changes in angular resolution also affect the masses of sources. If the surface density were uniform, and without blending of sources, twice larger sizes would imply sources four times more massive. In \object{Aquila}, where decreasing the angular resolution by a factor of two only blends together about 30\,\% of the sources, this degradation of the resolution leads to sources that are, on average, two times more massive, which suggests that sources are centrally peaked. In \object{NGC\,6334}, the downgrading of the resolution by a factor of four produces one extremely massive source
(${\gtrsim\,}5{\,\times\,}10^4$\,$M_{\sun}$) that bias the mean mass estimate. If we ignore this object, then the resolution lowered by successive factors of two corresponds to sources being consecutively three and five times more massive (see Fig.~\ref{af:ratio}). This is a joint effect of the inclusion of the background emission into the source emission (see below) and of the blending of sources. The resolution dependence of the measured source mass is reflected in the derived source mass function: Figure~\ref{f:reseffect-obs} shows a clear shift of the SMF towards higher masses for lower resolutions in both star-forming regions, \object{NGC\,6334} and \object{Aquila}. 

The cumulative form of the SMFs (Fig.~\ref{f:reseffect-obs}) exhibits the same shift towards higher mass for lower angular resolutions. Interestingly, the slope of the high-mass tail of the SMF is almost unaffected by the resolution changes: the best fit of the high-mass tail of the cumulative SMFs of \object{NGC\,6334} is very similar for the three angular resolutions, with $\alpha{\,\simeq\,}{-}1.06{\,\pm\,}0.05$. The slope is flatter than the $\alpha{\,=\,}-1.35$ for the canonical IMF, in line with the recent estimates for high-mass star-forming regions, that give flatter slopes than those for the low-mass star-forming regions \citep[see, e.g.][]{mottenat,liu18, cheng18, sanhueza19, massi19, kong19, servajean19, moser20}. The fits to the cumulative CMFs for \object{Aquila} are also almost invariant with respect to the angular resolution, with an index $\alpha{\,\simeq\,}{-}1.32{\,\pm\,}0.07$ indistinguishable from the slope of the canonical IMF, as previously reported by \cite{konyves15}.

\begin{figure*} 
    \centering
\includegraphics[trim=0cm 6cm 8cm 0cm, width=1.0\linewidth]{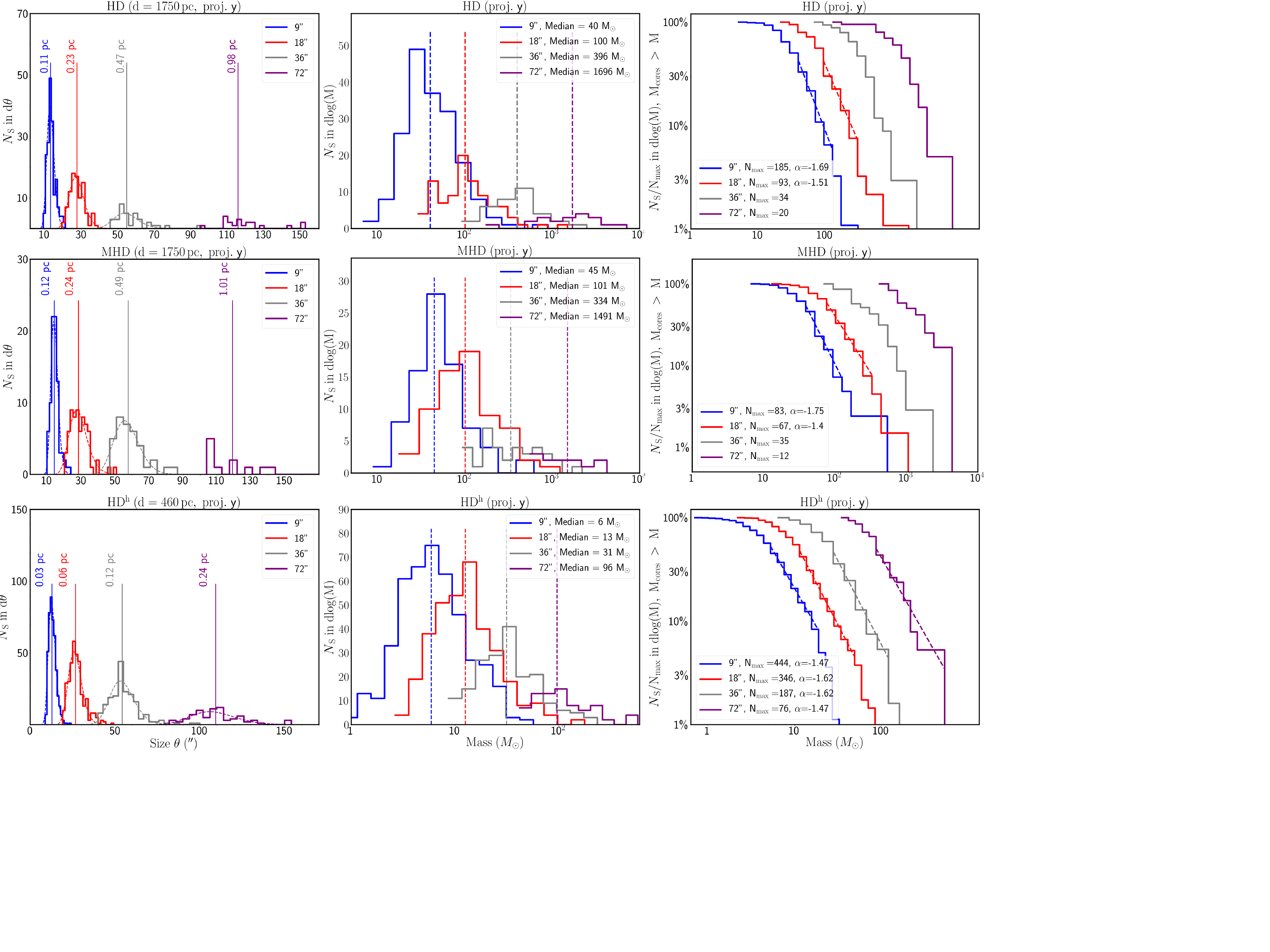}
\caption
{ 
Effects of different angular resolutions on the sizes and masses of bound sources in the simulated star-forming regions HD (\emph{top}), MHD (\emph{middle row}), and HD$^\mathrm{h}$ (\emph{bottom}). We show the source size function (\emph{left}), the source mass function (\emph{middle}), and the cumulative mass function (\emph{right}), obtained for the \emph{y} projection of the respective column density maps. The other projections on the \emph{x} and \emph{z} directions are displayed in Figs.~\ref{f:reseffect-simu-x} and \ref{f:reseffect-simu-z}. The fits of the cumulative SMFs were done only for extractions with more than $40$ bound sources. We fitted the cumulative SMFs in the mass range corresponding to 5-50\% of the source samples; it corresponds to, from top to bottom: 40-128, 39-125, 5-19 \Msun~for the blue curves, 94-304, 76-329, 11-49 \Msun~for the red curves, 28-120 \Msun~for the grey curve, and 87-537 \Msun~for the purple curve.
} 
\label{f:reseffect-simu}
\end{figure*}


\subsection{Angular resolution effects in simulations}
\label{sss:simu}

Figure~\ref{f:reseffect-simu} displays the sizes, masses, and cumulative source mass functions, obtained from the source extractions for the \emph{y} projection of the HD, MHD, and HD$^\mathrm{h}$ simulations; the \emph{x} and \emph{z} projections are presented in Appendix~\ref{a:plots}. Similarly to the observed maps, the number of sources decreases for all simulations when the resolution is degraded by a factor of two. This decrease is all the more pronounced the larger the spatial scales we probe (see table~\ref{t:core-extraction} and Fig.~\ref{af:ratio}).

The simulations show the same behaviour as the observed regions with respect to the measured sizes of the extracted sources. For all angular resolutions, the mean size of sources systematically peaks at approximately 1.5 times the beam size, with a spread that follows a lognormal distribution with a standard deviation of 5{\%} (Fig.~\ref{f:reseffect-simu}). Therefore, the degradation
of the angular resolution by a factor of two leads to sources whose average half-maximum sizes are twice larger. This effect leads to an increase in source masses.

As in the observed regions, downgrading the angular resolution by a factor of two makes the average mass of the extracted sources larger, by a factor of two at spatial scales smaller than 0.03\,pc and up to a factor five a scales greater than 0.3\,pc (see Fig.~\ref{af:ratio}). The smaller statistics at lower resolutions and the increased masses of the sources strongly affect the peak of the SMF, shifting the latter towards higher masses (Fig.~\ref{f:reseffect-simu}). The high-mass slope of the SMF is almost unaffected by the resolution changes (variations within ${\sim\,}$10{\%}) and no clear trends as long as the numbers of sources $N_{\rm S}{\,\gtrsim\,}40$.


\subsection{Viewing angle effects in the simulations}
\label{ss:vieweffect}

To understand whether the viewing angle at which a 3D molecular cloud is observed affects the numbers, masses, and sizes of
the extracted sources, we compared the source extractions for the \emph{x}, \emph{y}, and \emph{z} projections for all simulated regions (HD, MHD, and HD$^\mathrm{h}$) and angular resolutions (9, 18, 36, and 72{\arcsec}). For an illustration, we
selected the HD simulation at the resolution of 18{\arcsec} in Fig.~\ref{f:vieweffect_18as}, well representing the results; the other
simulations and projections are presented in Appendix~\ref{a:plots}.
 
\begin{figure*} 
    \centering
    \includegraphics[trim=0cm 19cm 3cm 0cm, width=1.0\linewidth]{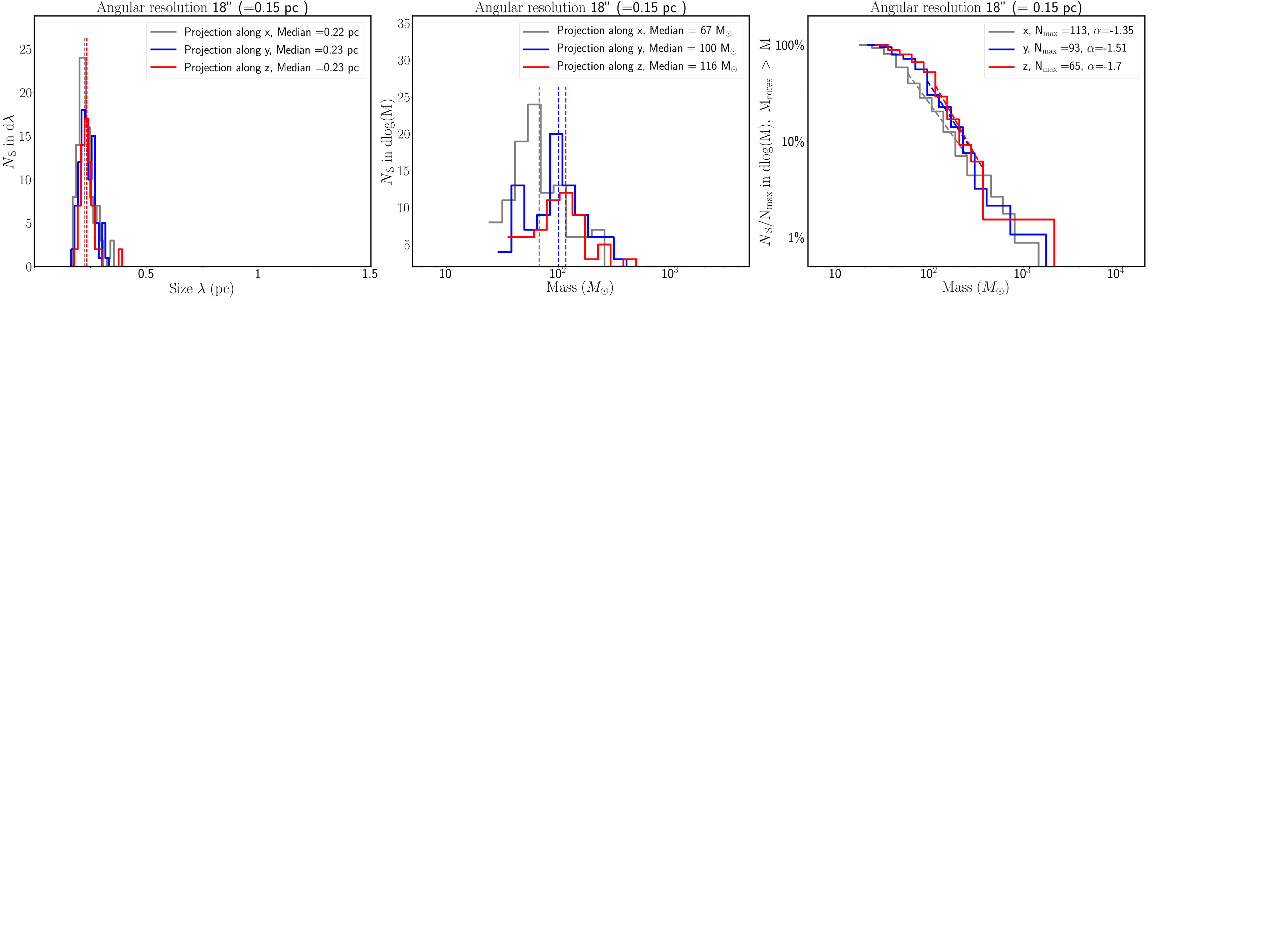}
\caption
{ 
Derived size and mass functions of the bound cores extracted by \textsl{getsf} from the column densities at 18{\arcsec} resolution, obtained from the HD simulation along the \emph{x}, \emph{y}, and \emph{z} axes. The vertical dashed lines indicate the 
median size or mass; the cumulative form of the mass functions were fit by power laws in the range of 5${-}$50{\%}.
} 
\label{f:vieweffect_18as}
\end{figure*}

The half-maximum sizes of the extracted sources show little variations (${<\,}$5{\%}) from one projection to another
(Fig.~\ref{f:vieweffect_18as}), as evidenced by the source size functions (SSFs) in the reference case (HD simulation at
18{\arcsec} resolution). The number of sources and their mean masses vary within $20{-}25${\%} between projections. For the
reference case, we extracted an average number of 90${\,\pm\,}$24 sources with a median mass of 94${\,\pm\,}$25\,$M_{\sun}$. Such
variations induce small displacements between the SMFs of the different projections (Fig.~\ref{f:vieweffect_18as}). A power-law fit
of the high-mass tail of the cumulative form of the SMF (Fig.~\ref{f:vieweffect_18as}) yields the mean slope $\alpha{\,\simeq\,}{-}1.58$ that varies by ${\sim\,}$20{\%} between the projections.

The variations in the numbers of extracted sources and mean masses increase as the resolution degrades. This is a pure statistical
effect caused by the disappearance of unresolved sources, diluted by insufficient angular resolutions (Sects.~\ref{sss:obs} and
\ref{sss:simu}, and Table~\ref{t:core-extraction}). The differences in the properties of sources from one projection to another are
less prominent when considering only the extractions with more than 40 sources. In such extractions, the number of identified bound sources varies by ${\sim}$15{\%}, their mean mass by ${\sim}$20{\%}, and the SMF slope by ${\sim}$10{\%}. It is reasonable to assume that the projection effects are moderate, but not negligible, for statistically significant samples of sources.


\section{Discussion}
\label{s:disc}

Below, we discuss the results of our study, compare them with previous works, and analyse the reasons
behind the very strong dependence of the properties of extracted sources on angular resolution.


\subsection{Comparison between observed and simulated regions}
\label{sss:compa}

Apart from one apparent discrepancy (see below) the effect of angular resolution on the derived properties of bound sources is the same in observations and numerical experiments. Among the resolutions common to the observations and simulations (18, 36, and 72\arcsec), the number of bound sources and their mean size and mass strongly depend on the resolution at which the sources are extracted. As a consequence, the peak of the source mass function is also resolution-dependent, whereas the slope of the SMF is only marginally affected. Interestingly, in the numerical experiments at the highest angular resolution (9\arcsec), exceeding the resolutions probed in the observed regions, we witness the same displacement of the SMFs.
The number of extracted sources and their size and mass distributions respond to the increase of angular resolution in exactly the same way as at lower angular resolutions (Sect.~\ref{sss:simu}).

\begin{figure*} 
    \centering
\subfloat{\includegraphics[trim=3cm 2cm 0cm 5cm, width=0.34\linewidth]{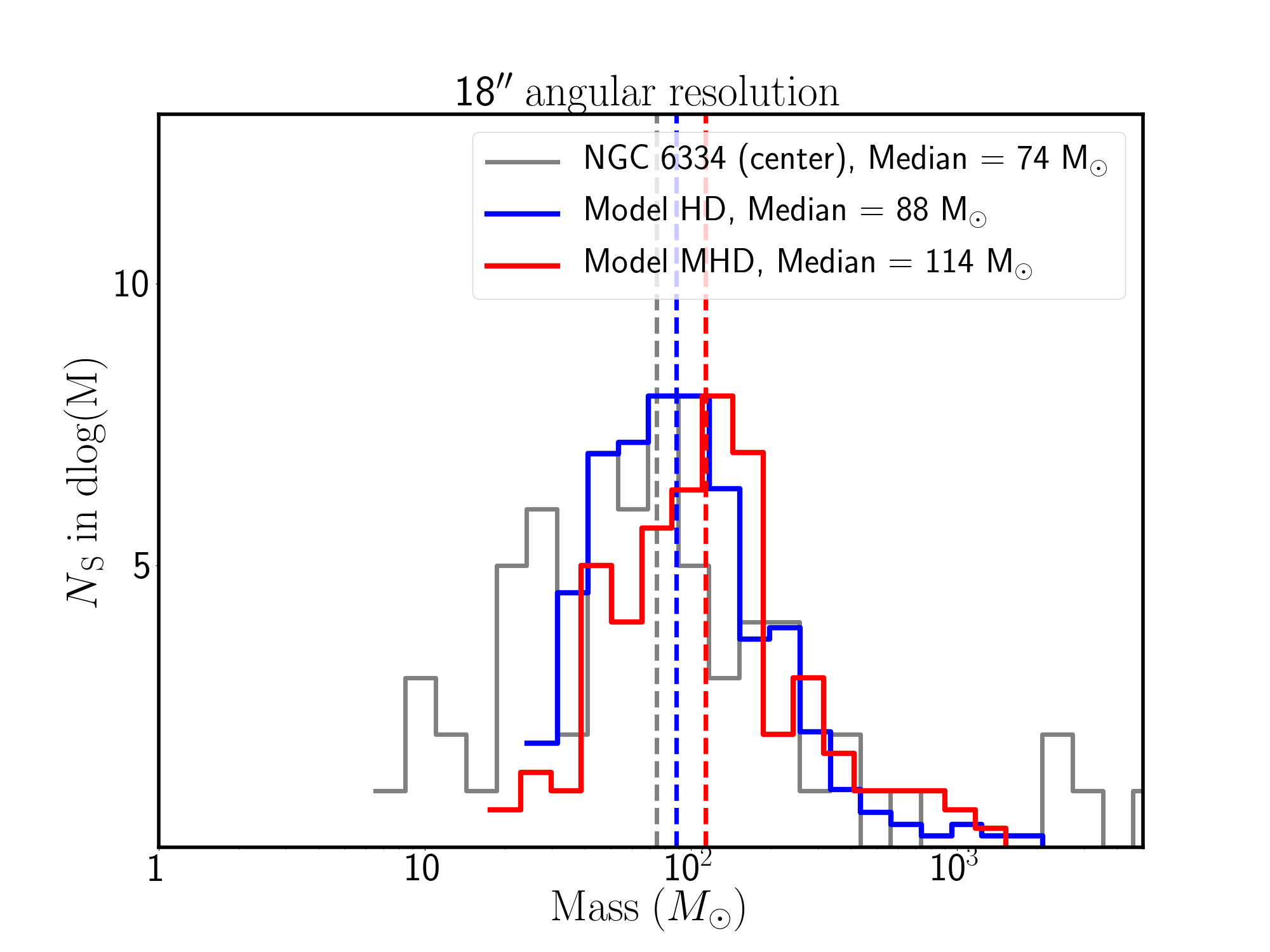}}
\subfloat{\includegraphics[trim=3cm 2cm 0cm 5cm, width=0.34\linewidth]{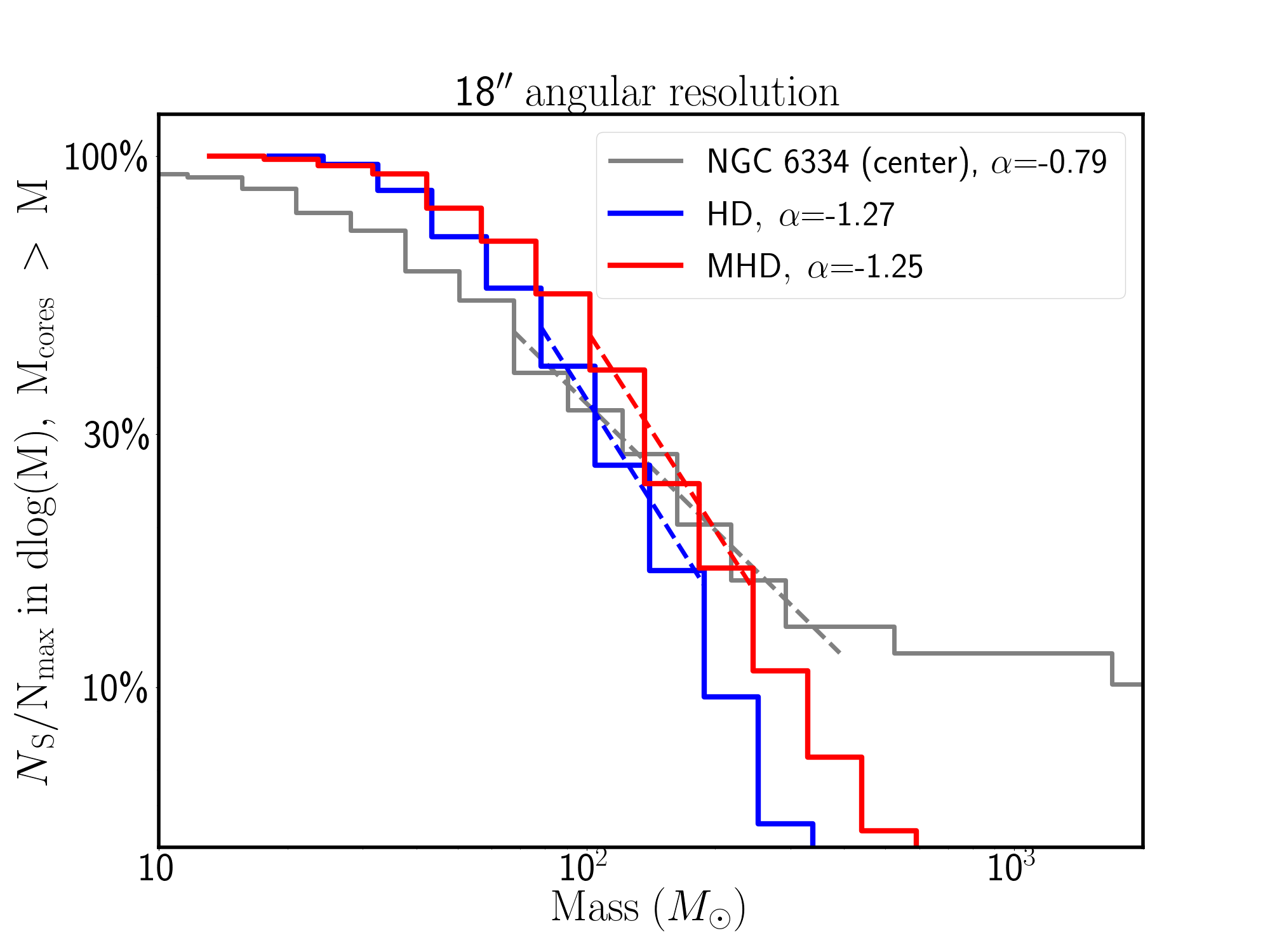}}
\caption
{ 
Comparisons of the mass functions at 18\arcsec of the bound sources extracted by \textsl{getsf} in the star-forming regions \object{NGC\,6334} with those extracted in the HD (blue) and MHD (red) simulations. In case of \object{NGC\,6334}, only the sources from the central region (red box in Fig.~\ref{f:ngc}) were used. The numbers of sources were normalised to the peak of the mass function for \object{NGC\,6334} and the fits of the cumulative mass function were done within the range 10$-$50{\%}. In all model histograms, the numbers of sources were averaged between the \emph{x}, \emph{y}, and \emph{z} projections.
} 
\label{f:compa-obs-simu}
\end{figure*}

Comparing our results for NGC\,6334 with those for HD and MHD at the same angular resolutions, the main difference between the simulated and observed regions is in the source mass. For instance, the median mass of sources in \object{NGC\,6334} at 18{\arcsec}
is 15\,$M_{\sun}$, whereas the median mass in HD and MHD are ${\sim\,}$95 and ${\sim\,}$110\,$M_{\sun}$, respectively (Table~\ref{t:core-extraction}). This difference is mainly caused by the dissimilar sizes of the regions: the observed region is roughly 6 times larger and includes a large fraction of low- and intermediate column densities. Restricting the analysis for \object{NGC\,6334} to its densest region (the red rectangle in Fig.~\ref{f:ngc}) with the same area as in the numerical
experiments, we derive a median source mass of ${\sim\,}$75\,$M_{\sun}$, compatible with those obtained in the simulations
(Fig.~\ref{f:compa-obs-simu}). The high-mass slope of the SMF $\alpha{\,\simeq\,}{-}1$ is estimated in \object{NGC\,6334}, whereas $\alpha{\,\simeq\,}{-}1.35$ is
found in the simulated regions. The difference is even larger, when we restrict the analysis to the central dense area of
\object{NGC\,6334}, thereby removing a large fraction of the low- and intermediate-mass sources from the sample
(Fig.~\ref{f:compa-obs-simu}). Statistically, there is an excess of massive sources in \object{NGC\,6334} with respect to those
found in the HD and MHD simulated regions. This discrepancy might expose missing physics in the simulations, because their initial
conditions match the properties of \object{NGC\,6334} (Sects.~\ref{ss:ngc} and \ref{ss:simus}).

This difference is unimportant for this study, that does not aim at reproducing all details of the observed regions with numerical simulations. We systematically investigated how different angular resolutions affect the derived parameters of extracted sources in the observed and simulated regions across similar physical scales. In that aspect, our results demonstrate full consistency between observations and simulations.


\subsection{Comparisons with previous source extractions}
\label{ss:d:compa}

\cite{tige17} analysed \object{NGC\,6334} during the HOBYS project \citep{motte10} with an effective angular resolution of 18{\arcsec}. Their strategy differs from ours: they detected sources with \textsl{getsources} \citep{sasha12} using the \textit{Herschel} images from $160{-}500$\,$\mu$m plus a column density map to detect sources, and measured the source fluxes from all available data: \textit{Herschel} bands from $70{-}500$\,$\mu$m, JCMT, APEX, SEST, \textit{Spitzer}, WISE, and MSX. Then, they built SED spectrum for each of the 4733 sources \textit{getsources} detected, and applied different criteria to refine the source selection. Therefore the source catalogue comparisons given below must be considered as indicative. They classified the 490 most massive sources, with a median mass of 32\,$M_{\sun}$, which is consistent with the 940 sources of our sample with a median mass of 15\,$M_{\sun}$ taking into account the differences in the two approaches. Indeed, considering only the 490 most massive sources of our sample, the median mass becomes about 30\,$M_{\sun}$.

\cite{konyves15} investigated \object{Aquila} based on \textit{Herschel} observations. They applied \textsl{getsources} to the \textit{Herschel} images at $160{-}500$\,$\mu$m and to a column density image. Several criteria were used to classify the extracted sources between YSOs, starless cores, and protostellar cores. In total, they reported 650 starless cores and 60
protostellar cores in the \object{Aquila} molecular cloud. To associate the sources with the self-gravitating starless cores, they used the ratio $\alpha_{\rm BE}$ between the source mass and the critical Bonnor-Ebert mass (see eq.~\ref{e:BE}). With their condition $\alpha_{\rm BE}{\,<\,}2$ they have found 290 prestellar cores, in good agreement with the 255 sources we found. Besides, they have reported a slope of the high-mass tail of the SMF of $\alpha{\,\simeq\,}{-}1.33$, very similar to the mean slope $\langle\alpha\rangle{\,=\,}{-}1.32$ that we found between the 72 and 18{\arcsec} resolutions (Sect.~\ref{ss:aquila}). We stress, nevertheless, that their detection and measurement strategies differed from ours. They detected the sources using all the individual bands of \emph{Herschel} besides the column density image in the one hand and computed the mass of the sources from SED fitting in the other. Hence the source catalogue comparisons must be considered as indicative
only.


\subsection{Interpretation of the angular resolution effects}
\label{ss:inter}

\begin{figure*}
\centering
\centerline{\resizebox{0.3415\hsize}{!}{\includegraphics{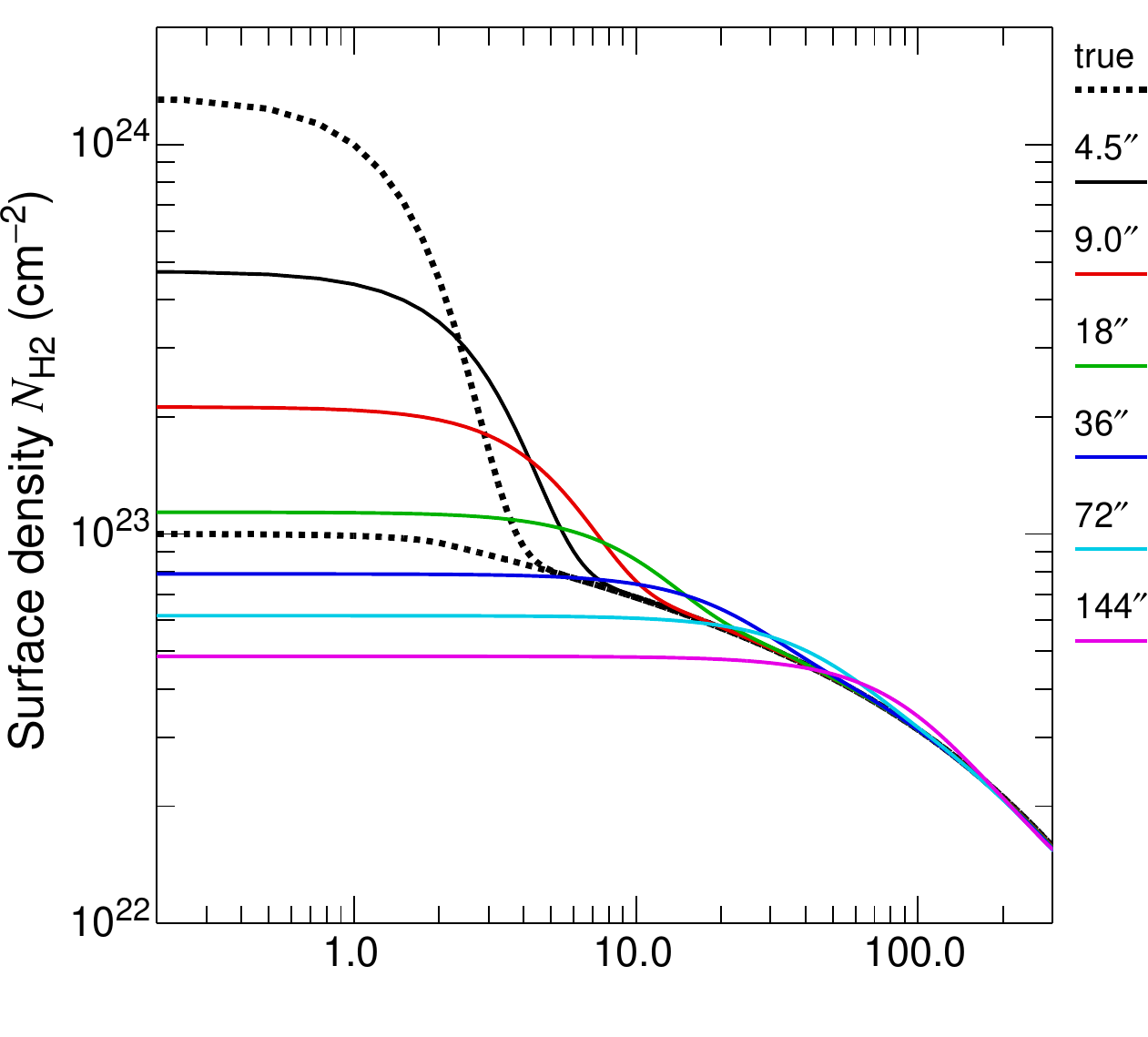}}
            \resizebox{0.3200\hsize}{!}{\includegraphics{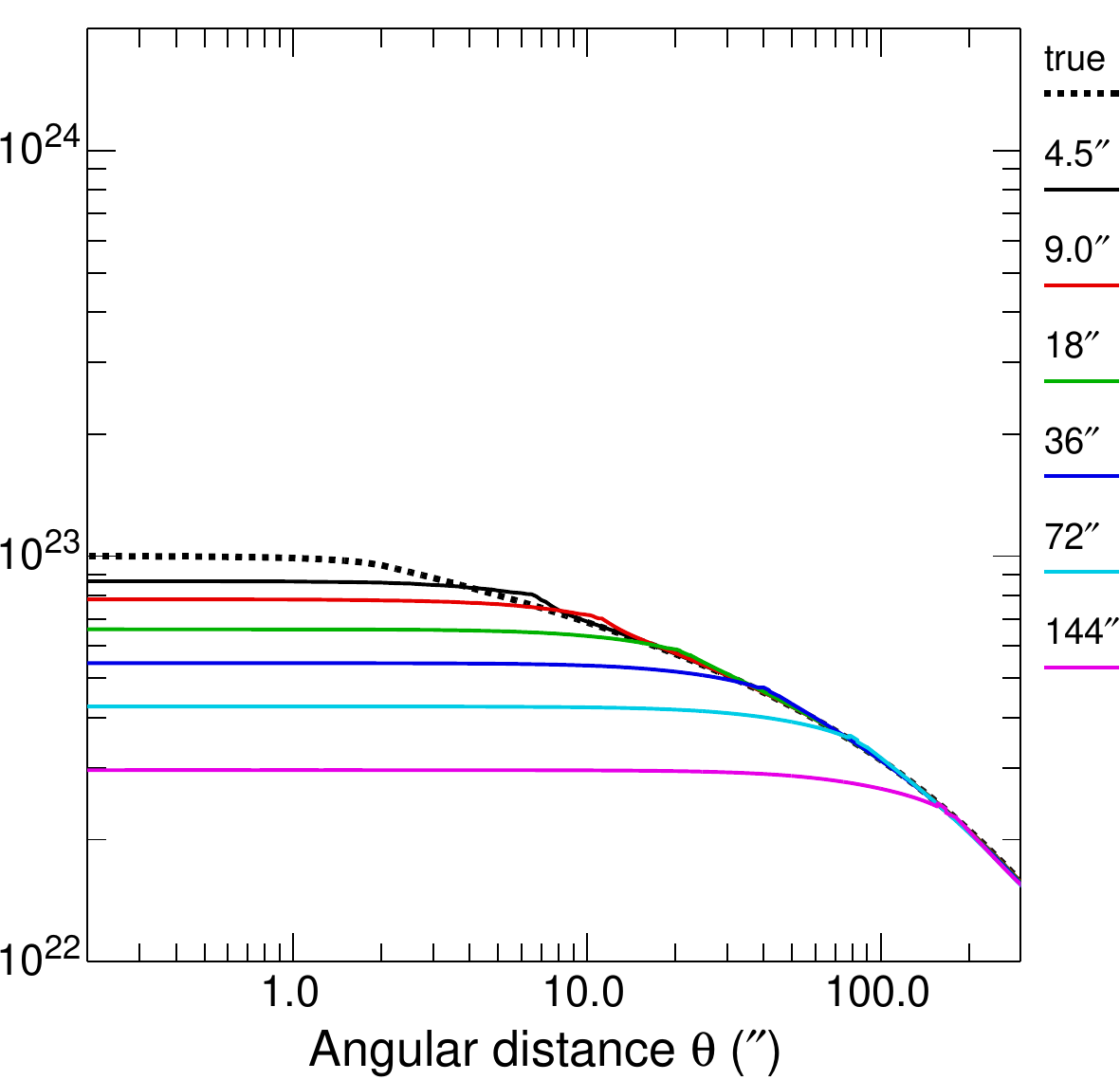}}
            \resizebox{0.3200\hsize}{!}{\includegraphics{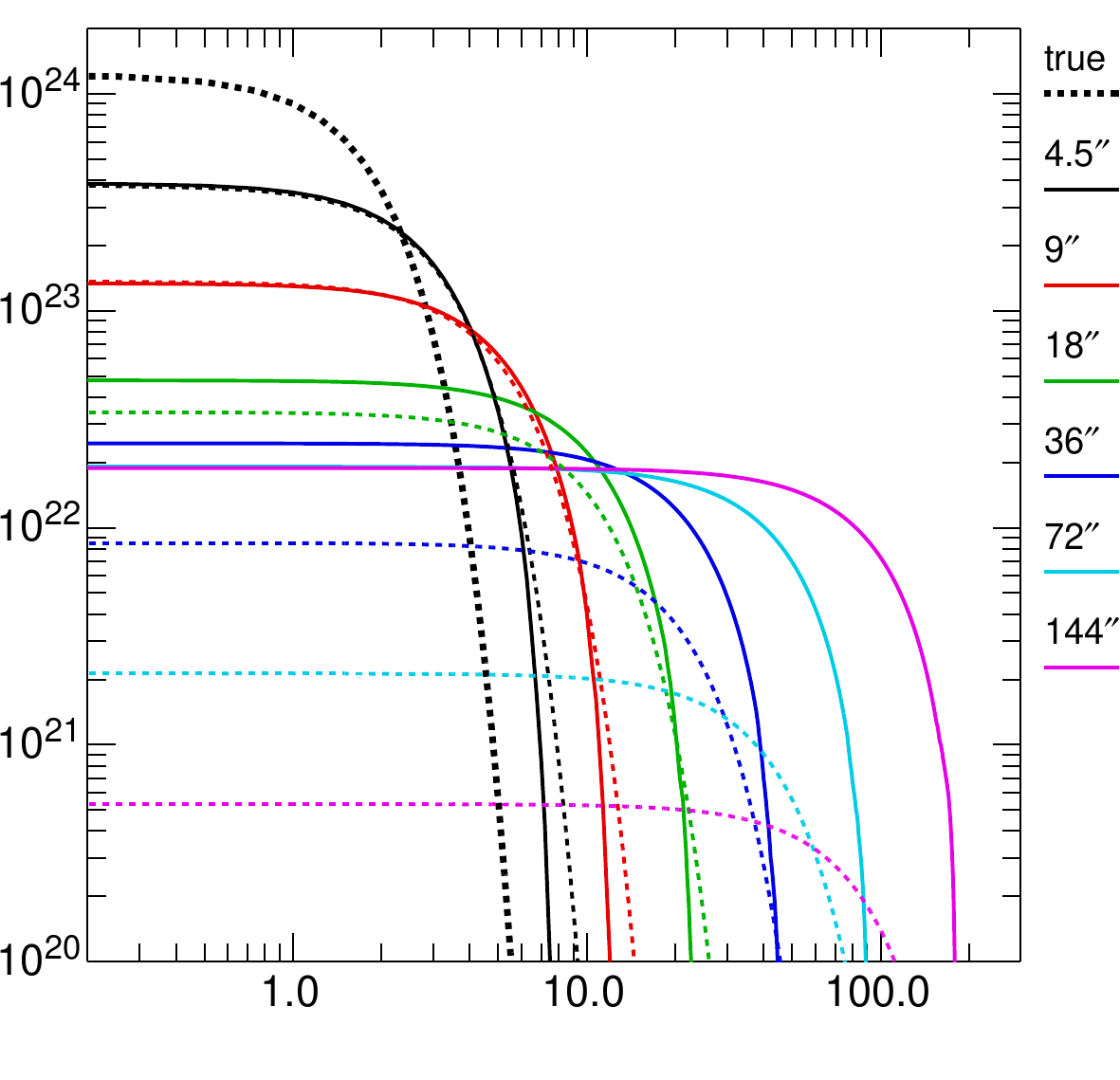}}}
\caption
{ 
Simple model explaining the angular resolution effects on the derived masses and sizes of the sources located on top of filaments or other background fluctuations. The model image, consisting of an unresolved source on top of a
filament, was convolved to different resolutions using Gaussian beams (FWHM), indicated in the panels. Shown are the radial profiles of the model across the filament (\emph{left}), the \textsl{getsf}-interpolated background of the source (\emph{middle}), and the background-subtracted source (\emph{right}). The true radial profile of the model source and filament is shown by the thick dotted line. The background was interpolated within a radius of 1.25 times the resolution, which fully encircles the true source convolved to the resolution. Derived mass of the background-subtracted source increases by the factors 1.08, 1.55, 2.45,
3.48, and 4.12, when degrading the resolution by a factor of two. At each resolution, the measured half-maximum size $(AB)^{1/2}$ of the source is larger than the size of the true convolved source by the factors 1.005, 1.077, 1.13, 1.31, 1.45, and 1.50, correspondingly.
} 
\label{explanation}
\end{figure*}

The results of our systematic investigation of the properties of sources extracted at different angular resolutions may seem surprising, if we do not make a clear distinction between sources of emission, observed with a certain angular resolution, and decoupled core in space that may be contributing to the observed peak. If we follow one of the standard observational practices and equate a bound source whose size is smaller than $\sim$0.05\,pc with the pre- or proto-stellar core, its mass and size should no longer vary when increasing the telescope resolution. However, a careful analysis of the dependencies found in both observed and simulated regions suggests a coherent and natural interpretation\footnote{For brevity, we use the term `emission', although we mostly discuss surface densities, not intensities. We also use the term `background' to denote either structureless or filamentary background of sources.}.

Comparing the images from the highest to the lowest resolution, we see that the number of sources drops faster and faster when the resolution is successively degraded by factors of two (cf. Table~\ref{t:core-extraction}). The sources dilute within the larger beams and their emission merges with the background emission (Figs.~\ref{f:ngc}--\ref{f:cores-projx-hr}, \ref{f:cores-projy-projz}--\ref{f:cores-mhd}). Therefore, the sources extracted at 72{\arcsec} resolution in Aquila or in the model HD$^\mathrm{h}$ contain the emission of their background, and the contribution from the blended clusters of sources identified in the higher-resolution images. 

Comparing the images from the lowest to the highest resolution, we see that the very large sources disappear or split into several sources and that many new sources appear. The appearance of new sources and the splitting of existing sources span from 72 to 18{\arcsec} in both observations and simulations (Figs.~\ref{f:ngc}--\ref{f:cores-projx-hr}, \ref{f:cores-projy-projz}--\ref{f:cores-mhd}). In the numerical simulations, where we also had 9{\arcsec} synthetic observations, this behaviour continues from the 18 to the 9{\arcsec} angular resolution.

Large numbers of filaments in the observed regions and produced by the simulations also experience similar transformations. The filaments are the widest at the 72{\arcsec} resolution and they become increasingly sub-structured into a finer web of narrower filaments at higher angular resolutions. The increasing structural complexity of the filamentary background continues down to the smallest spatial scales we studied: 0.02\,pc in the numerical simulations, 0.04\,pc in \object{Aquila}, and 0.15\,pc in \object{NGC\,6334} (see Fig.~\ref{af:ratio}).

Our results show that the majority of extracted sources appear more extended than the observational beam (by a factor of 1.5). If cores with definite boundaries do exist they must actually be unresolved in all images and for all resolutions, because the average size of the extracted sources closely follows the angular resolution (Table~\ref{t:core-extraction}). If the sources were resolved, their sizes would remain invariant with respect to the increasing resolution. Furthermore, the measured masses of the extracted sources also follow the angular resolution, changing by a factor of 2 even when the number of sources seems to approach convergence (Table~\ref{t:core-extraction}). This means that the integrated emission of a source must contain a major contribution of either the dense and variable background, or of an unresolved cluster of smaller objects, or both. 

The true background under extracted sources is fundamentally unknown: any information about the physical background of an unresolved source is diluted within the observational beam. This leads to major uncertainties in the measurement of source parameters\footnote{\cite{sasha12, sasha13, sasha16, sasha17, sasha21} emphasise the background-related inaccuracies on the basis of the benchmark source extractions with \textsl{getsources} and \textsl{getsf}.}, especially their masses (or integrated fluxes). Self-gravitating prestellar cores form in the dense parts of the molecular clouds that fluctuate on all spatial scales, including the scale of the cores and below. In theory, they are expected to appear in the local density enhancements of the clouds or in dense filaments, hence, the environment of prestellar cores must resemble a hill. Therefore, source extraction tools would underestimate their background, because they interpolate it based on the values just outside the source(see Sect.~\ref{ss:d:algo}). This means that the background-subtracted sources must contain some background contribution that could become quite large for faint unresolved sources. 

We did not find any clear sign of convergence in the mass values across the entire range of resolutions that we investigated (see Fig.~\ref{af:ratio}). Therefore, the currently available angular resolutions are probably insufficient to determine the correct mass of isolated prestellar cores, well decoupled from the background. If such decoupled cores exist in cluster-forming molecular clouds, they must be lighter than what we measured for unresolved sources.

We created a simple model (Fig.~\ref{explanation}) to verify and illustrate our interpretation. An unresolved round Gaussian source with a half-maximum $A{\,=\,}$3{\arcsec}, peak value $N_{\rm H_{2}}{\,=\,}1.2{\,\times\,}10^{24}$\,cm$^{-2}$, and mass $M_{\rm S}{\,=\,}0.13\,M_{\sun}$ was added to a long filament with a crest density of $10^{23}$\,cm$^{-2}$ and a relatively shallow, slowly varying radial profile. The exact profile, peak density, and width of the source are unimportant, provided that it raises above the background by a factor of several and remains unresolved at the 9{\arcsec} resolution. The filamentary shape of the background is unimportant: it may also be modelled as a round peak with a similar radial profile. The model image was convolved with Gaussian beams to the angular resolutions of 4.5, 9, 18, 36, 72, and 144{\arcsec}. For each image with different angular resolution, the source was background-subtracted with the algorithm used by \textsl{getsf} and its half-maximum size and mass were measured. The profiles of the model, of the background, and of the true and background-subtracted source are displayed in Fig.~\ref{explanation}. 

The sloping background of the model source leads to systematic, resolution-dependent inaccuracies of the derived background, which becomes progressively more underestimated towards lower resolutions (by the factors 1.15, 1.28, 1.52, 1.84, 2.35, and 3.37, respectively). The interpolated surface cuts too deep into the true background and, in effect, transfers substantial mass to the unresolved source and leads to a widening of the extracted sources. In the sequence of decreasing angular resolutions, the source becomes more extended than the original model source convolved to the same resolution, by the factors 1.005, 1.077, 1.13, 1.31, 1.45, and 1.50, respectively. At the lower resolutions (36, 72, and 144{\arcsec}), when the unresolved source becomes more strongly diluted within the beam, the factors become very similar to the factor of approximately 1.5 that we found for the observed and simulated regions. In the same sequence of decreasing resolutions, the  extracted source mass becomes 1.007, 1.089, 1.69, 4.14, 14.4, and 59.3 the mass of the source model. From one resolution to the next, it corresponds to a mass increase  by the factors 1.08, 1.55, 2.45, 3.48, and 4.12. The results of our simple model resemble those obtained in this work, telling us that the measured masses of unresolved sources, located on bright fluctuating backgrounds, may be very inaccurate.

This agreement of the toy-model with the results of this work confirms that our interpretation captures the essence of the resolution effects. However, this simple model cannot be considered as fully realistic in view of the presence of a large variety of complex structures and backgrounds in the observed and simulated images. For instance, it does not include the additional effects of unresolved background fluctuations, or blending of clusters into single sources at lower resolutions.


\subsection{Comparison to other extraction algorithms}
\label{ss:d:algo}

We solely used \textit{getsf} (see Sect.~\ref{ss:getsf}) to conduct the source extractions but various algorithms have been used in star-formation studies: \textsl{gaussclumps} \citep{stutzki90}, \textsl{clumpfind} \citep{williams94}, \textsl{dendrograms} \citep{rosolowsky08}, \textsl{cutex} \citep{molinari11}, \textsl{csar} \citep{kirk13}, and \textsl{fellwalker} \citep{berry15}. They employ different approaches: \textsl{clumpfind} and \textsl{dendrograms} analyse isointensity contours in the image; \textsl{cutex} analyses second derivative images to identify peaks; \textsl{csar} and \textsl{fellwalker} associate pixels one by one to local maxima. Two of these algorithms — \textsl{clumpfind} and \textsl{fellwalker} — do not subtract the background from sources. They are therefore irrelevant regarding angular resolution effects and background inclusion into the flux of sources. 

The individual background of sources is unknown and differs for sources lying in different areas of a molecular cloud. Therefore, the background can only be estimated from the pixels outside the source. \textsl{Gaussclumps}, \textsl{csar} and \textsl{dendrograms}\footnote{\textsl{Dendrograms} offers different ways to determine the source flux: either without background subtraction, or subtracting the value at the border of the source (default mode), or extrapolating the source profile down to the zero emission level.} subtract the value at the border of the source. \textsl{Cutex} fits a Gaussian plus an inclined plane to the peak of each source, thereby removing this planar background. \textit{Getsf} interpolates the source background along 4 diagonals linking the pixels just outside the sources, averages the interpolated values and estimates a non-planar background. Background of sources are highly uncertain, progressively more so for fainter sources. The problem is especially serious for embedded sources that are part of a cloud or part of a filament embedded in a cloud. 

The different methods of background subtraction are accompanied by other differences in the extraction methods, the most important of them being deblending of overlapping sources. Only \textsl{cutex} and \textit{getsf} deblend sources, whereas the other methods just partition the image between sources, not allowing them to overlap. These differences would engender dissimilarities in the results, such as the numbers of extracted sources and spurious sources, extraction completeness, inaccuracies in flux and mass measurements. 

Regarding our problem on how angular resolution affects the sources features, we can separate the algorithms in two families: those fitting the emission peaks (\textsl{cutex}, \textsl{gaussclumps}, \textit{getsf}), and those progressively associating the emission to local maxima (\textsl{csar}, \textsl{dendrograms}). For the first family, the smoothing of the map will \textit{i)} artificially increase the sources flux due to the inclusion of background emission (see Sect.~\ref{ss:inter}) and \textit{ii)} reduce the number of detected sources as they dilute into the background. In the second family, sources extend until they reach a user-defined intensity threshold, or until they meet another source.  When the angular resolution degrades, the faint sources will blend into the background. As a consequence, the sources that remain detectable will extend further away. These sources will appear even bigger, and more massive, than with fitting methods. The issue we report, that angular resolution affects source features, is generic, present at the data level, and independent of the source extraction algorithms. 


\subsection{Implications for the studies of star formation}
\label{ss:d:cons}

The results, presented in this paper, imply that the masses of cores, derived for various star-forming regions in the recent years are likely overestimated. There are various consequences of this finding for the standard approach used in the observational studies of star formation. Below, we touch upon some of the important issues.

Our study shows that extracted sources must be unresolved even in the nearby star-forming regions, although the measured sizes seem to indicate the contrary. As shown by our results and explained in Sect.~\ref{ss:inter}, the sources size at 1.5 times the beam is caused by an insufficient angular resolution in the presence of a complex sloping background. This means that the standard practice of determining the sizes of the physical cores by Gaussian beam deconvolution of the measured sizes cannot be applied. For the same reasons, the isolated prestellar cores must have lower masses than those obtained for the extracted sources at a certain angular resolution, if there is no convergence of the physical properties with respect to the angular resolution.

The standard practice of equating the observed sources with individual isolated objects is misleading. On the basis of our results, it makes sense to associate with single isolated prestellar cores only those whose measured properties (size, flux, mass) would remain almost invariant with respect to the angular resolution. 

We propose a practical approach to determine, whether an observed region is affected by angular resolution problems or not, based on a convergence test. Instead of a single extraction of sources in the images, observed with a certain resolution $O$, it would be necessary to perform a series of extractions with the resolutions $O$, $2O$, $4O$, and maybe $8O$ (if possible). The extractions would provide the measurements of source sizes and masses at each resolution and enable a conclusive assessment, whether there is a sign of convergence in the distribution of the measured values. 

With respect to the CMFs measured in various studies in the recent years, our results imply that the true CMF must shift to lower masses. The strong dependence of the masses on the telescope resolution prohibits direct comparisons of unconverged CMF with IMF. When the CMF peak is resolution-dependent, it is impossible to determine the efficiency $\epsilon$ of the mass conversion from the prestellar cores to the newly born stars. Using an SMF obtained for \object{Aquila} with an effective resolution of 18{\arcsec} and equating the extracted sources with the prestellar cores, \cite{konyves15} estimated $\epsilon{\,\simeq\,}$40{\%}. For comparison, our
source extractions done for \object{Aquila} at 36 and 72{\arcsec} resolutions provided SMFs that would yield $\epsilon$ of 25 and 10{\%}, respectively (Fig.~\ref{f:reseffect-obs}).

Our source extractions for the observed star-forming regions \object{NGC\,6334} and \object{Aquila} demonstrate that encapsulated bound sources exist on spatial scales from ${\sim\,}1$ to $0.02$\,pc (see Figs.~\ref{f:ngc}\,\&\,\ref{f:aquila}). These results show that the concept of an isolated core is very questionable and may point towards possible oversimplification of the core collapse models \citep[see
e.g.][]{padoan97,PN02,HC08,hopkins12}, in which the gravo-thermal fragmentation stops near the Jeans length, ${\sim\,}$0.2\,pc for the temperatures
$T{\,\simeq\,}$10\,K and volume densities $n_{\rm H_2}{\,\simeq\,}$10$^4$\,cm$^{-3}$, typical of dense parts of molecular clouds where cores form. The core
collapse model was criticised by \cite{smith09} in their study of the validity of the one-to-one relationship between the CMF and the sink mass function on the basis of numerical models, although they find a statistical relationship between the two mass functions \cite[see also][]{lomax14}. Similarly, \cite{pelkonen21} reported a statistical correspondence between the CMF and the sink mass function, but a weak correlation between the mass of the progenitor core and the final stellar mass. Only one-half of the mass of a low-mass star (${<\,}1$\,$M_{\sun}$) originated from its progenitor and this fraction dropped to
${\sim}$10{\%} for the higher-mass stars ($2\,{\,<\,}M{\,<\,}5$\,$M_{\sun}$).


\subsection{Possible problems}
\label{ss:d:limit}

Our analysis is based partly on numerical simulations. An important aspect of these numerical simulations is that they are scale-free, in the sense that they do not produce structures with certain
distinct spatial scales. It seems to contradict the numerous observations of the filamentary structures that thread the molecular
clouds, reportedly having a width of ${\sim\,}$0.1\,pc \citep[e.g.][]{arzoumanian11,arzoumanian19}. These filamentary structures, in which cores are preferentially found \citep[e.g.][]{andre10}, could host a dominant fraction of the gas of the molecular cloud \citep[e.g.][]{konyves15}. If the temperature and density are about homogeneous in filaments, that would create a population of cores dominating in numbers the
global population of cores, and create a peak in the prestellar CMF at the typical Jeans
mass associated with the physical conditions in marginally critical filaments \citep{andre14,andre19}.
To address this issue it is necessary to observe molecular clouds at higher angular resolution, and question observationally the 
peak of the CMFs reported so far. It may also be necessary to seek for missing physics in the numerical simulations that could 
explain the formation of the filamentary structures with a mean width of ${\sim\,}$0.1\,pc. We note that, so far, all studies aiming at studying filaments failed to reproduce a width of 0.1\,pc over nearly two order of magnitude in column densities \citep{H13,federrath15,ntormousi16,smith14}. This is pointing either towards a missing physical ingredient or observational biases.


\section{Conclusions}
\label{s:concl}

This paper presented a systematic investigation of the relationship between the properties of sources extracted in star-forming regions on one of the most important observational parameters: the angular resolution of the observations. Our analysis of 6 and 36 source extractions in observed and simulated star-forming regions with angular resolutions ranging from 9 to 72{\arcsec} allowed us to establish a clear and coherent pattern in the results. We found that the measured sizes and masses of the sources depend on the resolution, which means that these sources cannot be assigned to individual and well isolated cores — the objects of interest in the studies of star formation.

Our results demonstrate that bound sources are extracted for all adopted resolutions, which correspond to physical scales from 0.6 to 0.02\,pc (or 4000\,au). The sources remain unresolved in all the regions and at all resolutions; their average half-maximum size closely follows the angular resolution. The average mass of bound sources also scales with angular resolution. It increases by a factor greater than two when the angular resolution doubles when it causes blending of sources. At higher spatial resolution, where the blending of sources is limited, the average mass still increases by a factor of about two when the angular resolution is degraded by a factor of two. We interpret our findings as caused by the underestimated background of unresolved sources observed against the sloping, hill-like backgrounds of the fluctuating molecular clouds. We do not see any sign of convergence of the sizes and masses to their resolution-independent values. Therefore we conclude that isolated prestellar cores, if they exist in cluster-forming molecular clouds, must be significantly less massive and smaller in size than the values obtained in the measurements of the sources extracted at 0.04\,pc.

As a consequence, the peak of the source mass functions (SMFs) shifts towards lower masses when the angular resolution increases. By contrast, the slope of the high-mass tail of the SMFs remains almost invariant with respect to angular resolution. The near invariance of the high-mass
slope of the SMF may be explained by the fact that with varying beams we probe different scales of the same background with the same scale-free properties on all scales.

Our systematic study has various implications for studies of star formation. It demonstrates that the implicit assumption that measurements of observed sources give masses of isolated cores is invalid; a clear distinction must be made between sources and objects. In our study, all the sources we probed from 0.6\,pc to 0.02\,pc seem resolved, with a typical size near 1.5 the beam size, but do not correspond to single and coherent objects. In such configuration, the standard approach determining the sizes of the physical cores by Gaussian beam deconvolution of the measured source sizes cannot apply. Finally, with the resolution dependent CMF peaks, it is impossible to determine a constant efficiency of the mass conversion from prestellar cores to the stars. We propose a convergence test to determine whether an observed region is affected by angular resolution problems.


\begin{acknowledgements}
This research is funded by the Marie Curie Action of the European Union (project \textit{MagiKStar}, Grant agreement number 841276).
This work was supported by the Programme National de Physique Stellaire and Physique et Chimie du Milieu Interstellaire (PNPS and PCMI) of CNRS/INSU (with INC/INP/IN2P3) co-funded by CEA and CNES. This research used data from the Herschel Gould Belt survey (HGBS) project (http://gouldbelt-herschel.cea.fr). The HGBS is a Herschel Key Programme jointly carried out by SPIRE Specialist Astronomy Group 3 (SAG 3), scientists of several institutes in the PACS Consortium (CEA Saclay, INAF-IFSI Rome and INAF-Arcetri, KU Leuven, MPIA Heidelberg), and scientists of the Herschel Science Center (HSC). FKL thanks Philippe Andr\'e for numerous and in-depth discussions, Vera K\"{o}nyves for facilitating the comparison of our extractions in Aquila, No\"e Brucy for his help in handling the RAMSES outputs, and Bilal Ladjelate for his advises about the treatment of the column density map of NGC\,6334.
\end{acknowledgements}

\bibliographystyle{aa} 
\bibliography{cmf.bib}


\begin{appendix} 
\section{Angular resolution and projection effects}
\label{a:plots}

\begin{figure*}[!h]
\centering
\centerline{\resizebox{\hsize}{!}{\includegraphics{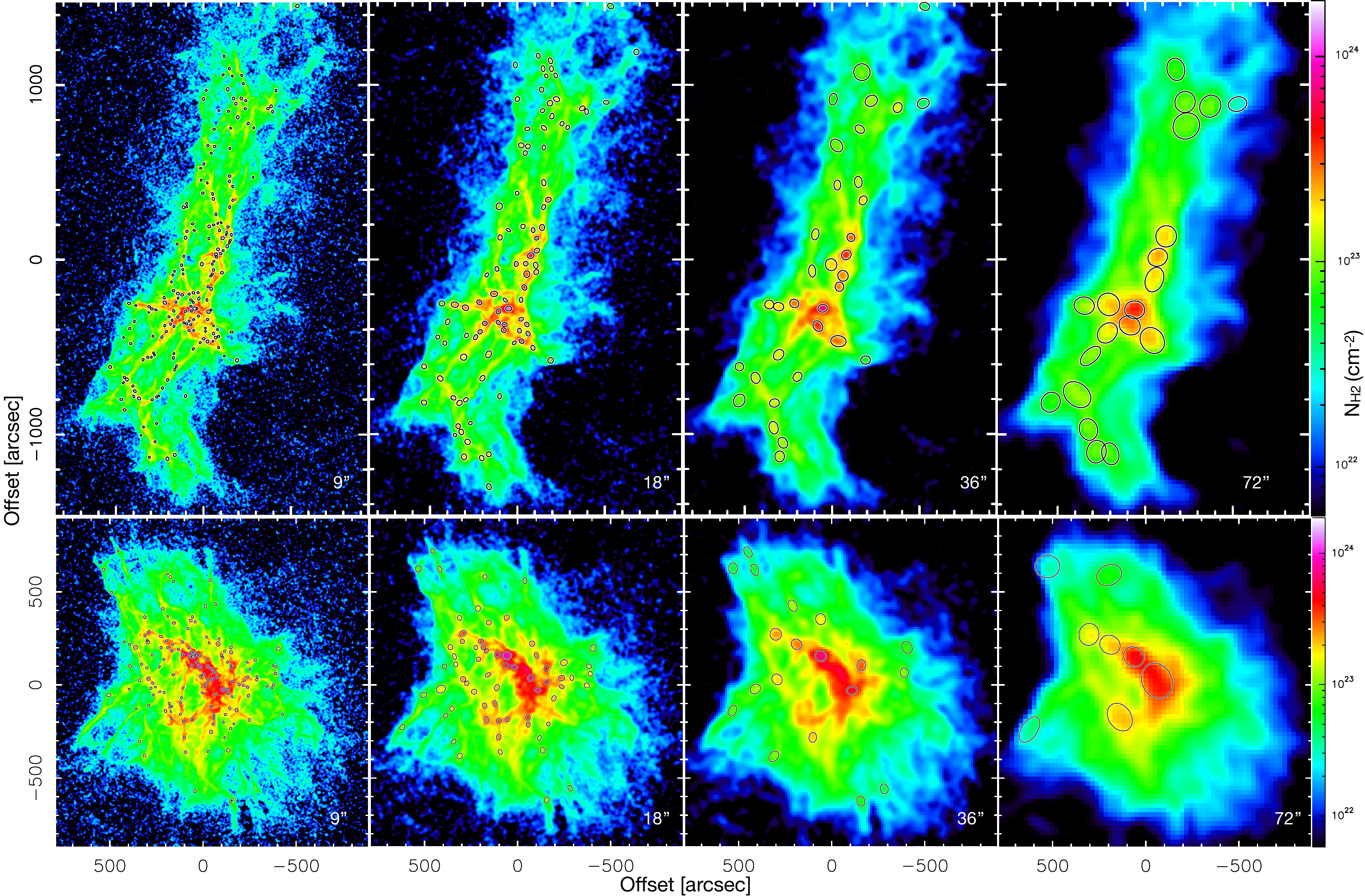}}}
\caption
{ 
Column densities in the HD simulation projected along the \emph{y} axis (\emph{top}) and \emph{z} axis (\emph{bottom}), with the
angular resolution indicated in the panels, overlaid with half-maximum ellipses representing the bound sources extracted by
\textsl{getsf}. The density projection along the \emph{x} axis is shown in Fig.~\ref{f:cores-projx}.
} 
\label{f:cores-projy-projz}
\end{figure*}

\begin{figure*} 
\centering
\centerline{\resizebox{\hsize}{!}{\includegraphics{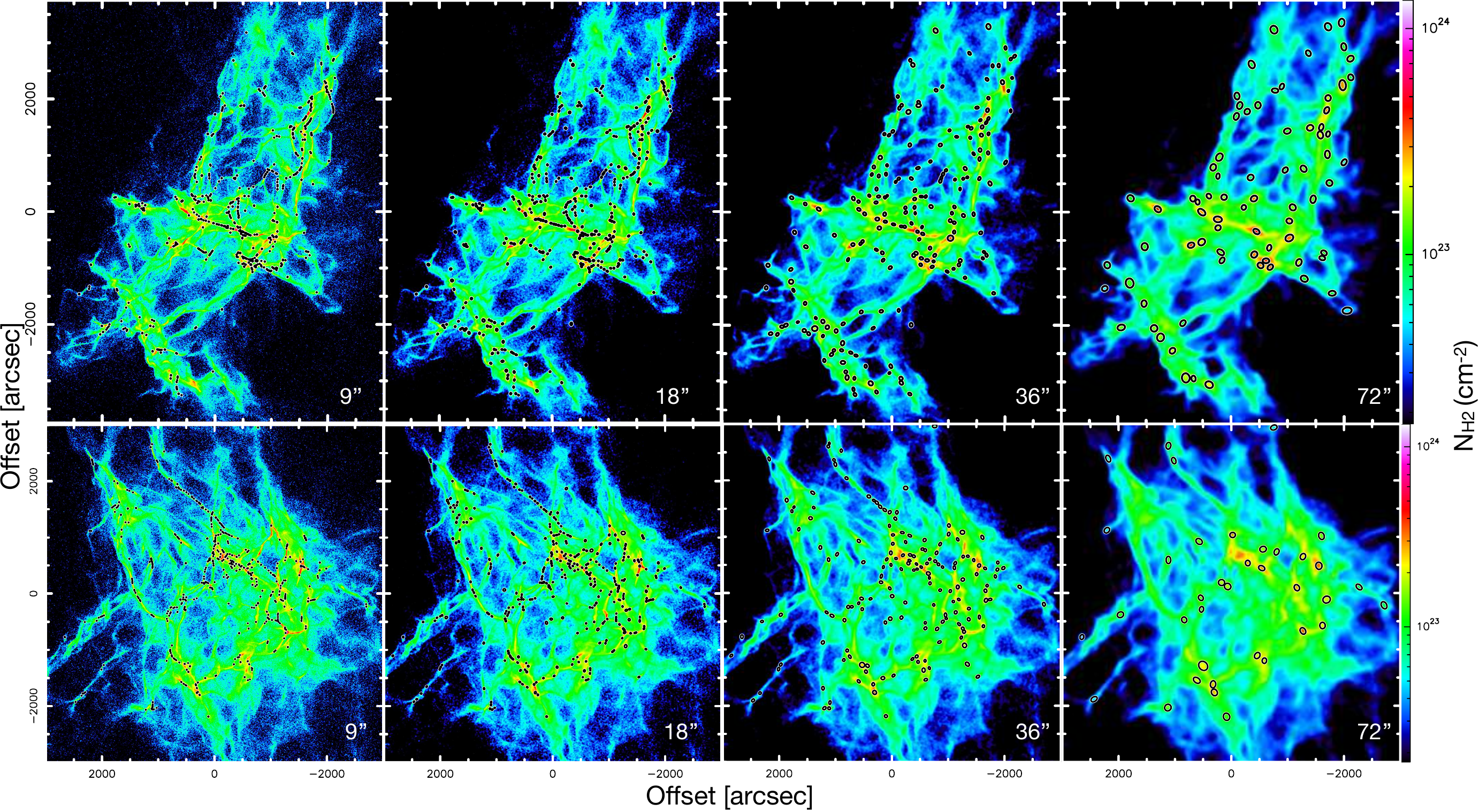}}}
\caption
{ 
Column densities in the HD$^\mathrm{h}$ simulation projected along the \emph{y} axis (\emph{top}) and \emph{z} axis
(\emph{bottom}), with the angular resolution indicated in the panels, overlaid with half-maximum ellipses representing the bound
sources extracted by \textsl{getsf}. The density projection along the \emph{x} axis is shown in Fig.~\ref{f:cores-projx-hr}.
} 
\label{f:cores-projy-projz-hr}
\end{figure*}

\begin{figure*} 
\centering
\centerline{\resizebox{\hsize}{!}{\includegraphics{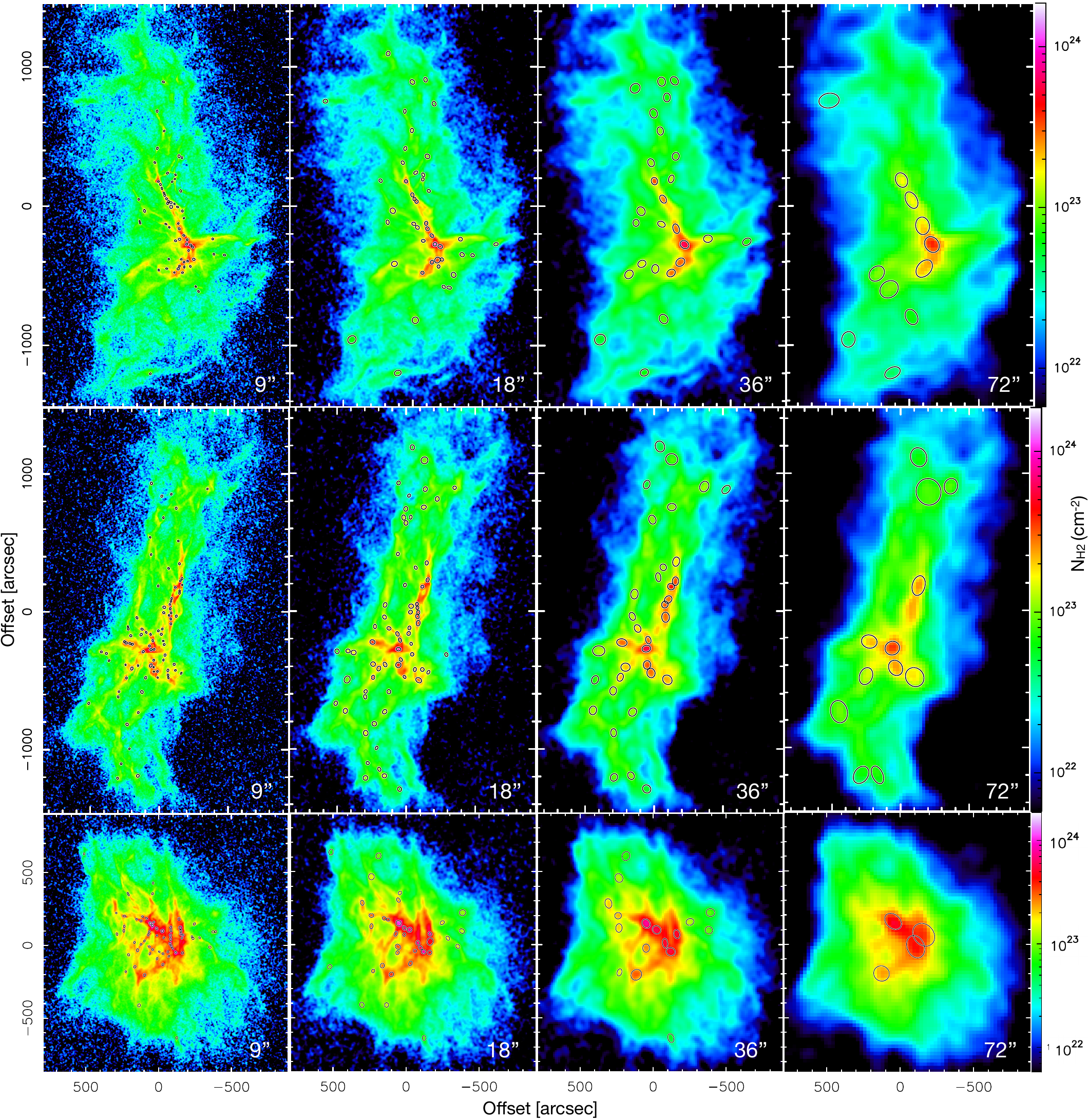}}}
\caption
{ 
Column densities in the MHD simulation projected along the \emph{x} axis (\emph{top}), \emph{y} axis (\emph{middle}) and \emph{z}
axis (\emph{bottom}) with the angular resolution indicated in the panels, overlaid with half-maximum ellipses representing the
bound sources extracted by \textsl{getsf}. The linear scales, corresponding to the resolutions, are 0.07, 0.15, 0.31, and 0.62\,pc.
} 
\label{f:cores-mhd}
\end{figure*}

\begin{figure*} 
    \centering
    \subfloat{\includegraphics[trim=3cm 0cm 0.8cm 5cm, width=0.34\linewidth]{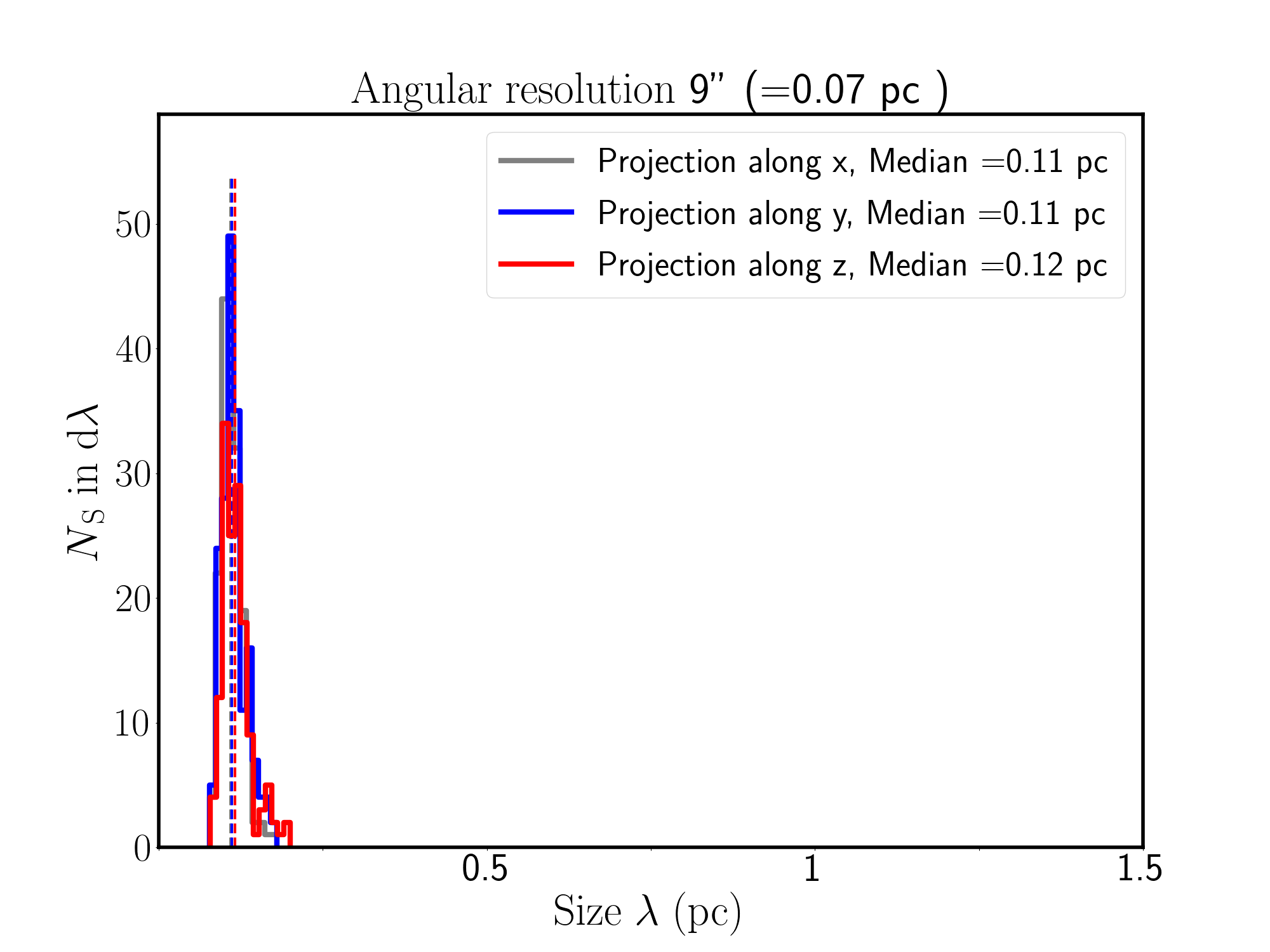}}
    \subfloat{\includegraphics[trim=3cm 0cm 0.8cm 5cm, width=0.34\linewidth]{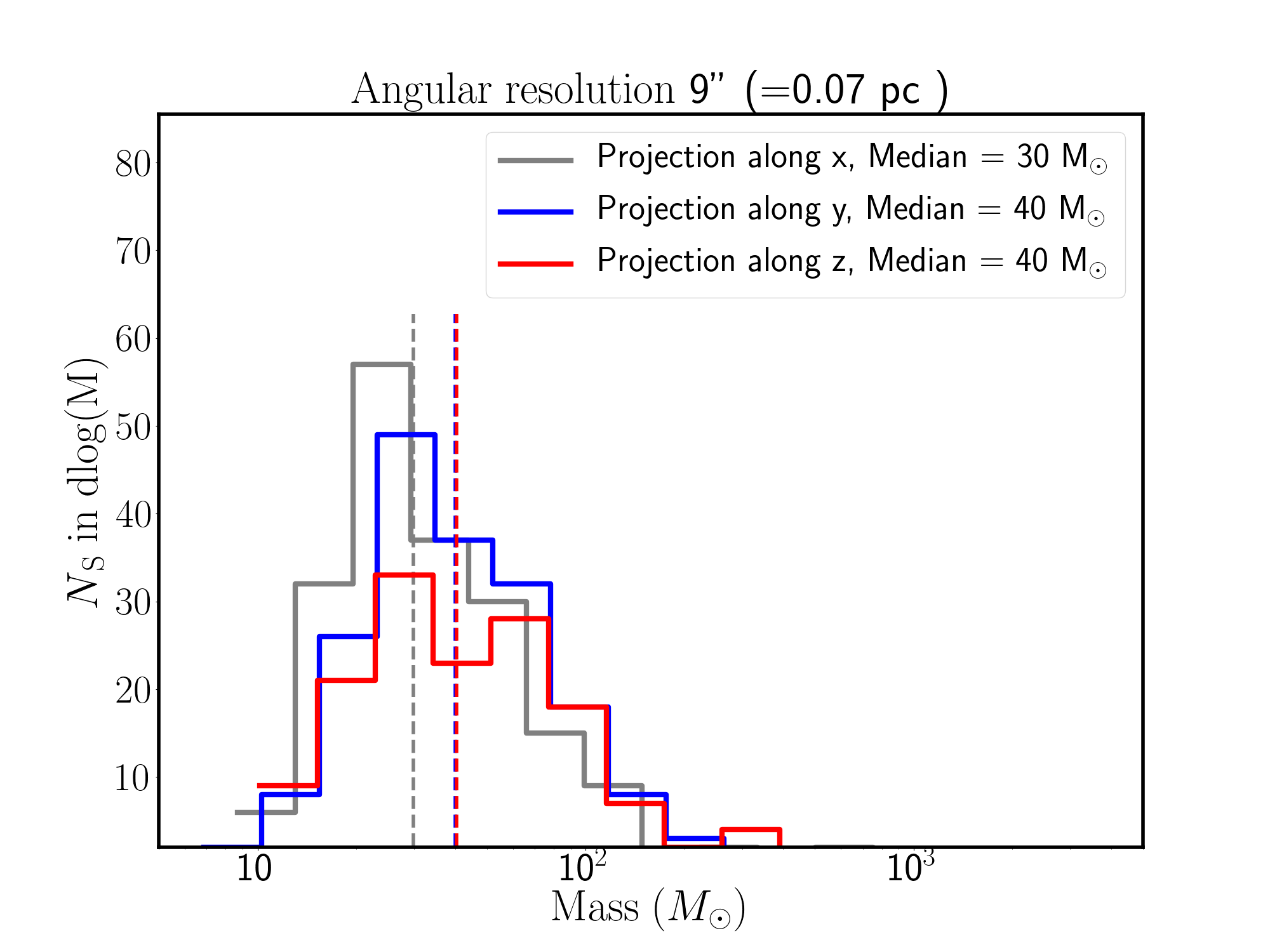}}
    \subfloat{\includegraphics[trim=3cm 0cm 0.8cm 5cm, width=0.34\linewidth]{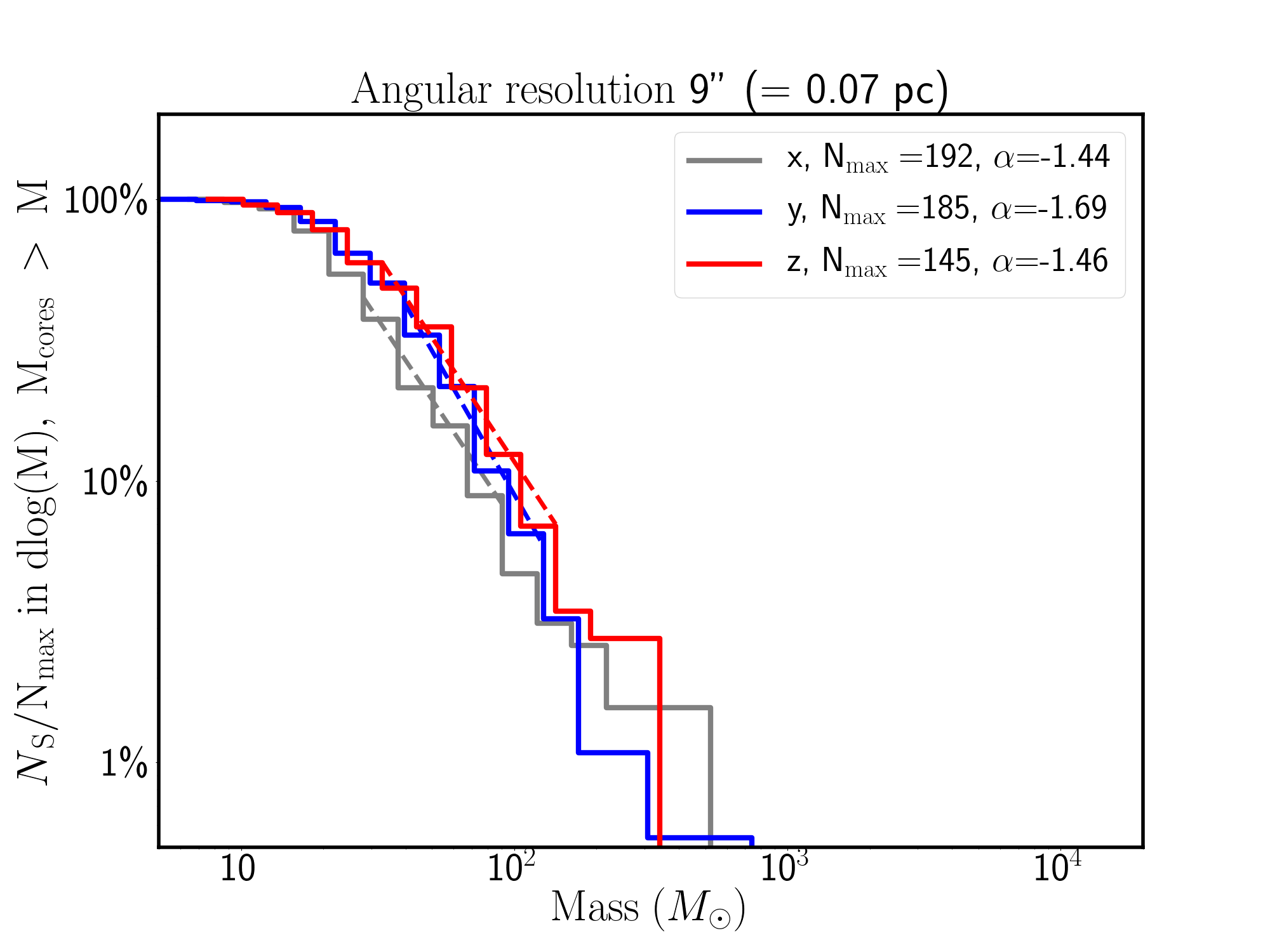}}\\
    \subfloat{\includegraphics[trim=3cm 0cm 0.8cm 5.5cm, width=0.34\linewidth]{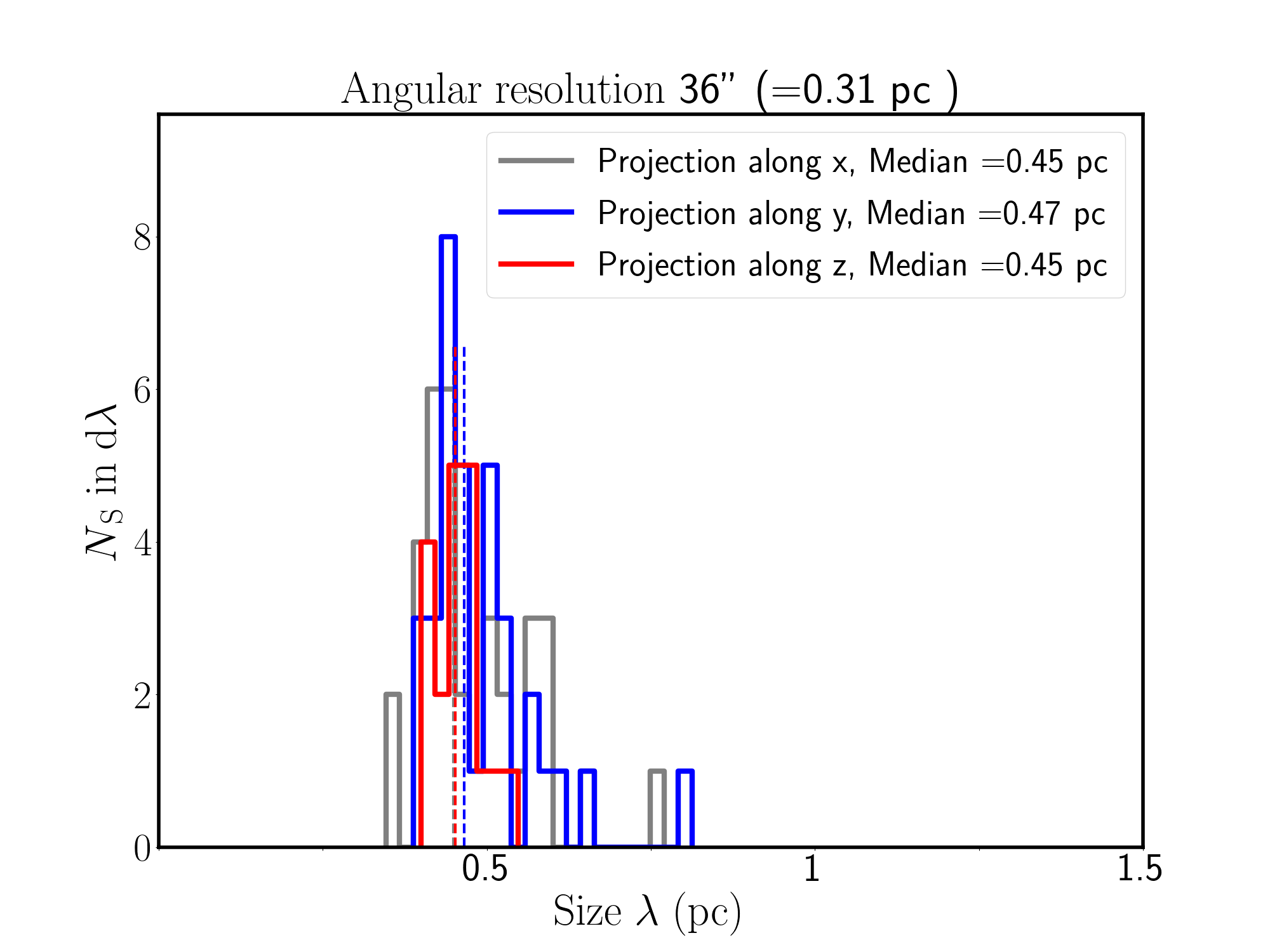}}
    \subfloat{\includegraphics[trim=3cm 0cm 0.8cm 5.5cm, width=0.34\linewidth]{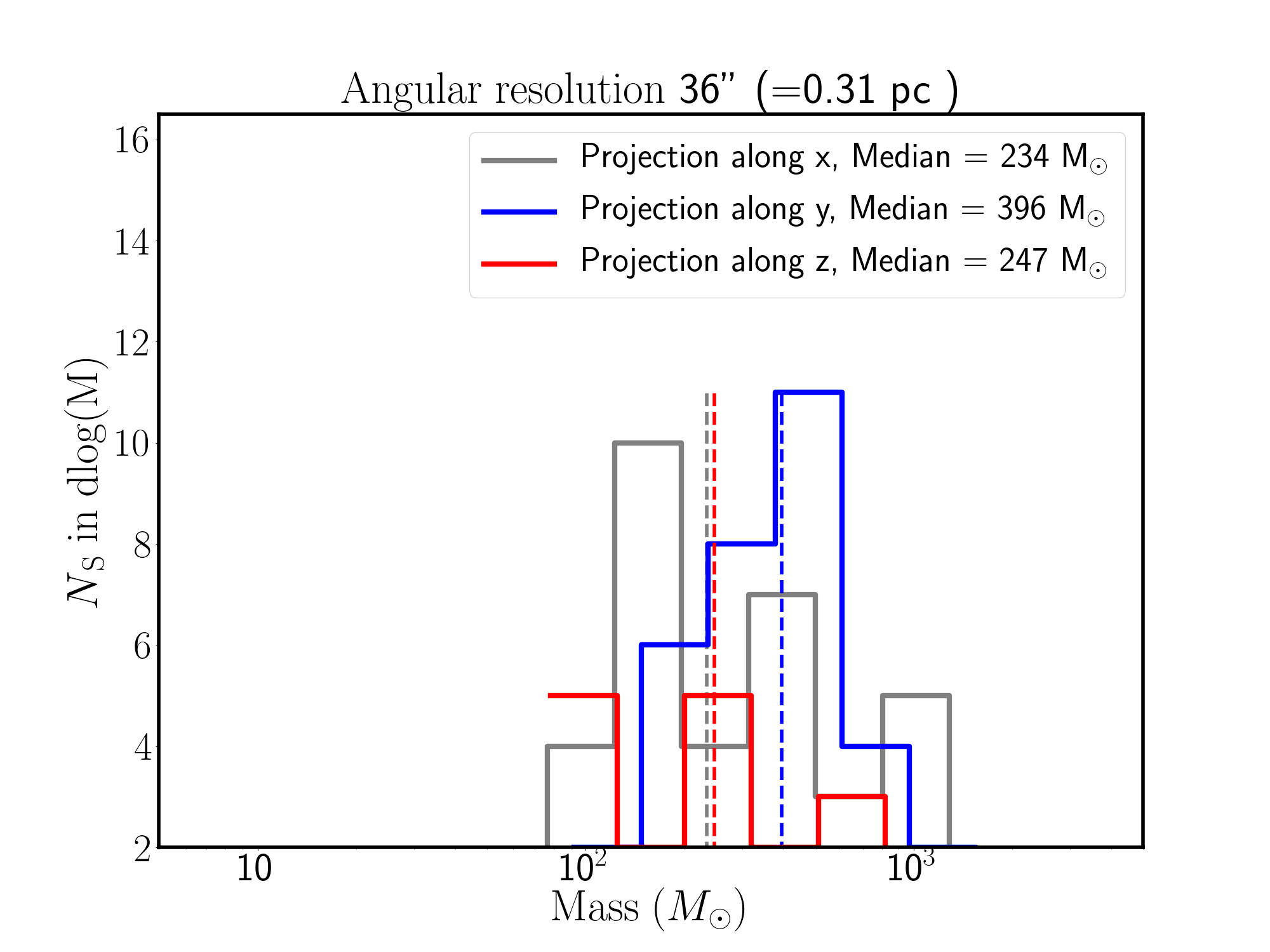}}
    \subfloat{\includegraphics[trim=3cm 0cm 0.8cm 5.5cm, width=0.34\linewidth]{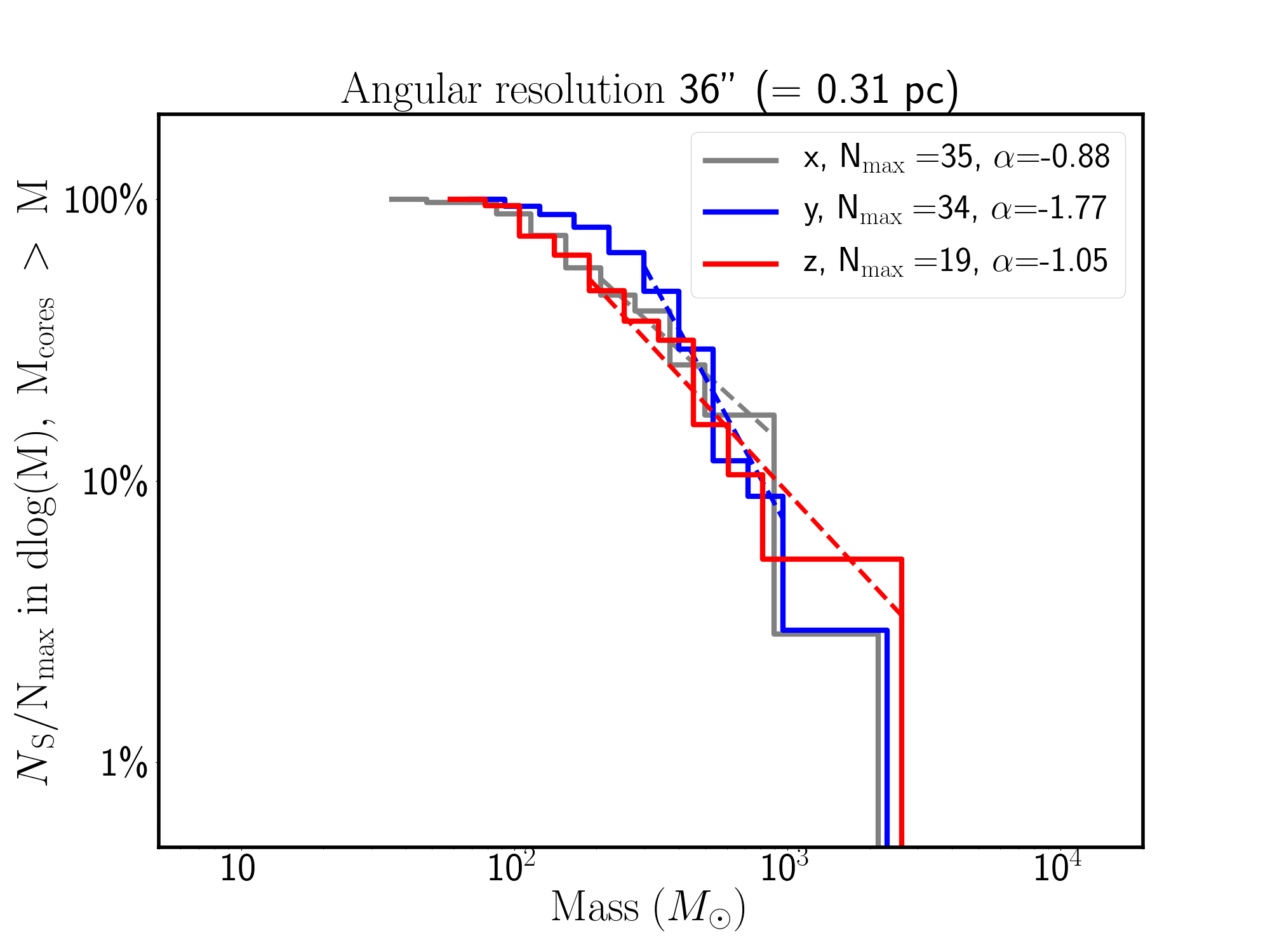}}\\
    \subfloat{\includegraphics[trim=3cm 0cm 0.8cm 5.5cm, width=0.34\linewidth]{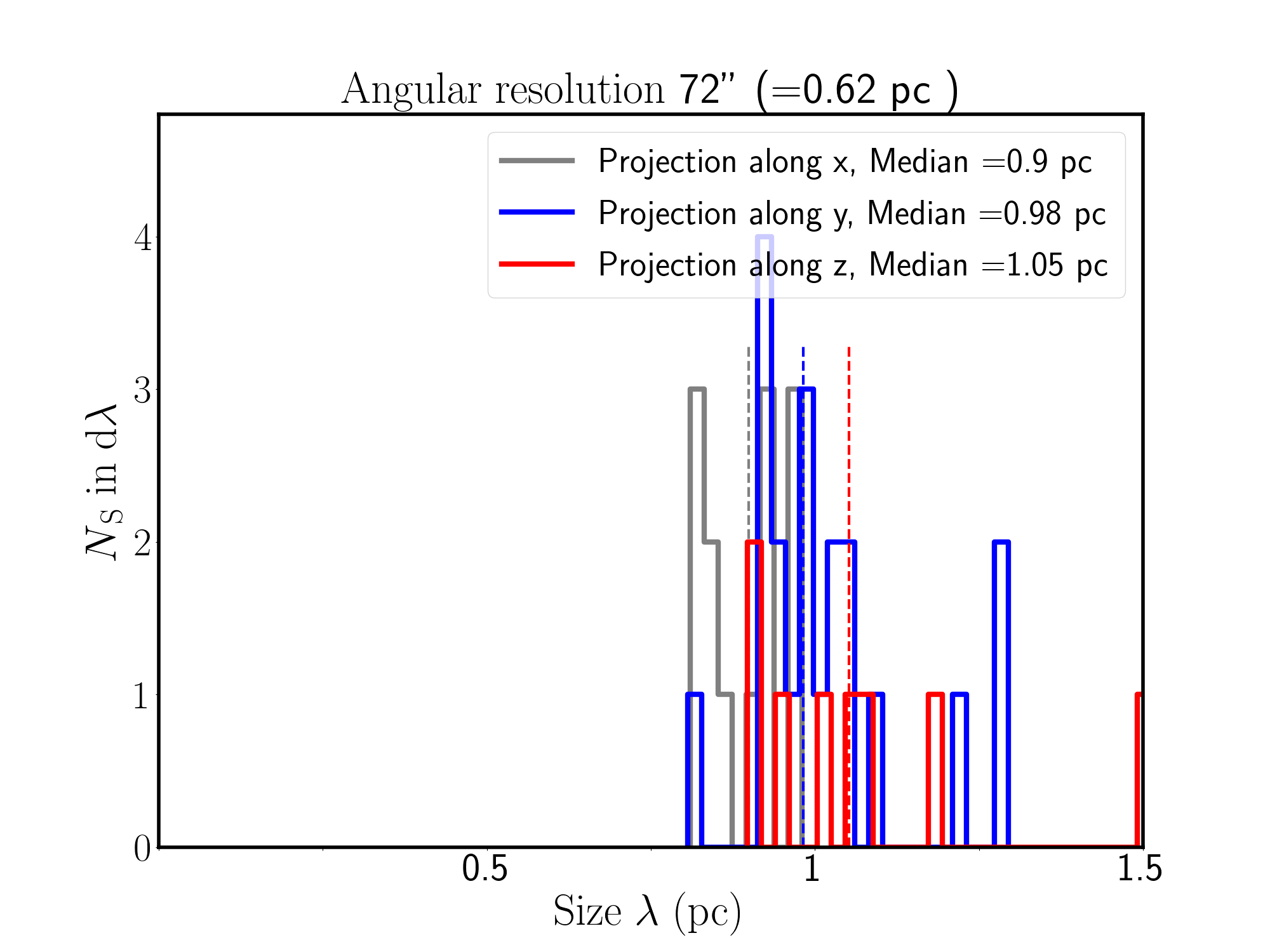}}
    \subfloat{\includegraphics[trim=3cm 0cm 0.8cm 5.5cm, width=0.34\linewidth]{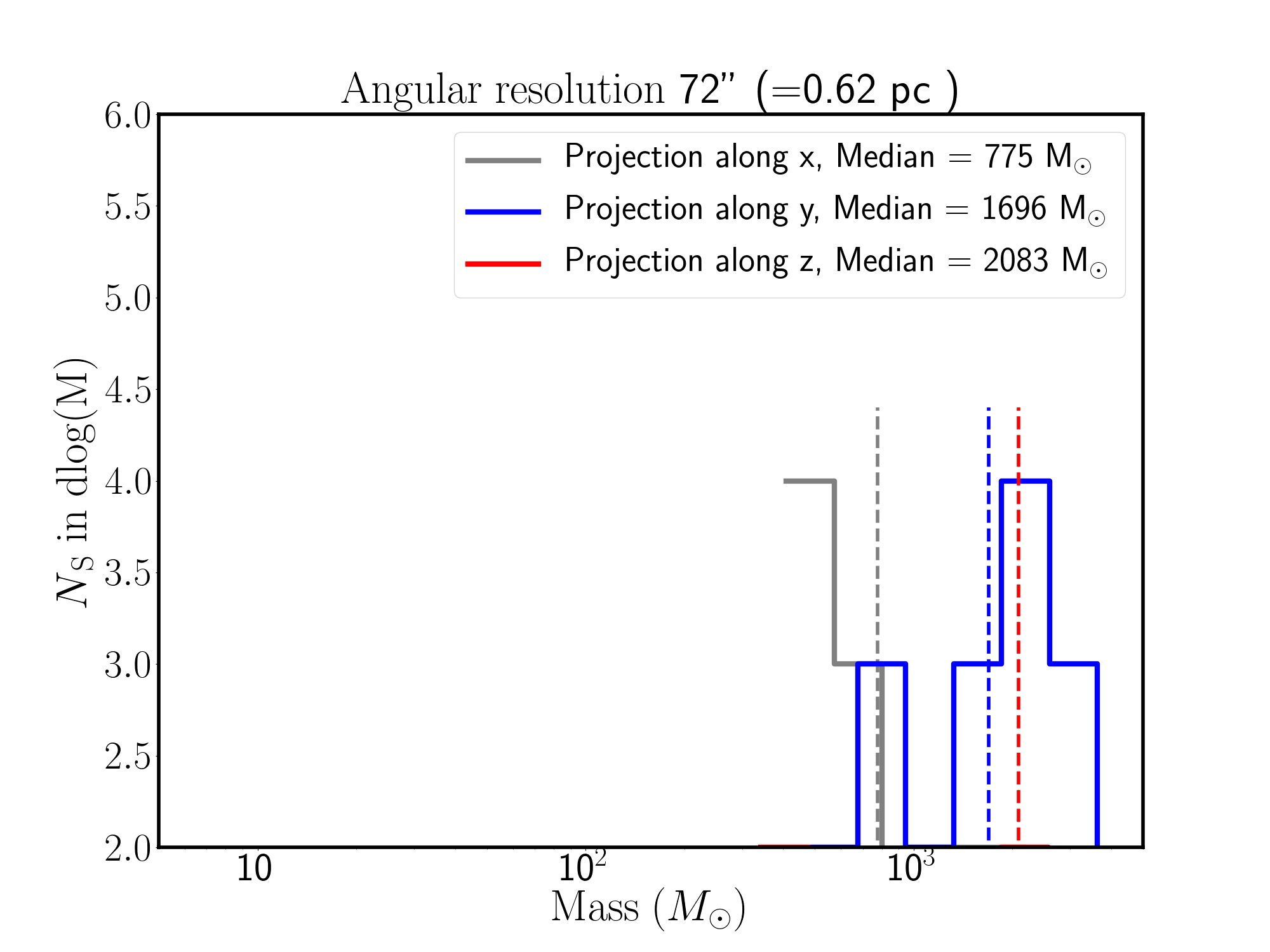}}
    \subfloat{\includegraphics[trim=3cm 0cm 0.8cm 5.5cm, width=0.34\linewidth]{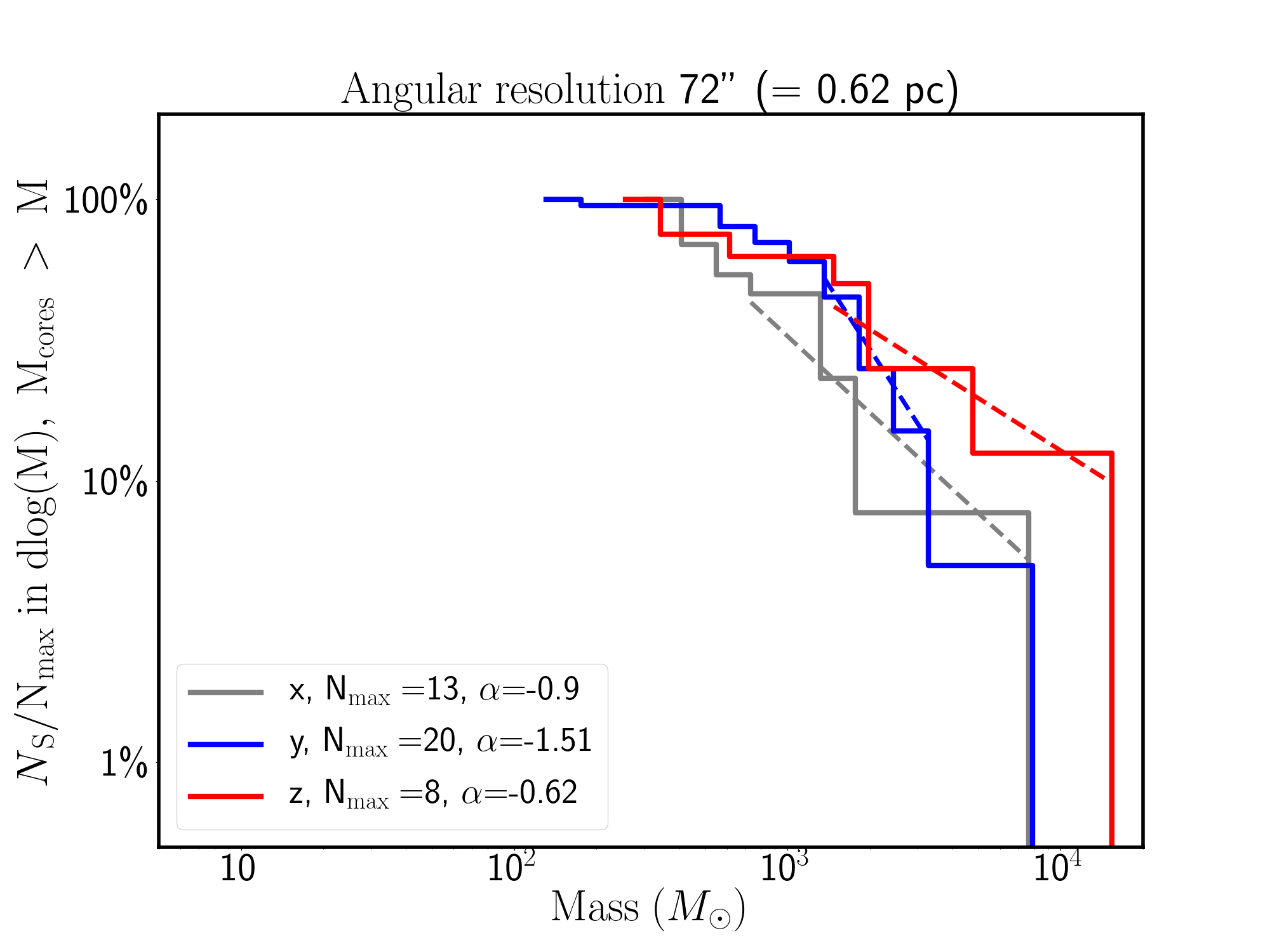}}
\caption
{ 
Same as in Fig.~\ref{f:vieweffect_18as}, but for the angular resolutions of 9{\arcsec} (\emph{top}), 36{\arcsec} (\emph{middle row}),
and 72{\arcsec} (\emph{bottom}).
} 
\label{f:vieweffect_9-18-36-72as}
\end{figure*}

\begin{figure*} 
    \centering
    \subfloat{\includegraphics[trim=3cm 3cm 0.8cm 5cm,   width=0.34\linewidth]{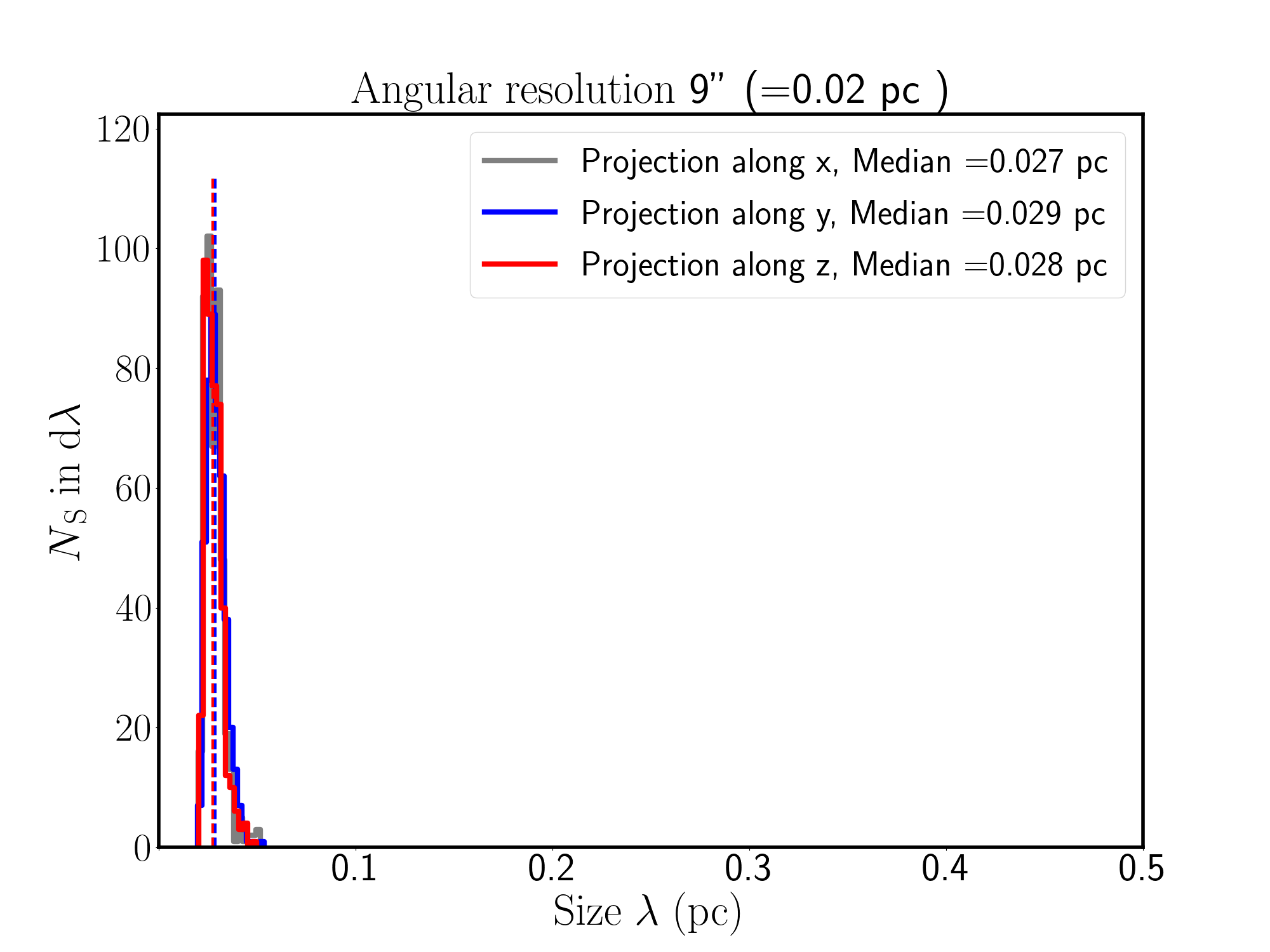}}
    \subfloat{\includegraphics[trim=3cm 3cm 0.8cm 5cm,   width=0.34\linewidth]{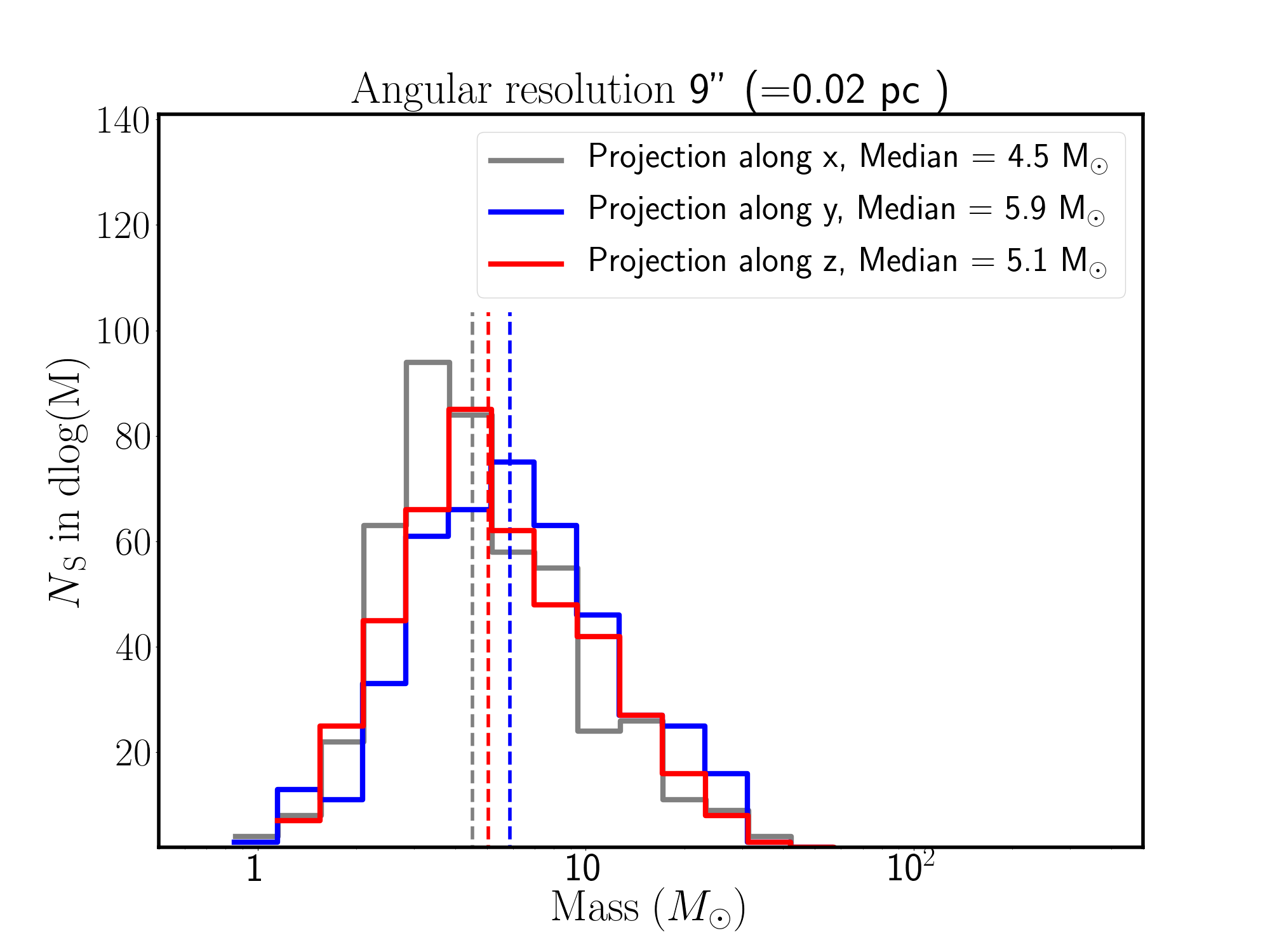}}
    \subfloat{\includegraphics[trim=3cm 3cm 0.8cm 5cm,   width=0.34\linewidth]{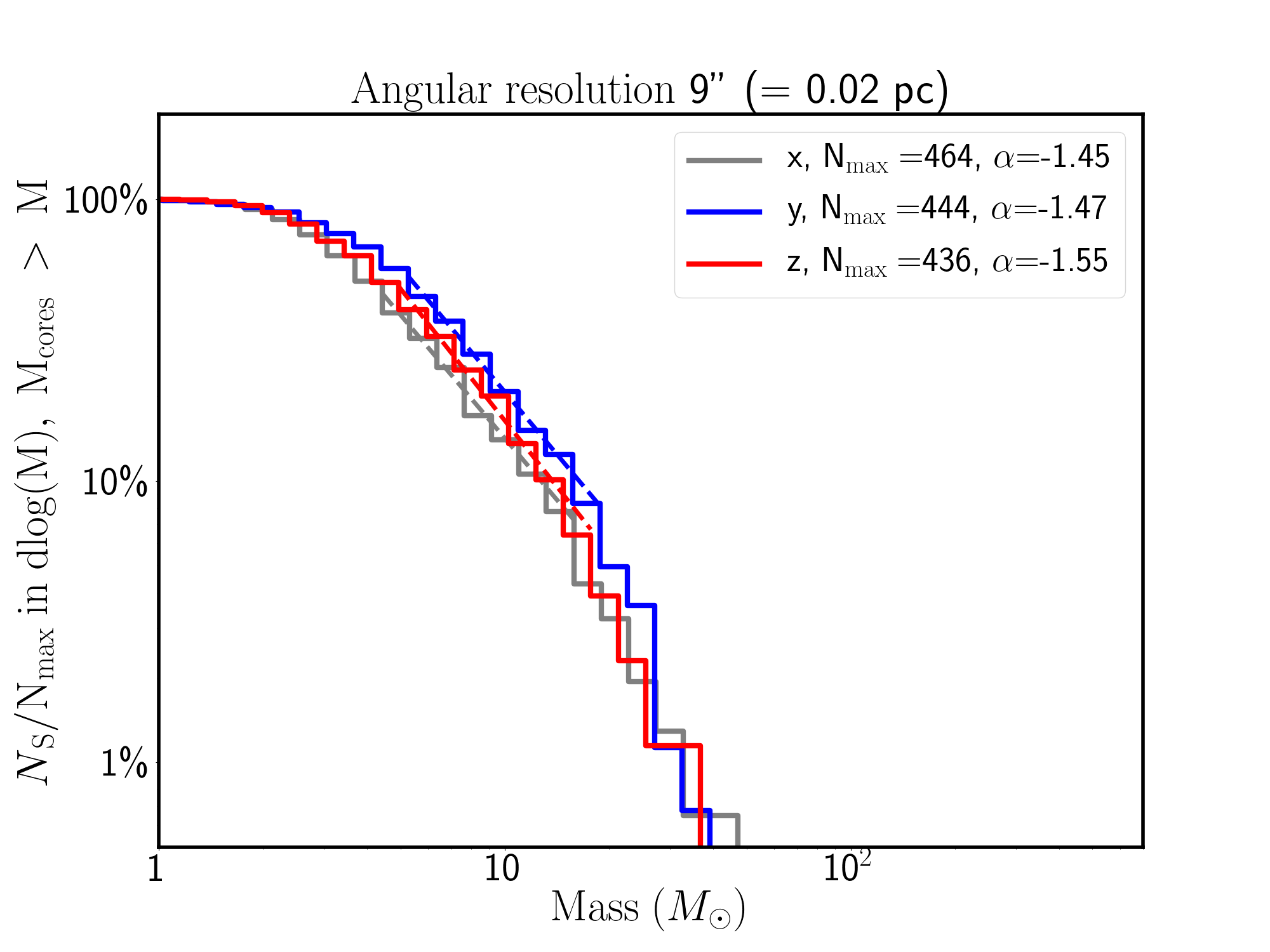}}\\
    \subfloat{\includegraphics[trim=3cm 3cm 0.8cm 2.5cm, width=0.34\linewidth]{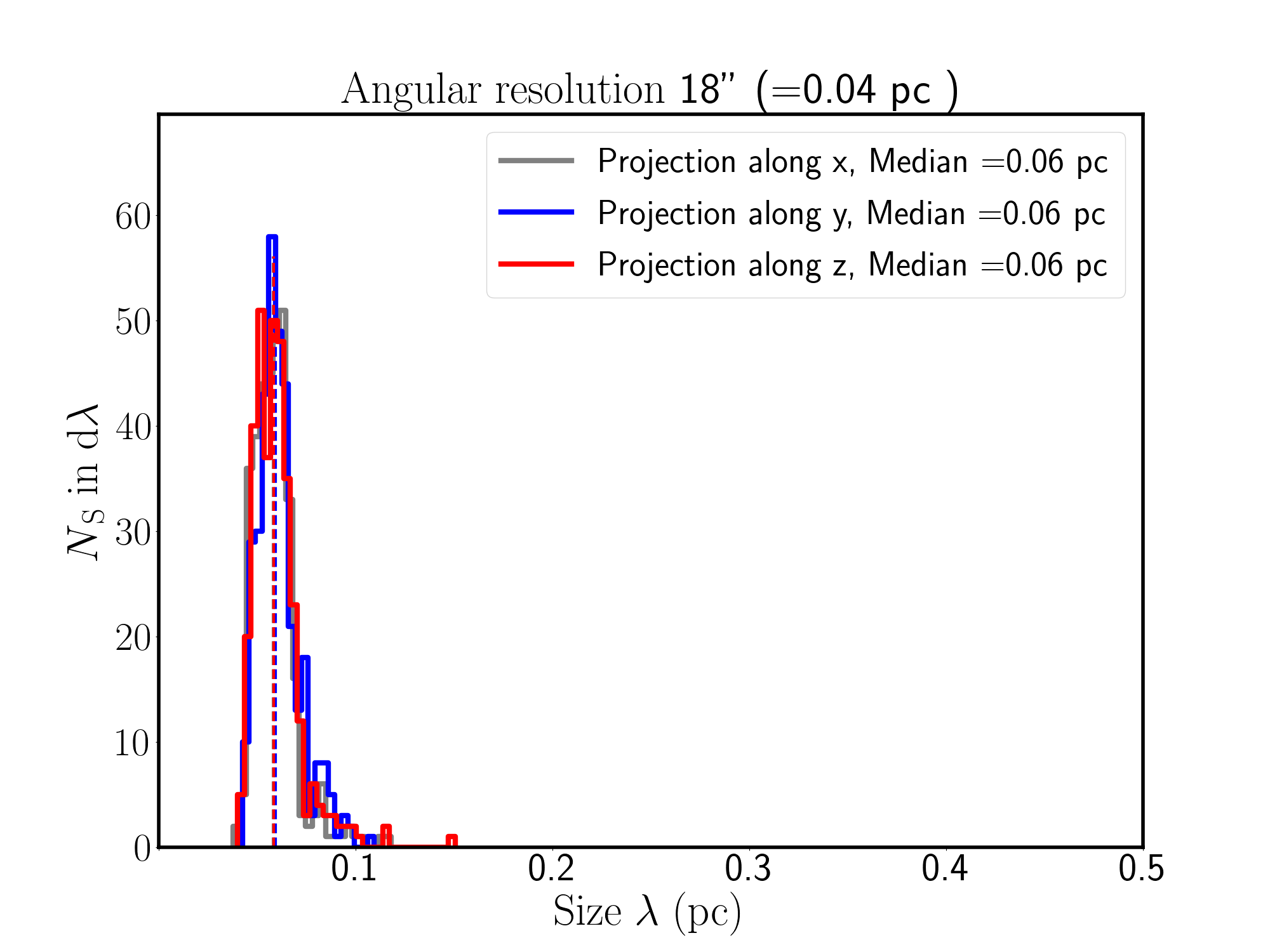}}
    \subfloat{\includegraphics[trim=3cm 3cm 0.8cm 2.5cm, width=0.34\linewidth]{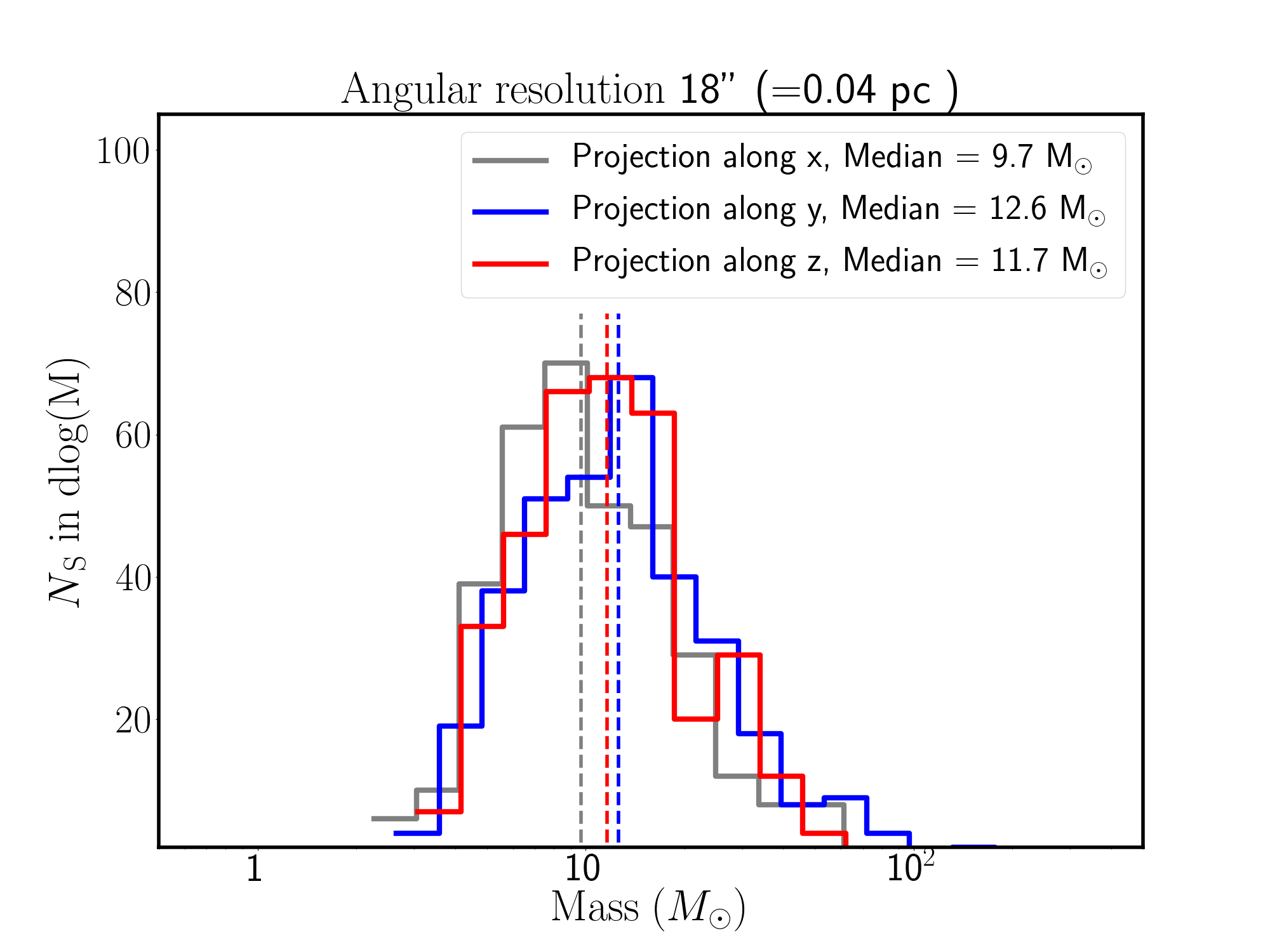}}
    \subfloat{\includegraphics[trim=3cm 3cm 0.8cm 2.5cm, width=0.34\linewidth]{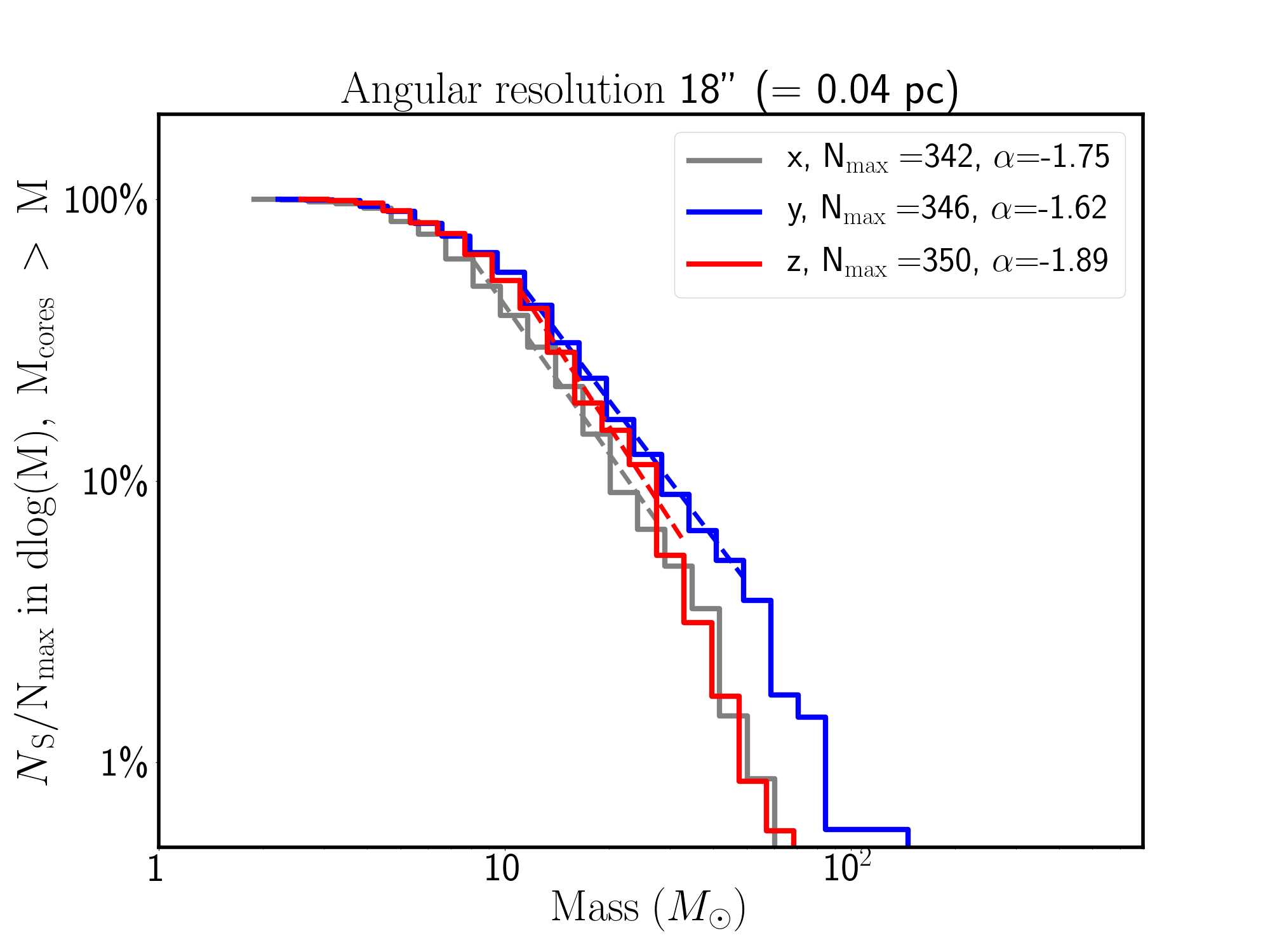}}\\
    \subfloat{\includegraphics[trim=3cm 3cm 0.8cm 2.5cm, width=0.34\linewidth]{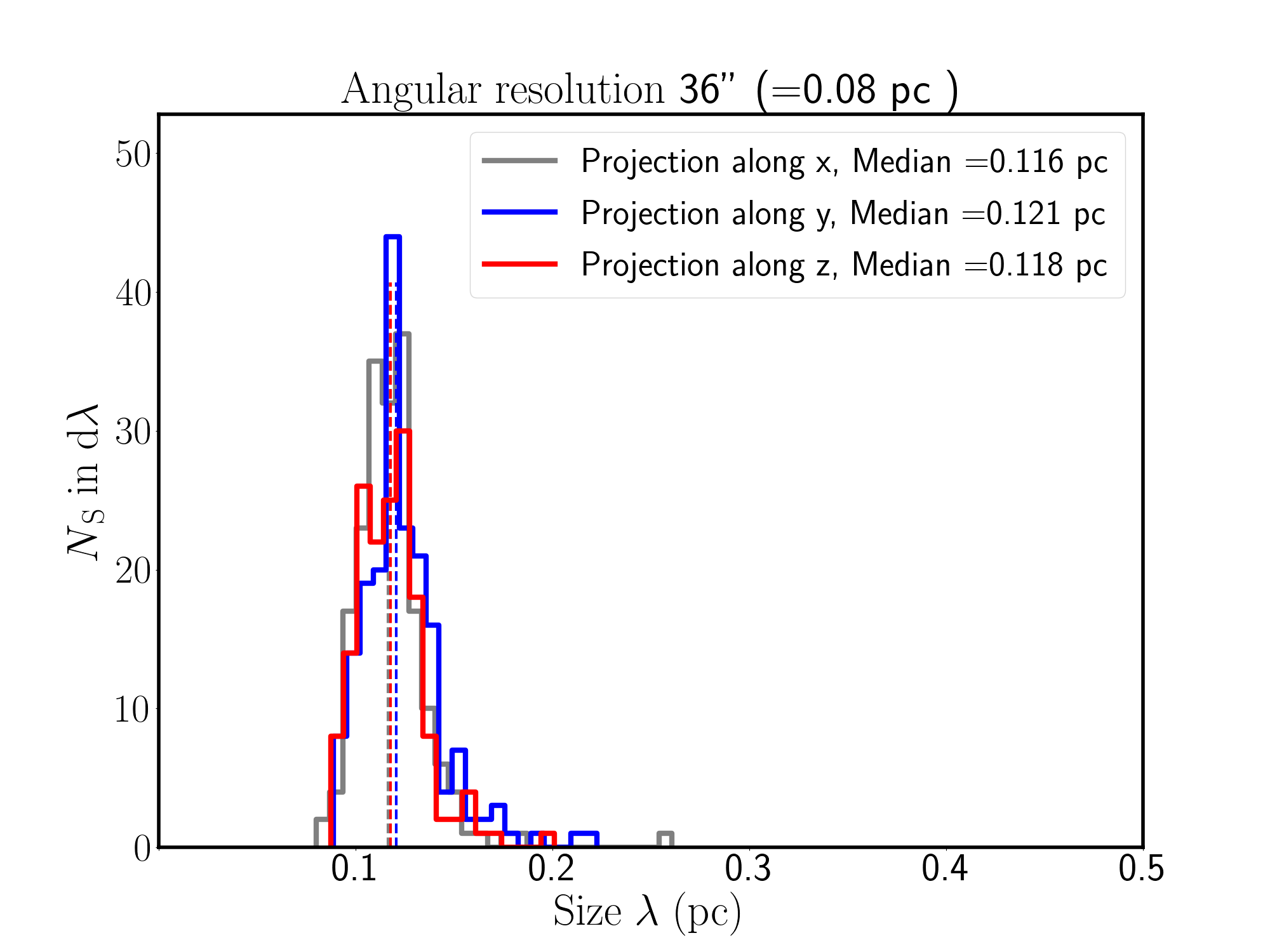}}
    \subfloat{\includegraphics[trim=3cm 3cm 0.8cm 2.5cm, width=0.34\linewidth]{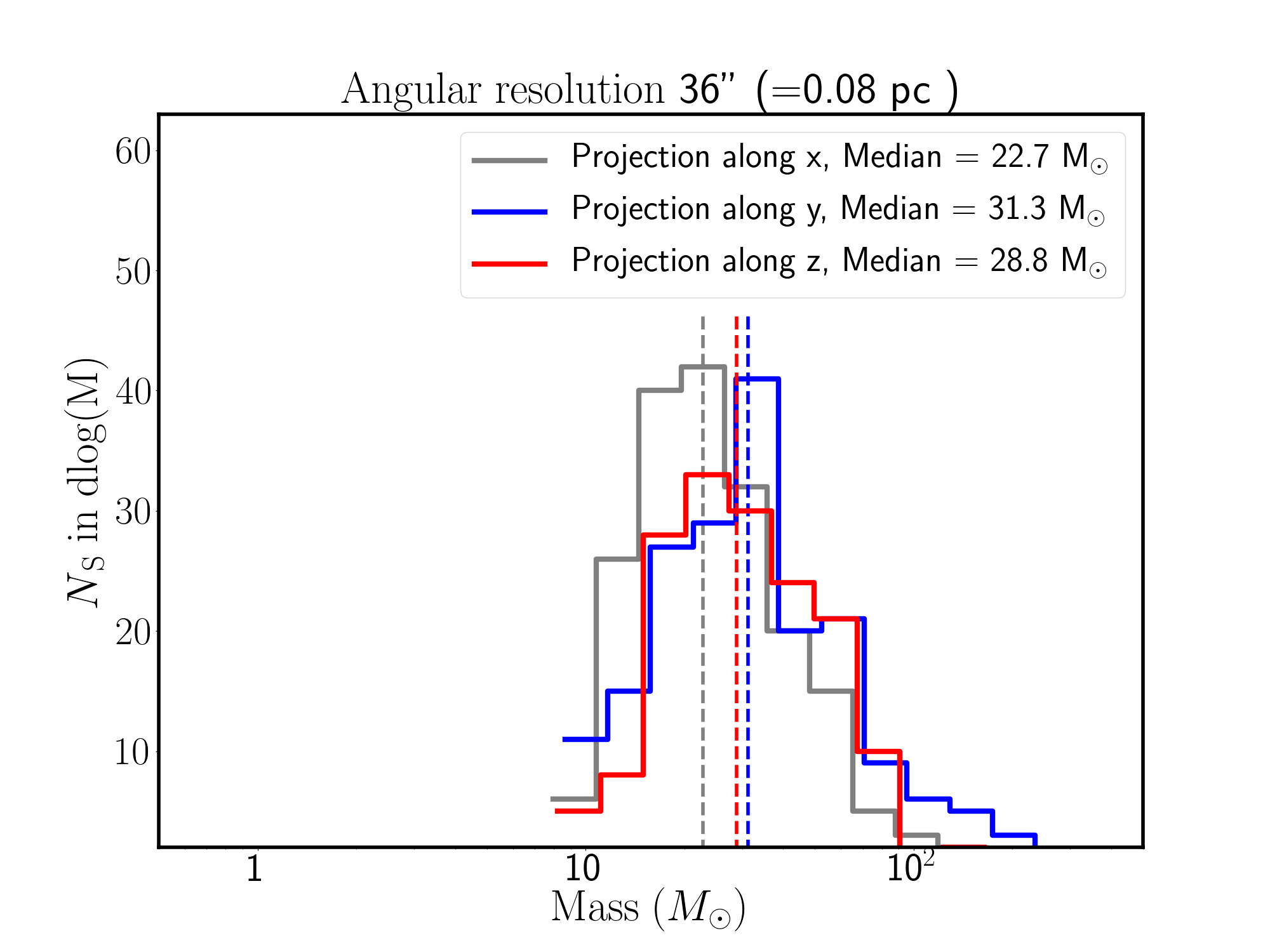}}
    \subfloat{\includegraphics[trim=3cm 3cm 0.8cm 2.5cm, width=0.34\linewidth]{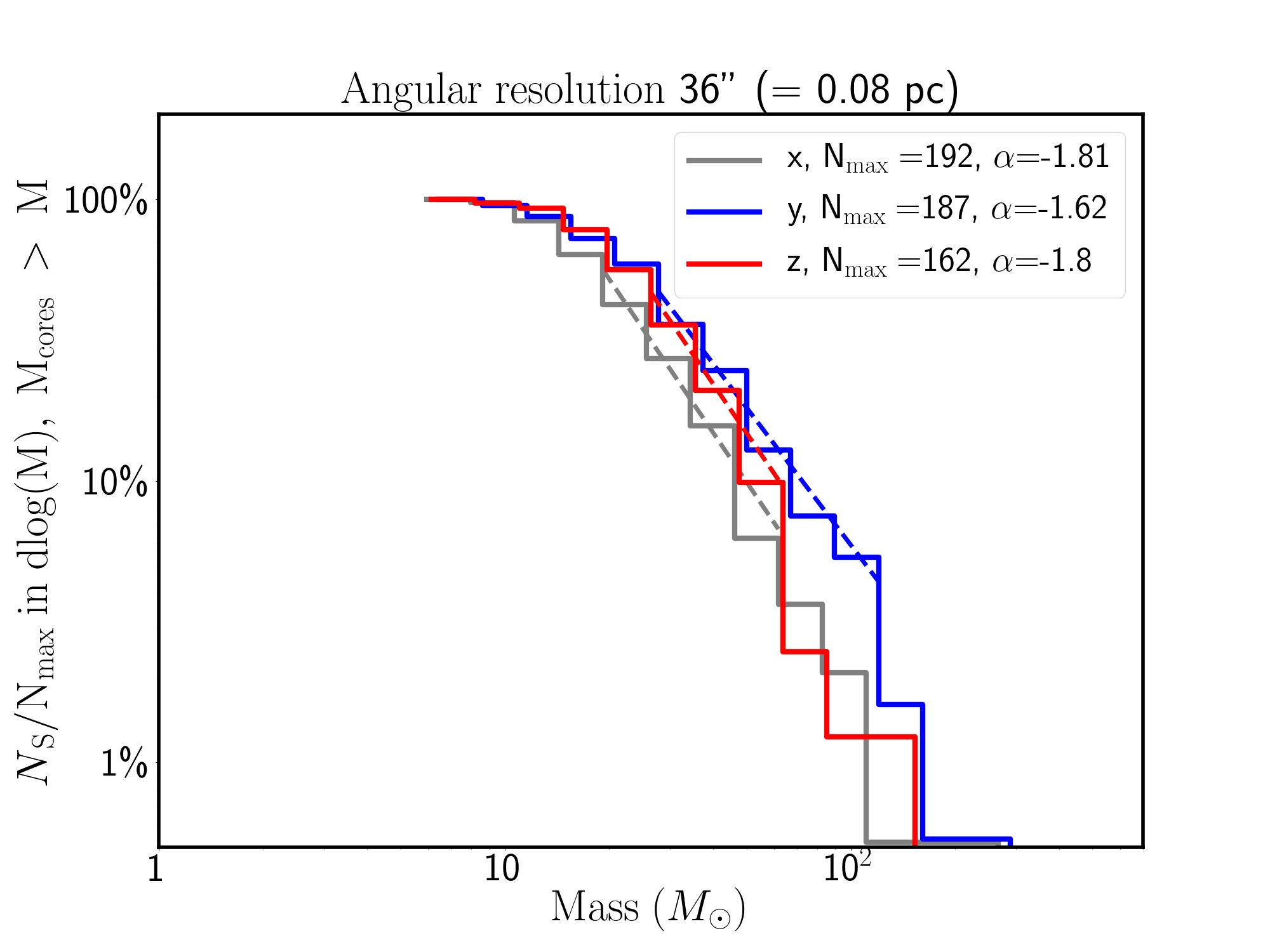}}\\
    \subfloat{\includegraphics[trim=3cm 3cm 0.8cm 2.5cm, width=0.34\linewidth]{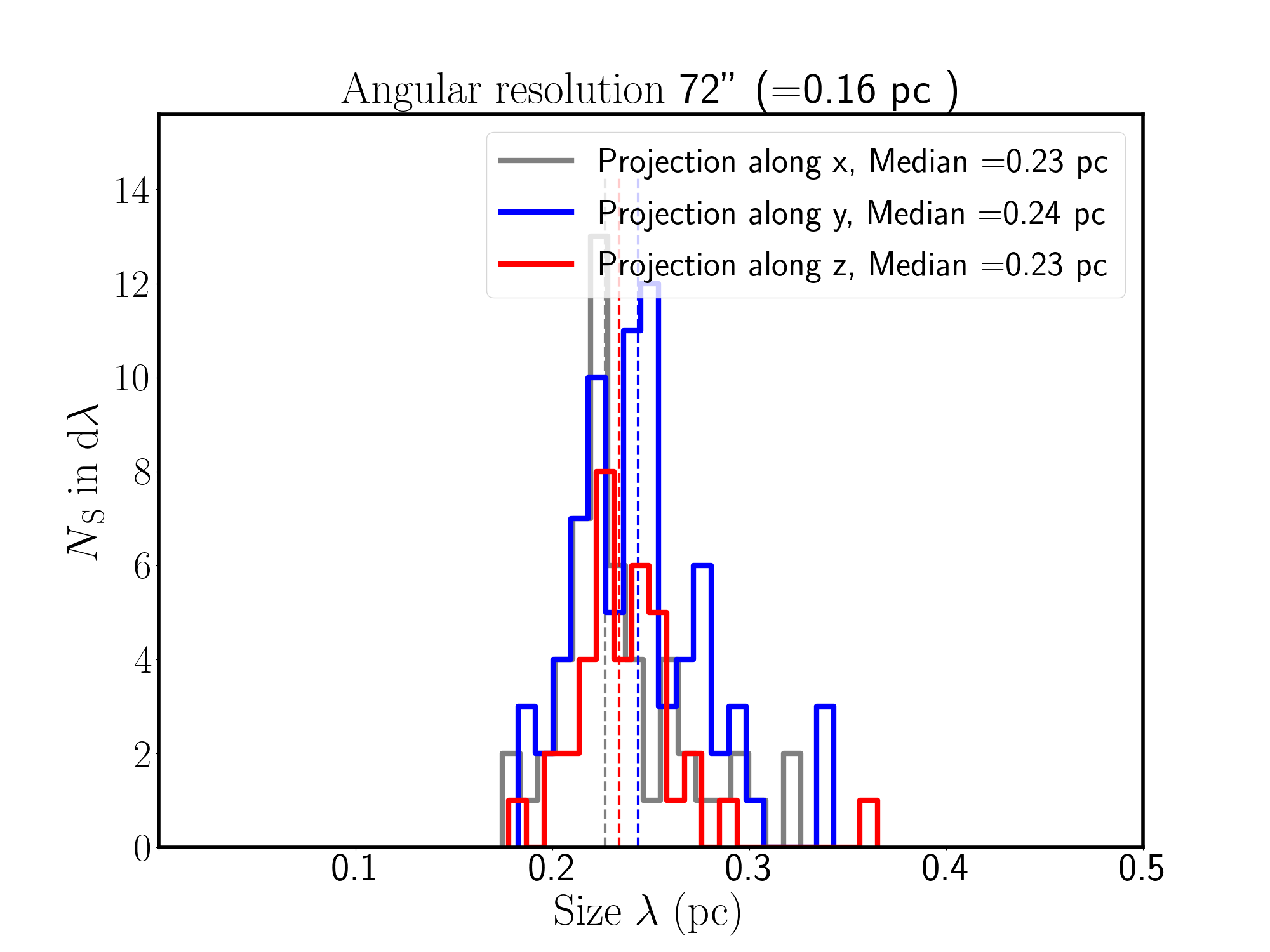}}
    \subfloat{\includegraphics[trim=3cm 3cm 0.8cm 2.5cm, width=0.34\linewidth]{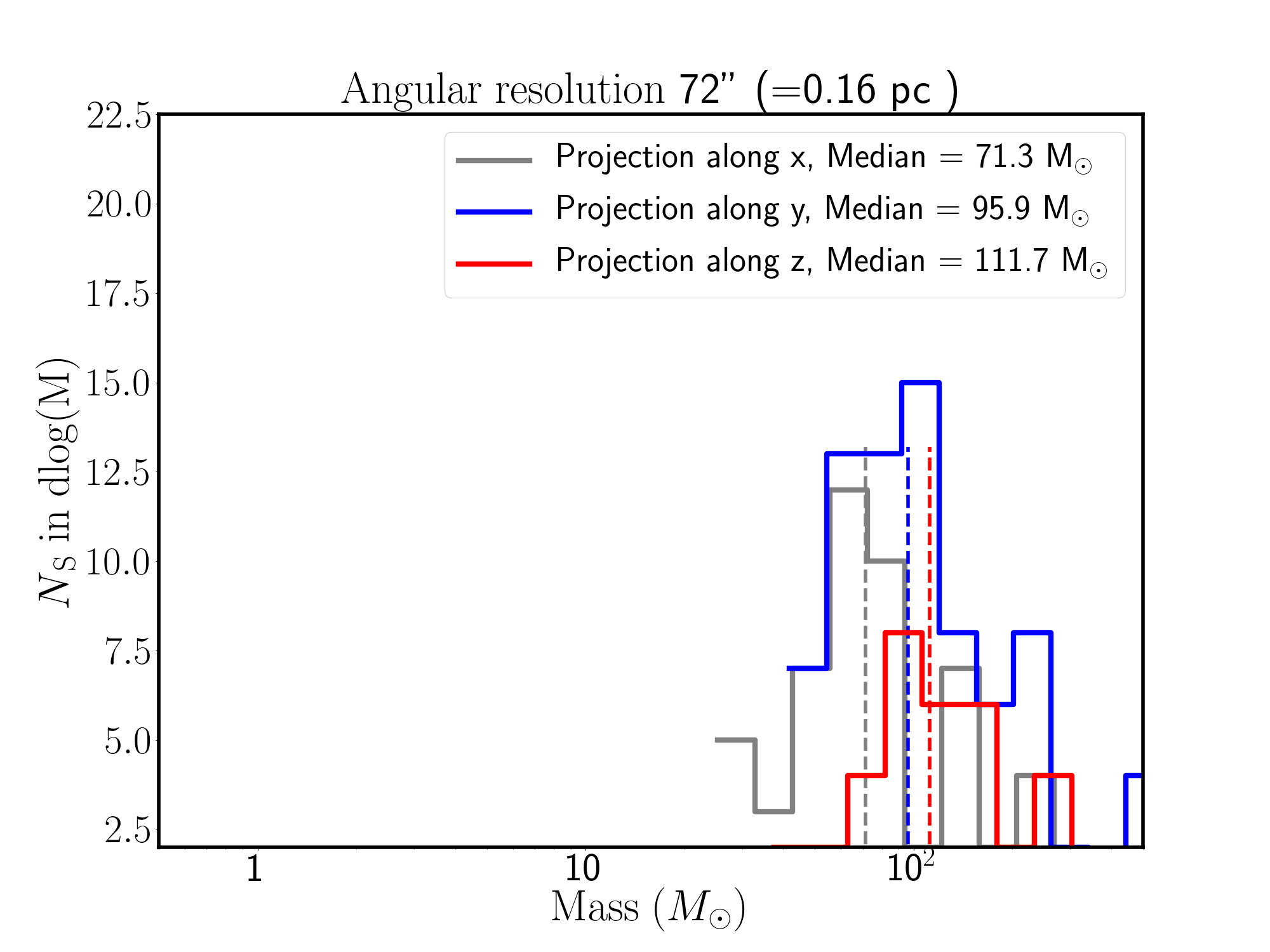}}
    \subfloat{\includegraphics[trim=3cm 3cm 0.8cm 2.5cm, width=0.34\linewidth]{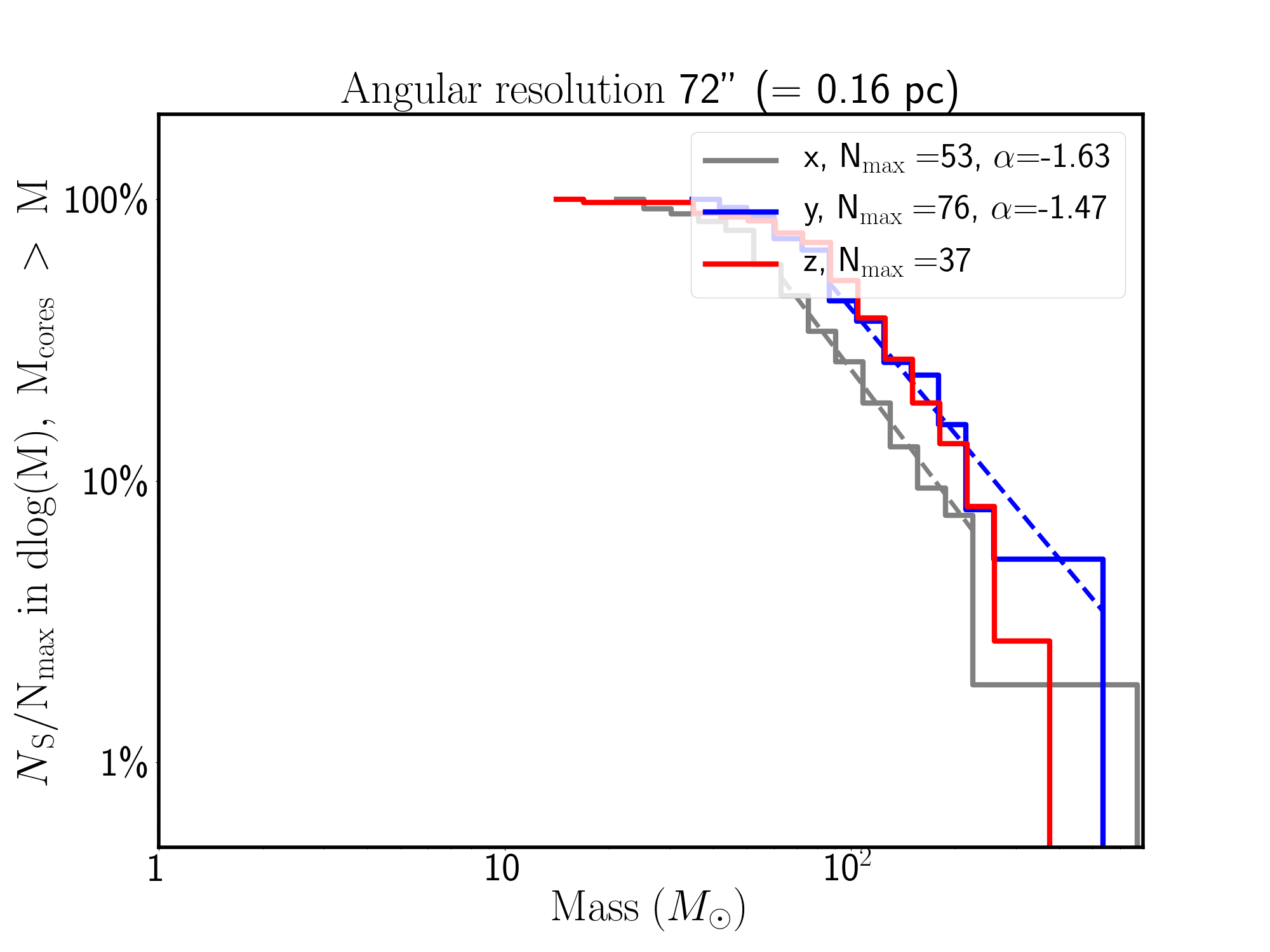}}
\caption
{ 
Same as Fig.~\ref{f:vieweffect_9-18-36-72as}, but for the HD$^\mathrm{h}$ model at 9{\arcsec} (\emph{top}), 18{\arcsec}
(\emph{second row}), 36{\arcsec} (\emph{third row}), and 72{\arcsec} (\emph{bottom}).
} 
\label{f:vieweffect_36-72-144as-aquila}
\end{figure*}

\begin{figure*} 
    \centering
    \subfloat{\includegraphics[trim=3cm 3cm 0.8cm 5cm,   width=0.34\linewidth]{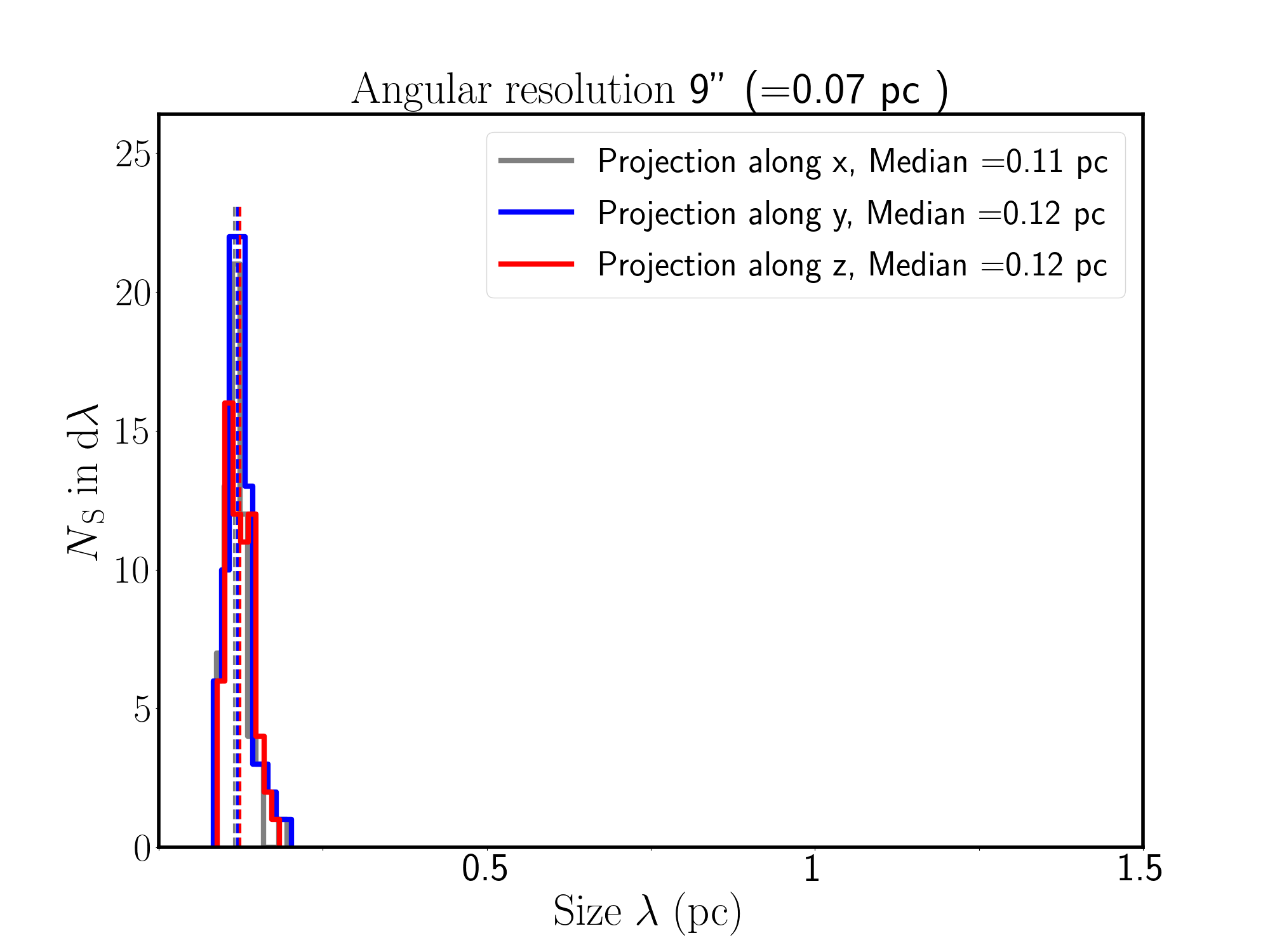}}
    \subfloat{\includegraphics[trim=3cm 3cm 0.8cm 5cm,   width=0.34\linewidth]{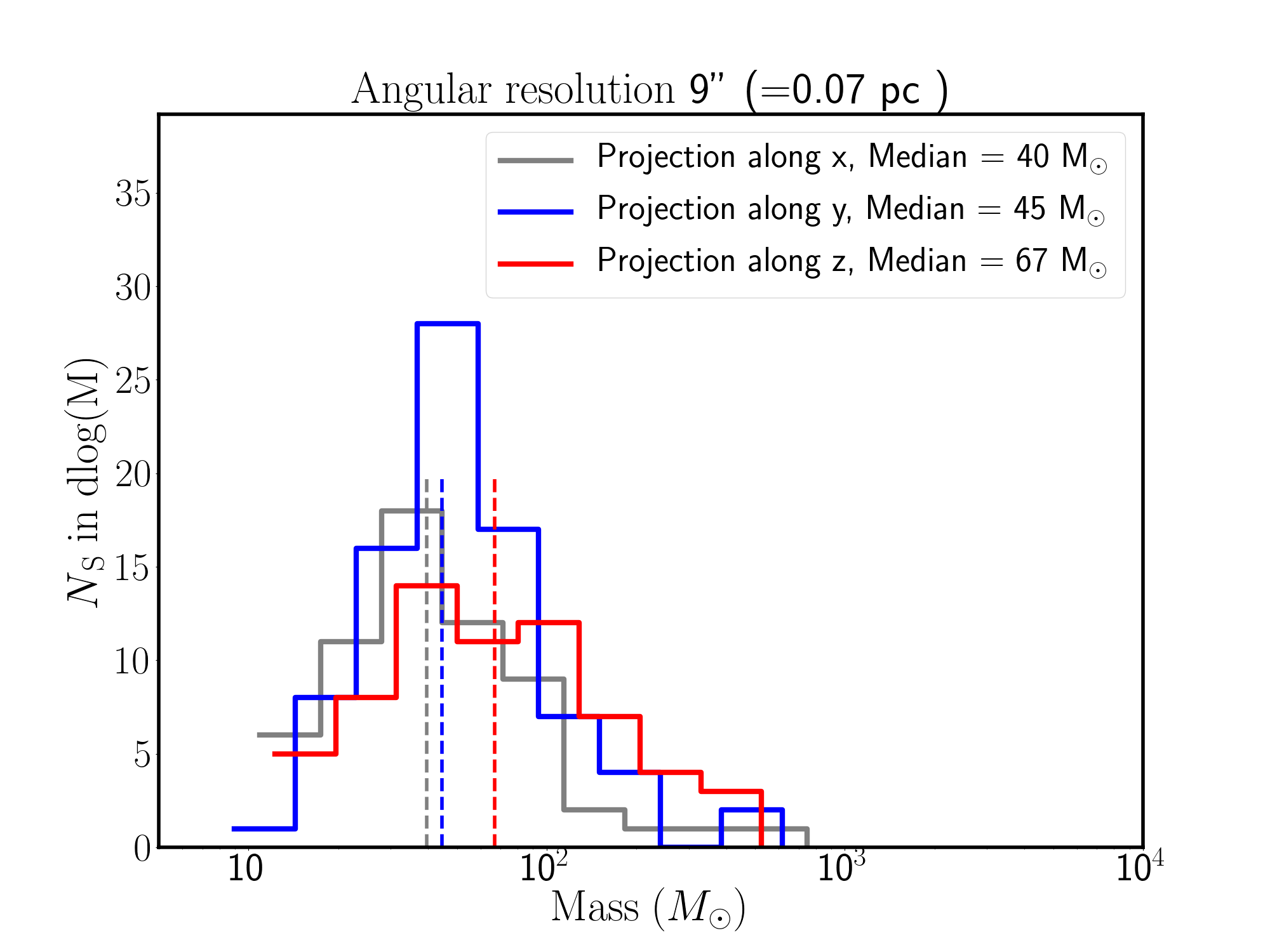}}
    \subfloat{\includegraphics[trim=3cm 3cm 0.8cm 5cm,   width=0.34\linewidth]{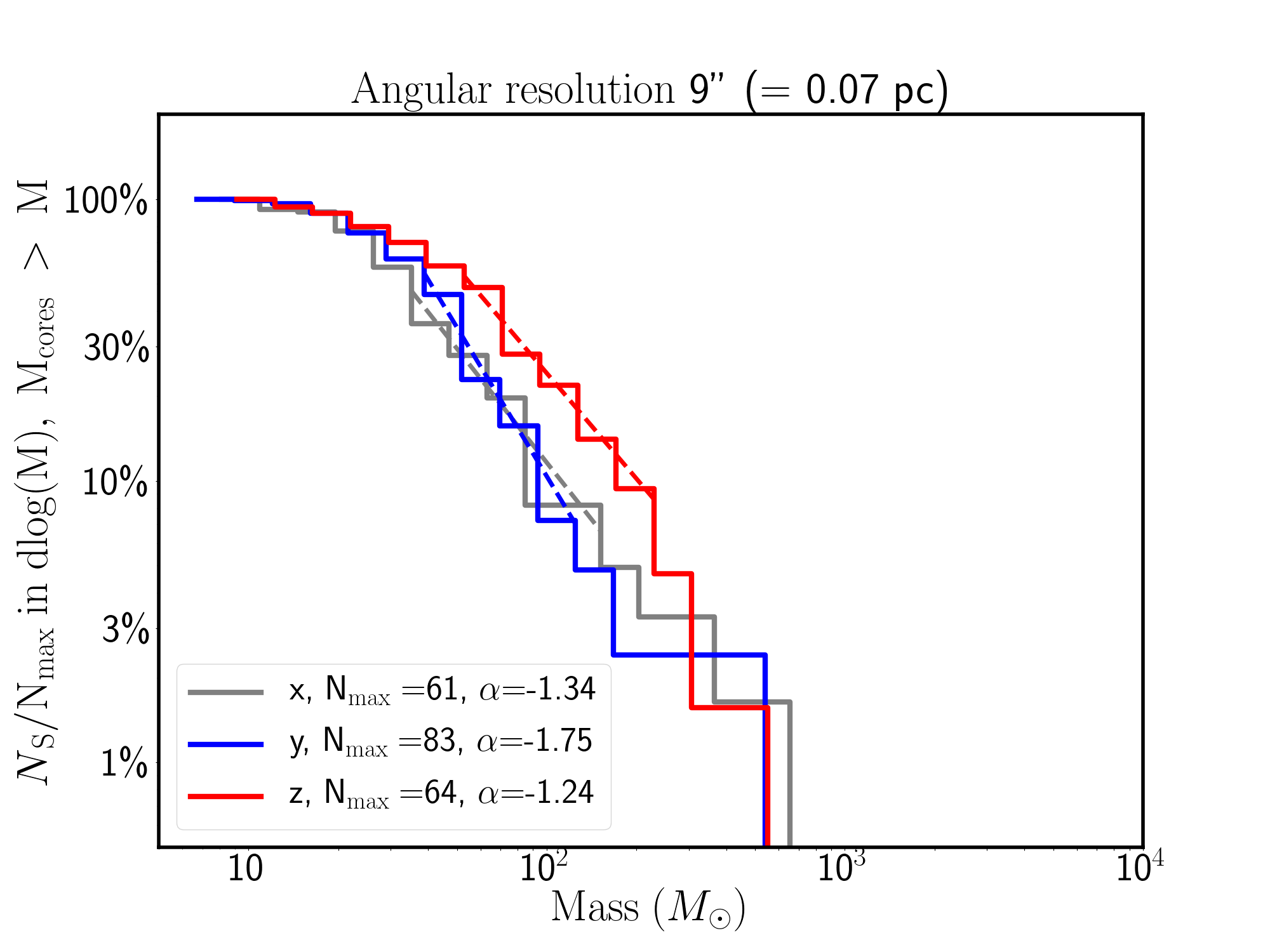}}\\
    \subfloat{\includegraphics[trim=3cm 1cm 0.8cm 2.5cm, width=0.34\linewidth]{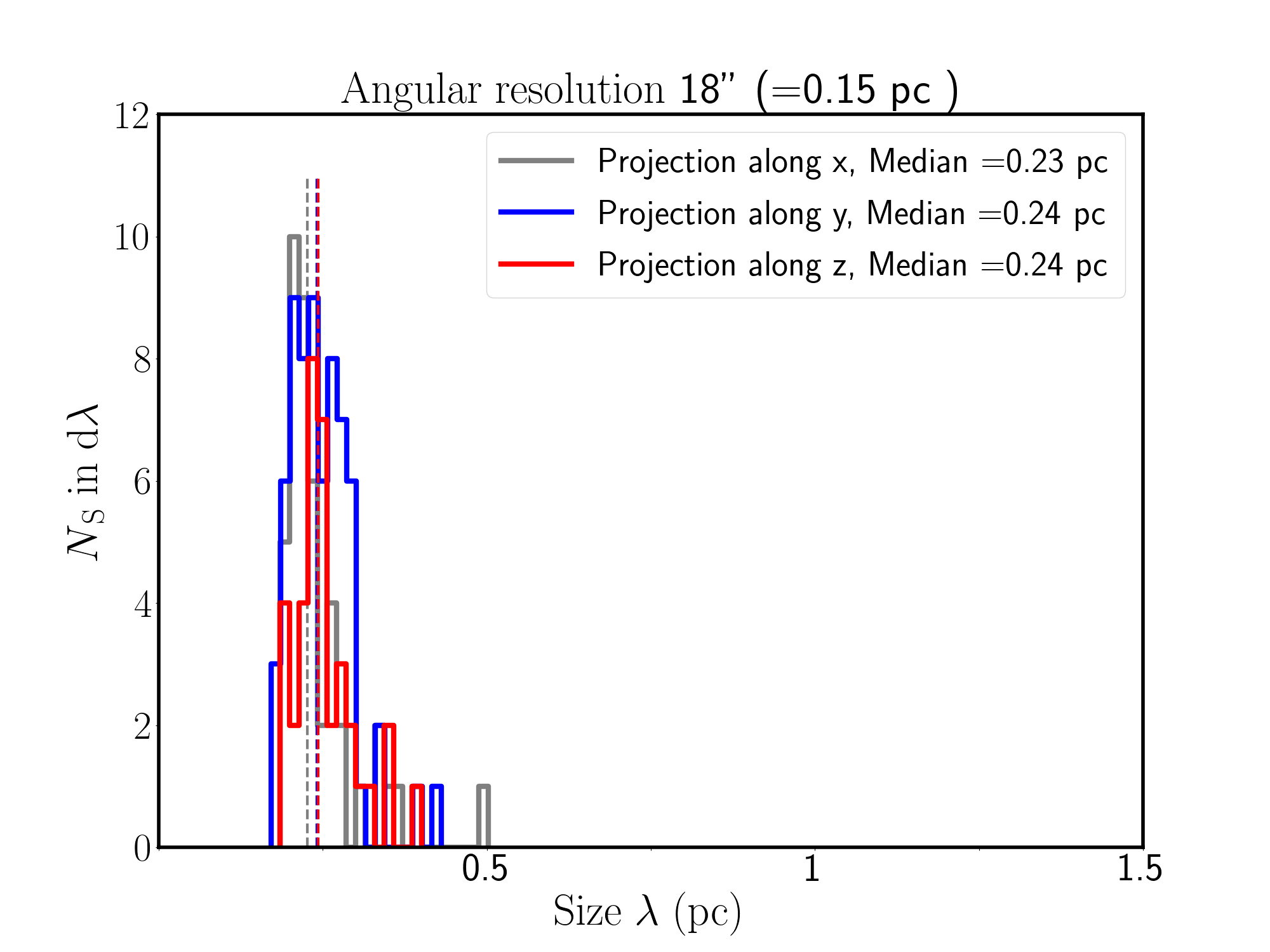}}
    \subfloat{\includegraphics[trim=3cm 1cm 0.8cm 2.5cm, width=0.34\linewidth]{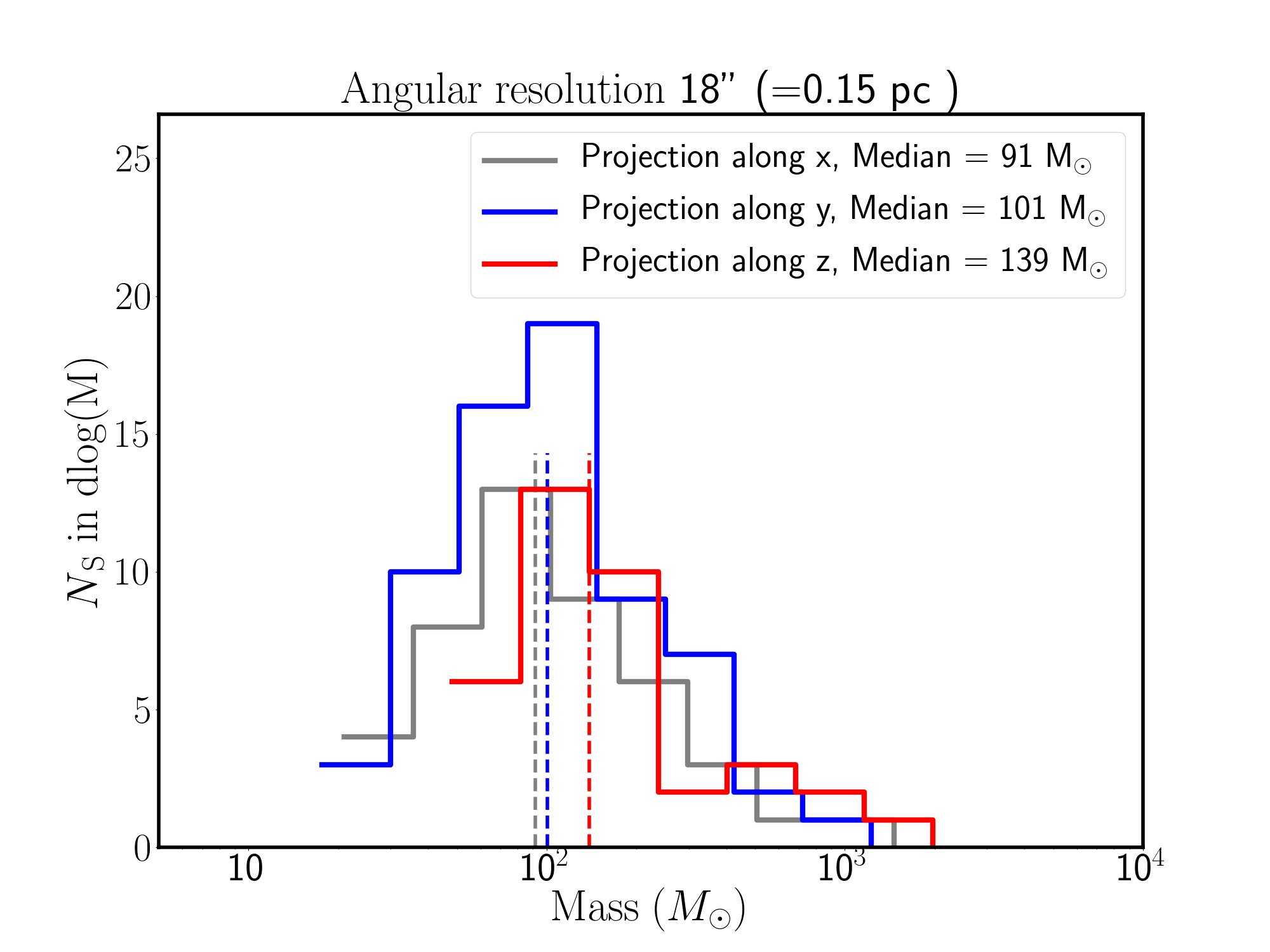}}
    \subfloat{\includegraphics[trim=3cm 1cm 0.8cm 2.5cm, width=0.34\linewidth]{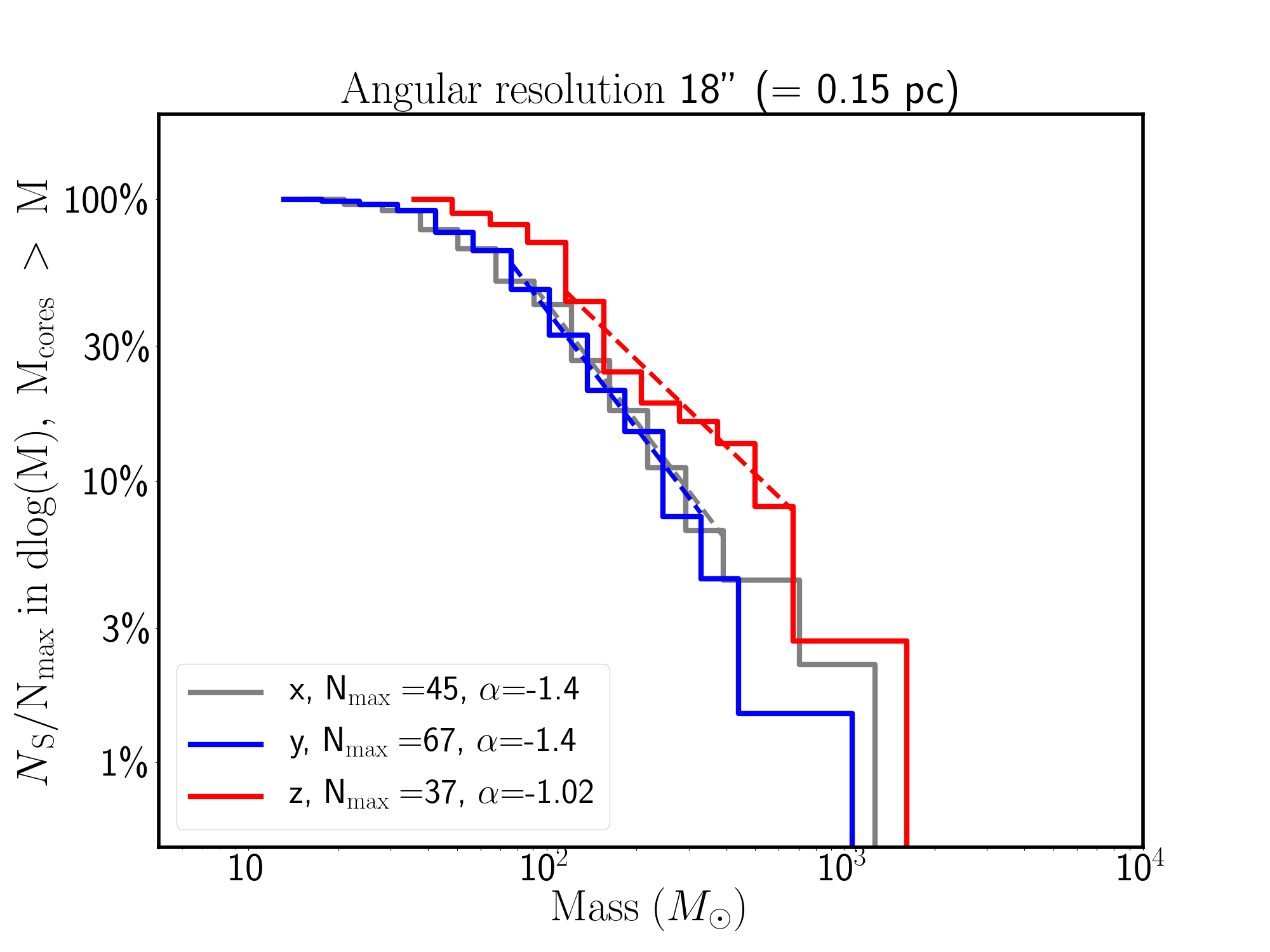}}\\
    \subfloat{\includegraphics[trim=3cm 1cm 0.8cm 2.5cm, width=0.34\linewidth]{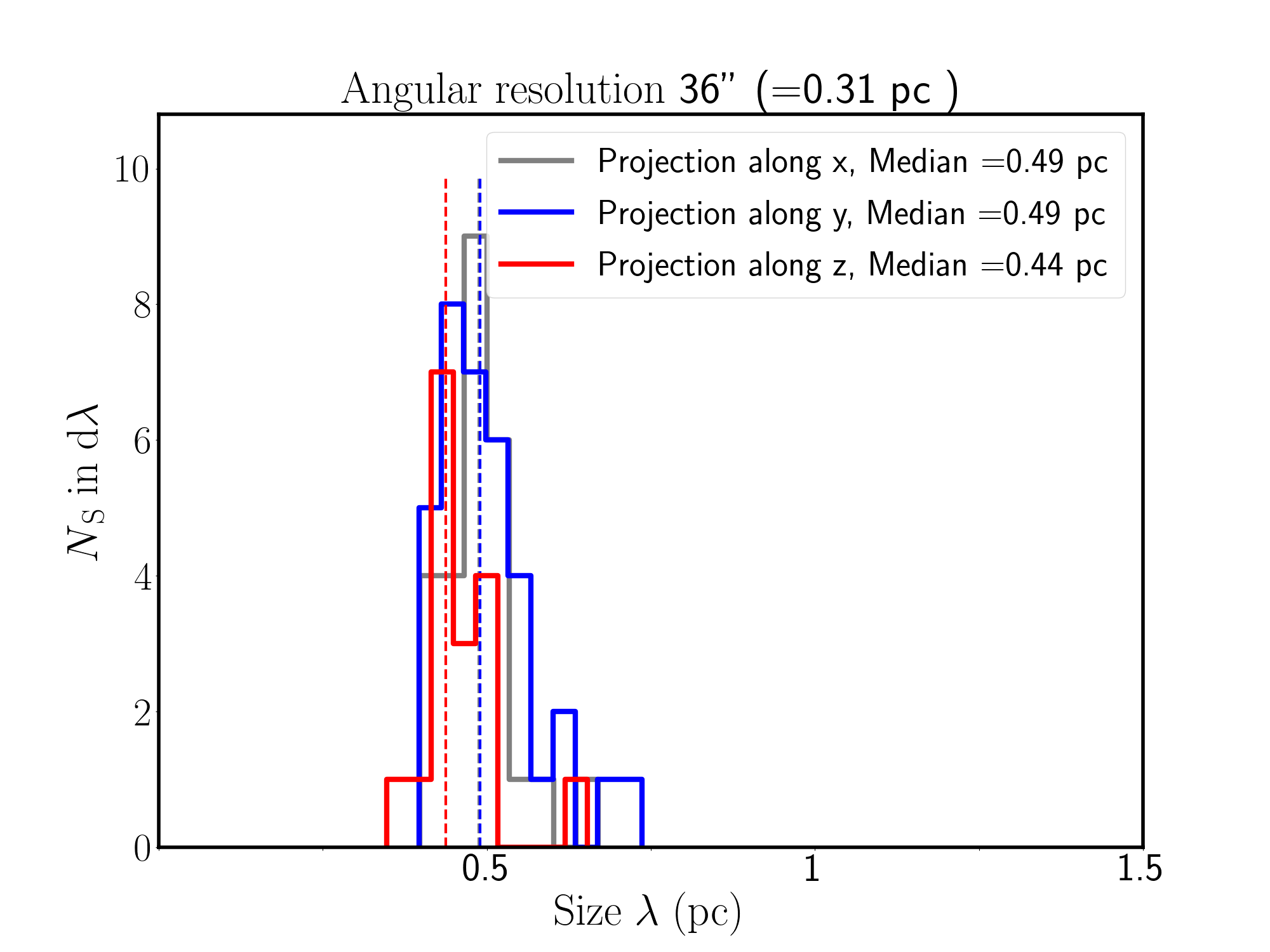}}
    \subfloat{\includegraphics[trim=3cm 1cm 0.8cm 2.5cm, width=0.34\linewidth]{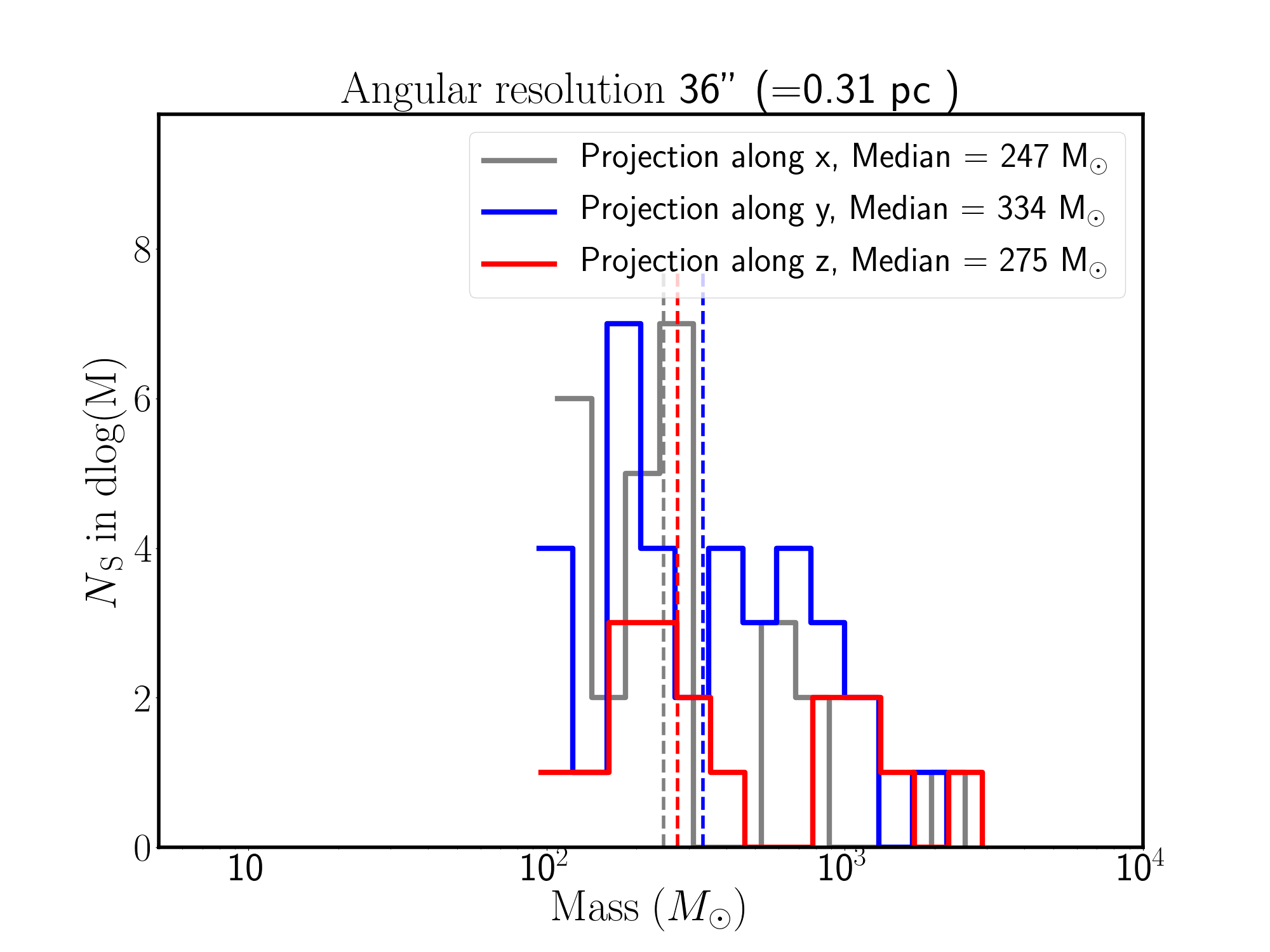}}
    \subfloat{\includegraphics[trim=3cm 1cm 0.8cm 2.5cm, width=0.34\linewidth]{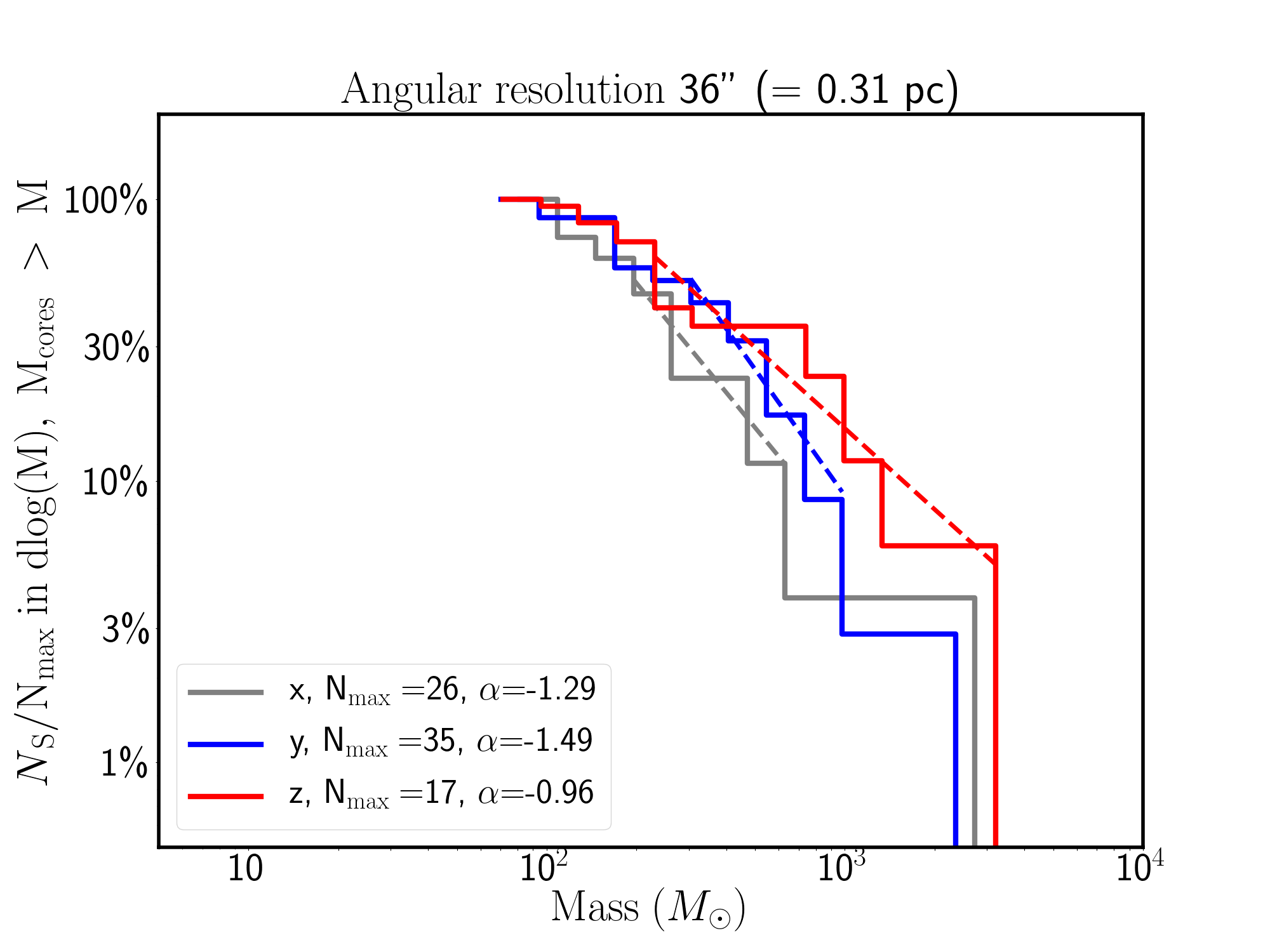}}\\
    \subfloat{\includegraphics[trim=3cm 1cm 0.8cm 2.5cm, width=0.34\linewidth]{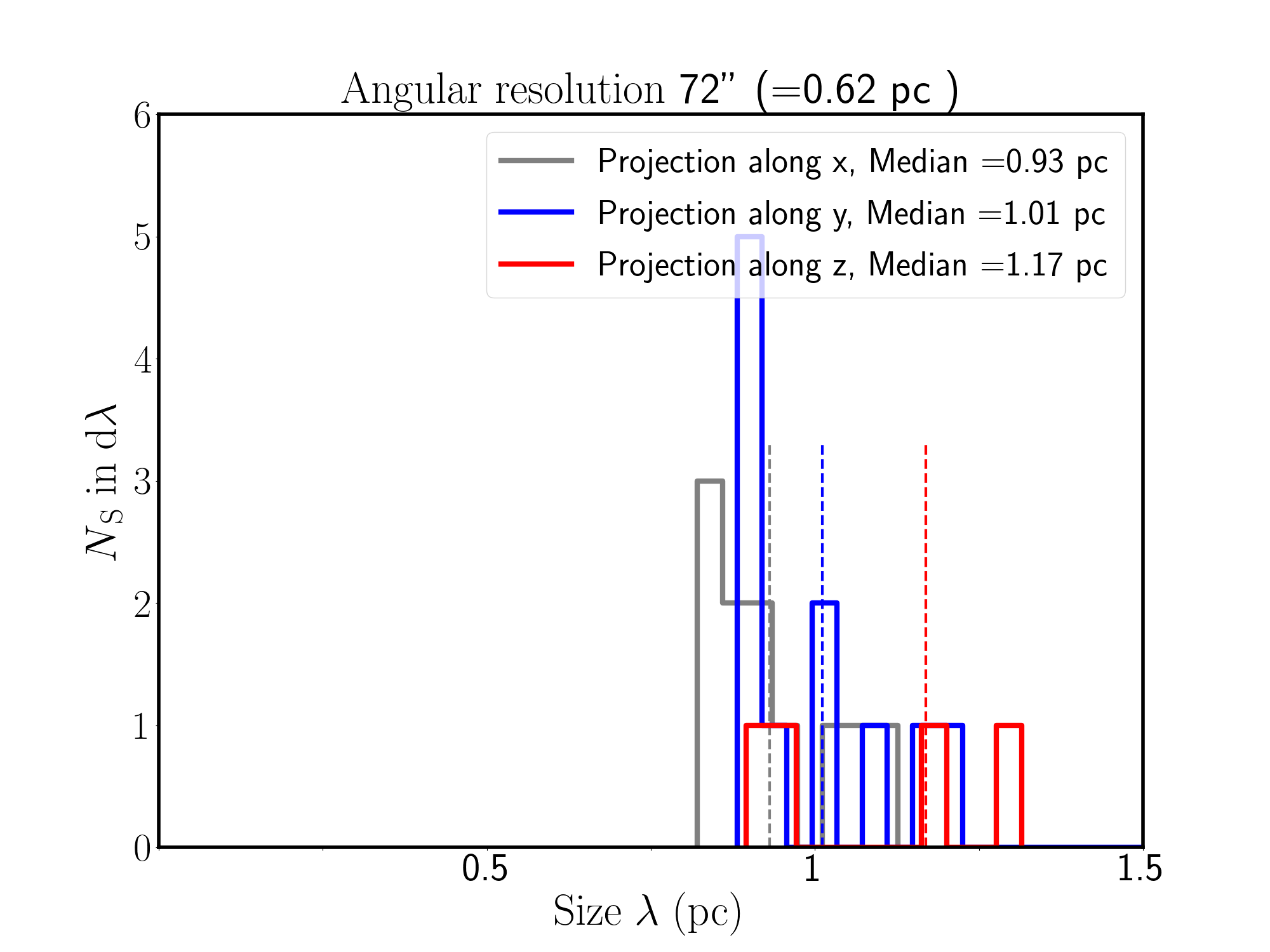}}
    \subfloat{\includegraphics[trim=3cm 1cm 0.8cm 2.5cm, width=0.34\linewidth]{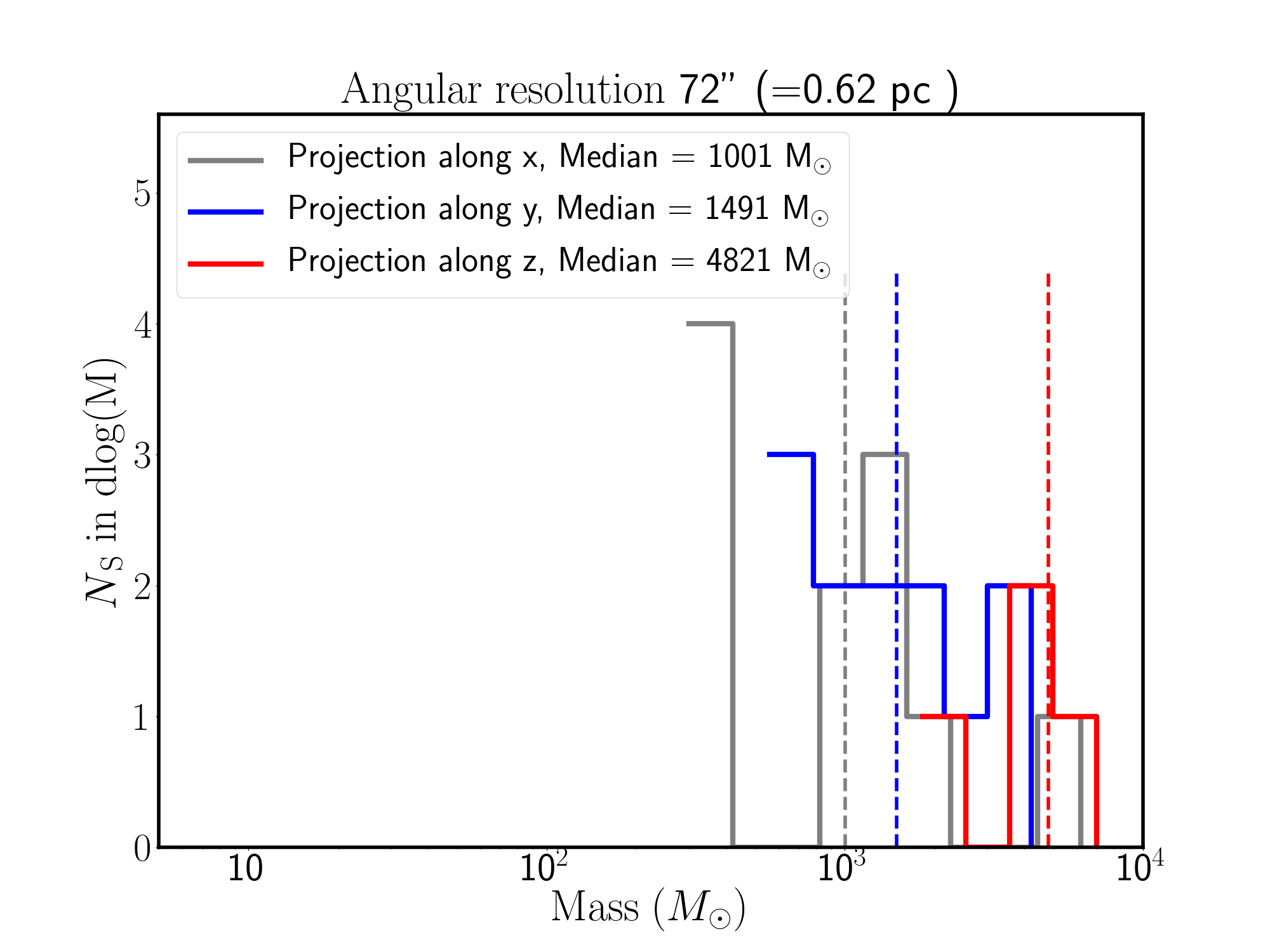}}
    \subfloat{\includegraphics[trim=3cm 1cm 0.8cm 2.5cm, width=0.34\linewidth]{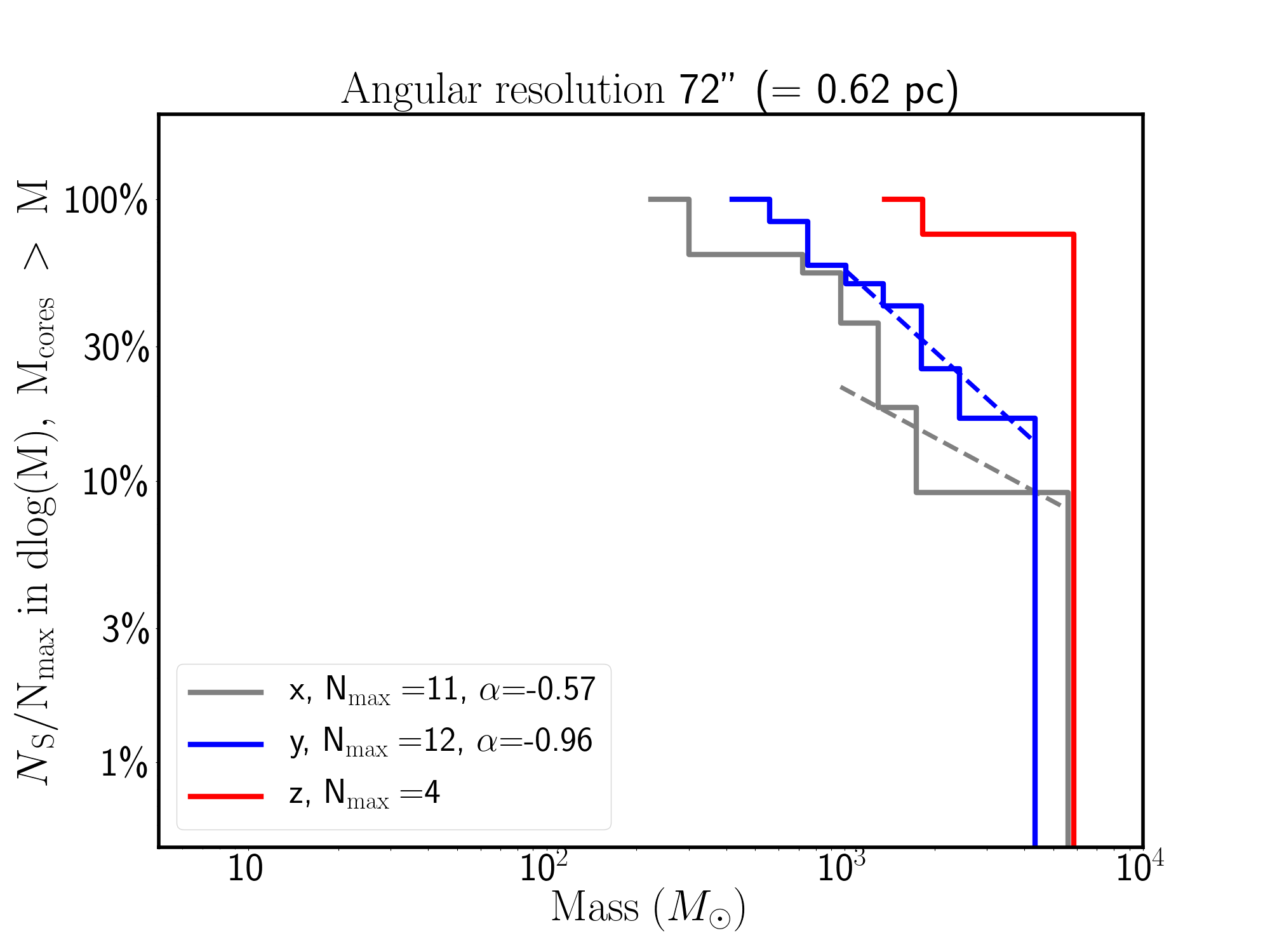}}
\caption
{ 
Same as Fig.~\ref{f:vieweffect_36-72-144as-aquila}, but for the MHD model.
} 
\label{f:vieweffect_mhd_9-18-36-72as}
\end{figure*}

\begin{figure*} 
    \centering
\subfloat{\includegraphics[trim=3cm 1.2cm 1cm 5cm, width=0.34\linewidth]{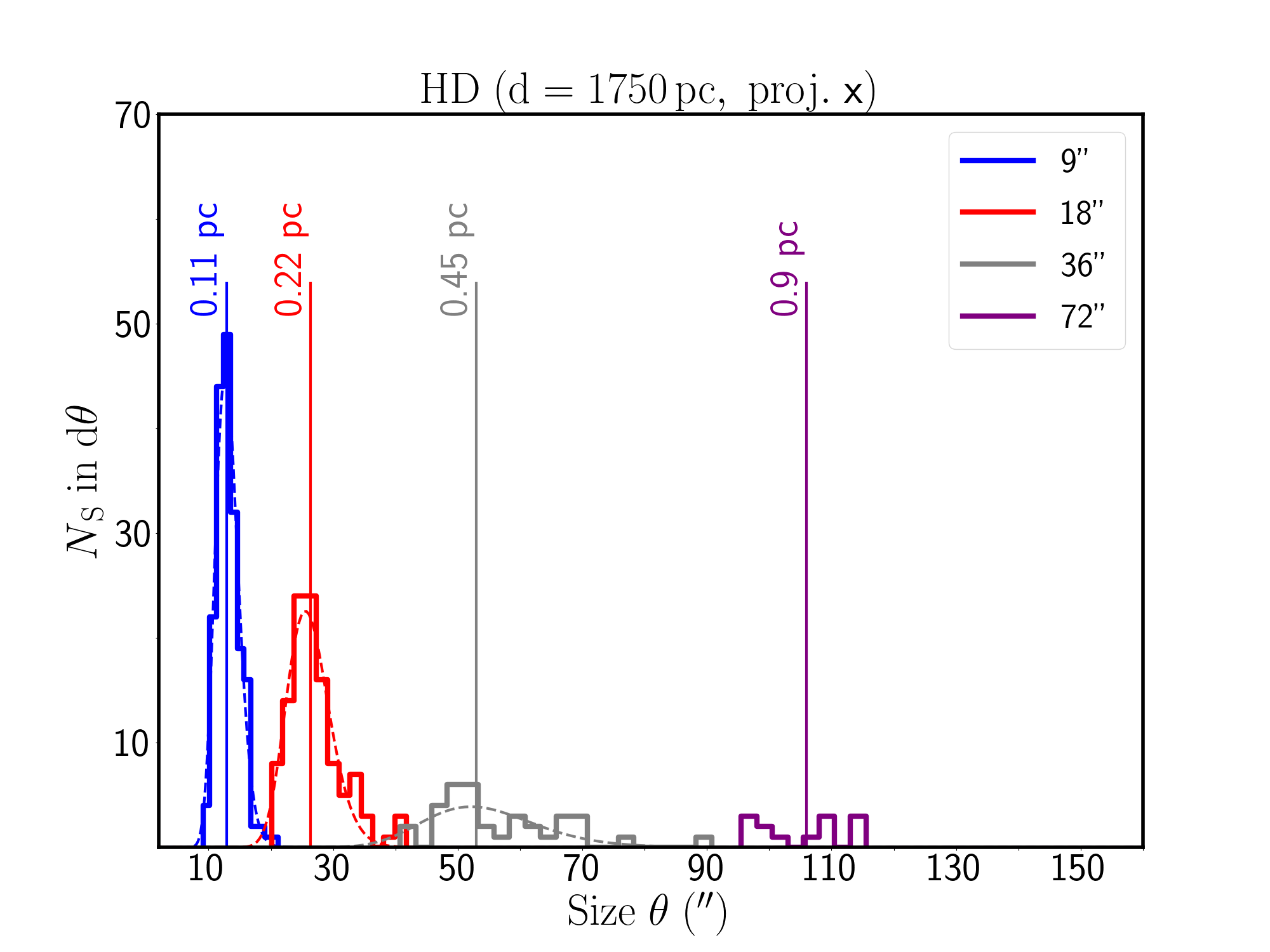}}
\subfloat{\includegraphics[trim=3cm 1.2cm 1cm 5cm, width=0.34\linewidth]{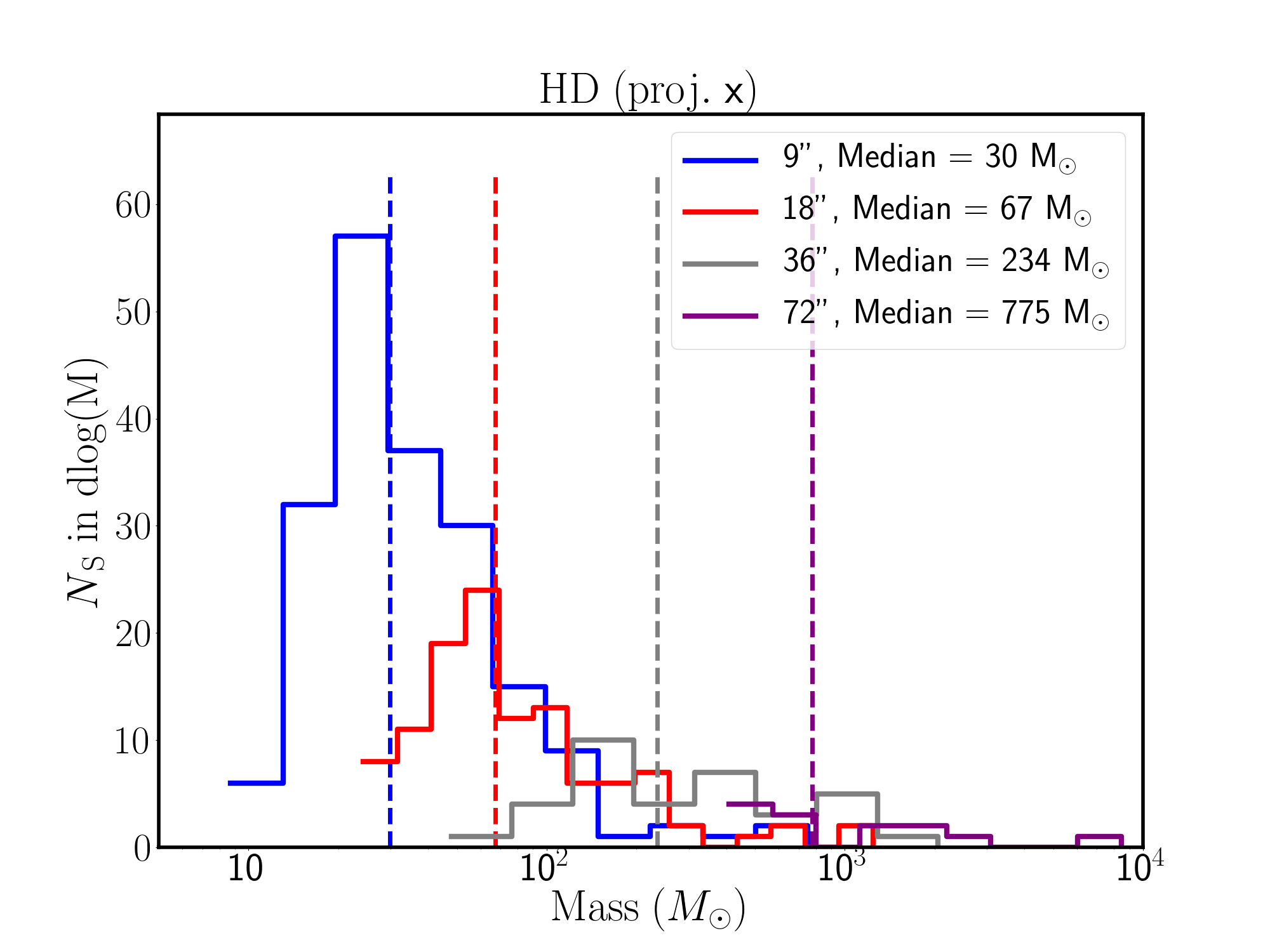}}
\subfloat{\includegraphics[trim=3cm 1.2cm 1cm 5cm, width=0.34\linewidth]{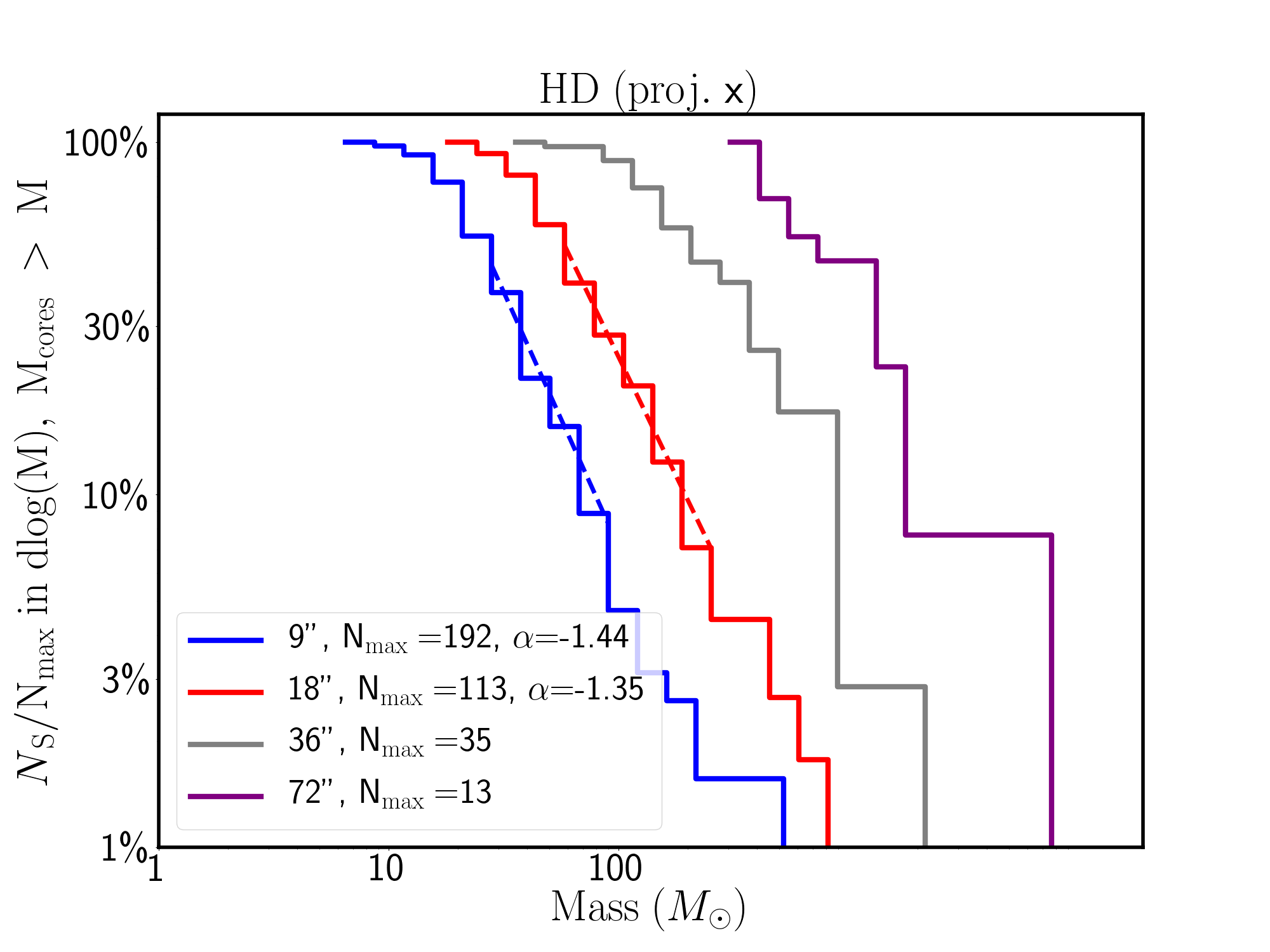}}\\
\subfloat{\includegraphics[trim=3cm 1.2cm 1cm 4.5cm, width=0.34\linewidth]{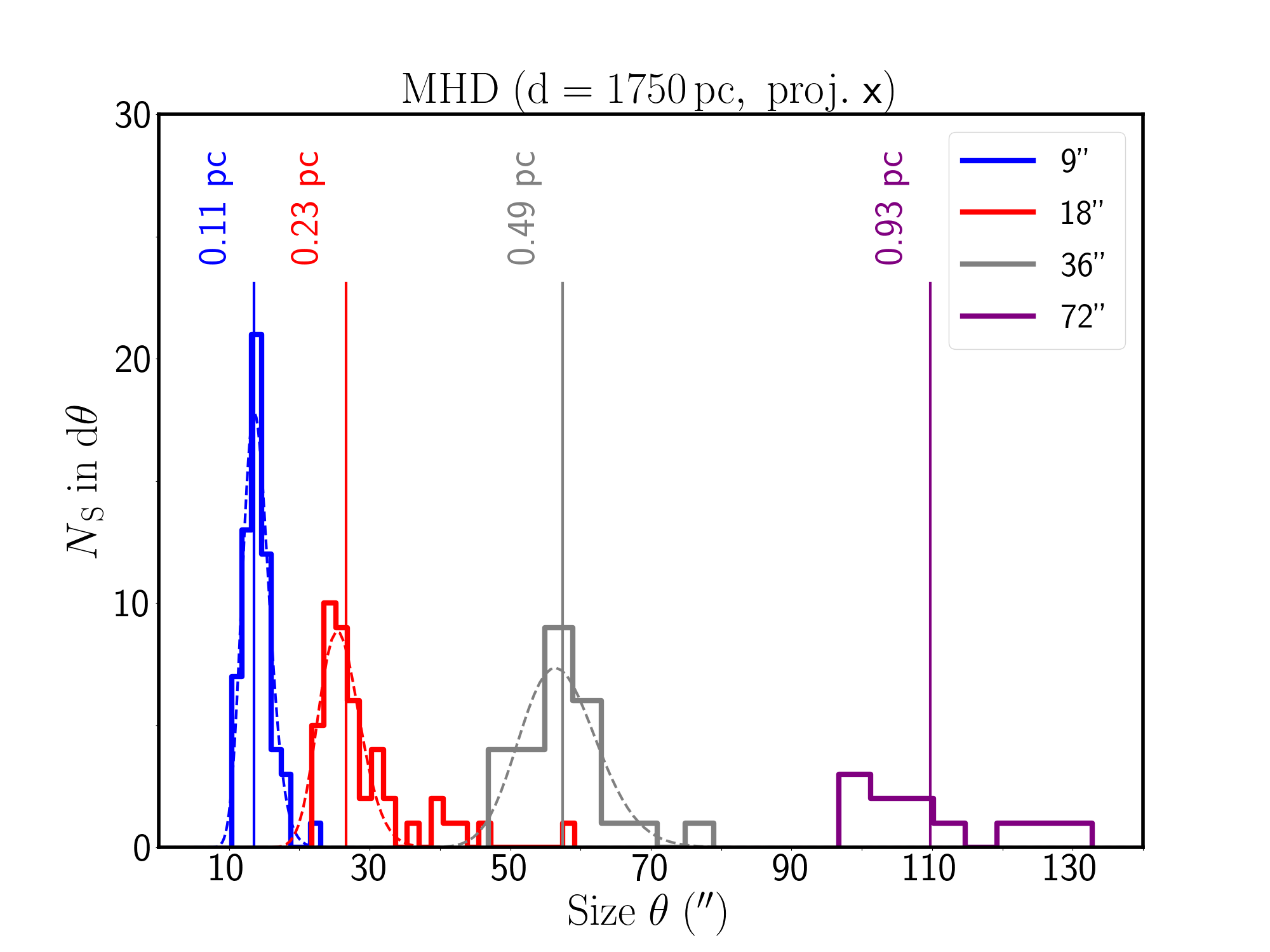}}
\subfloat{\includegraphics[trim=3cm 1.2cm 1cm 4.5cm, width=0.34\linewidth]{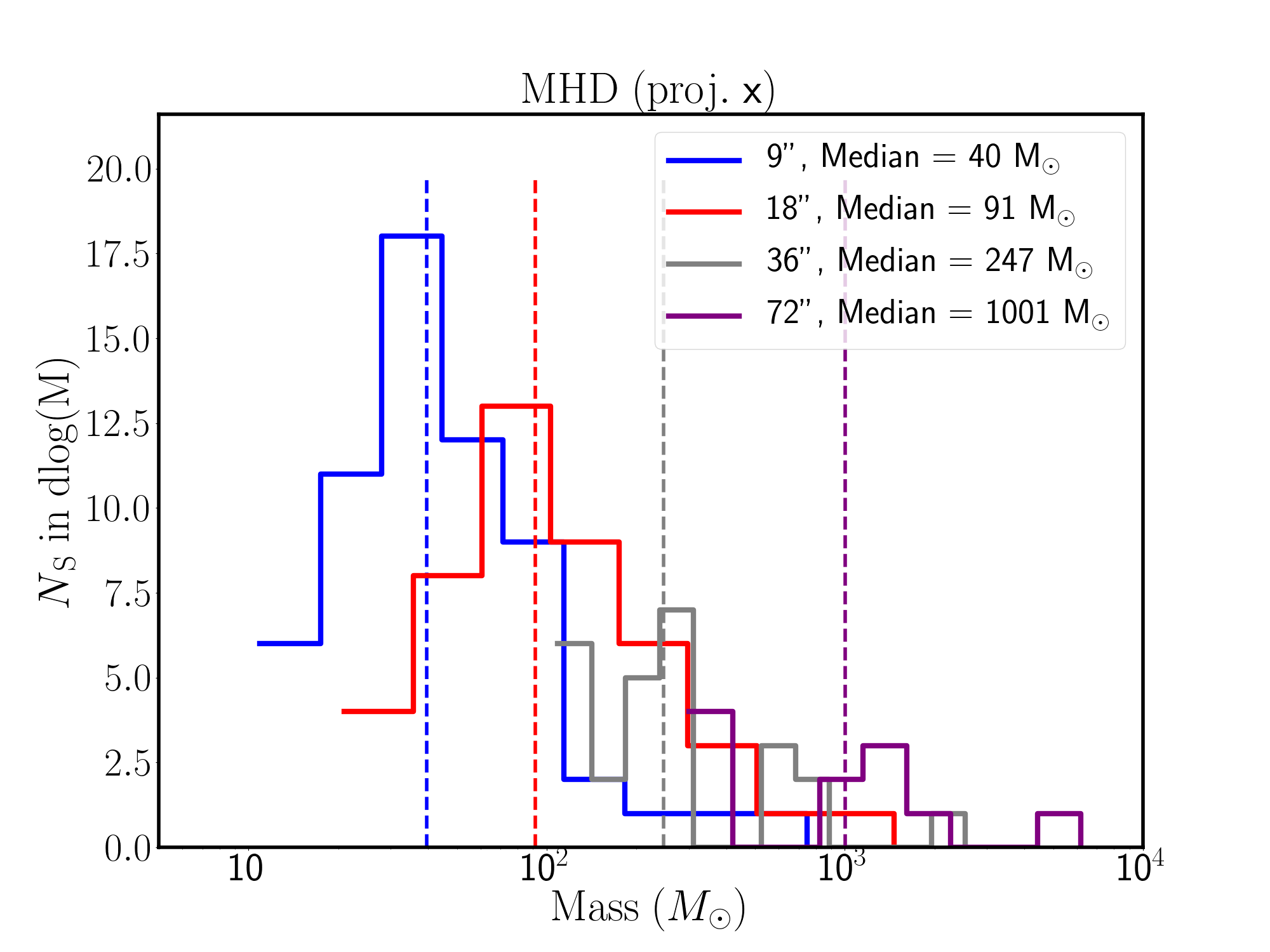}}
\subfloat{\includegraphics[trim=3cm 1.2cm 1cm 4.5cm, width=0.34\linewidth]{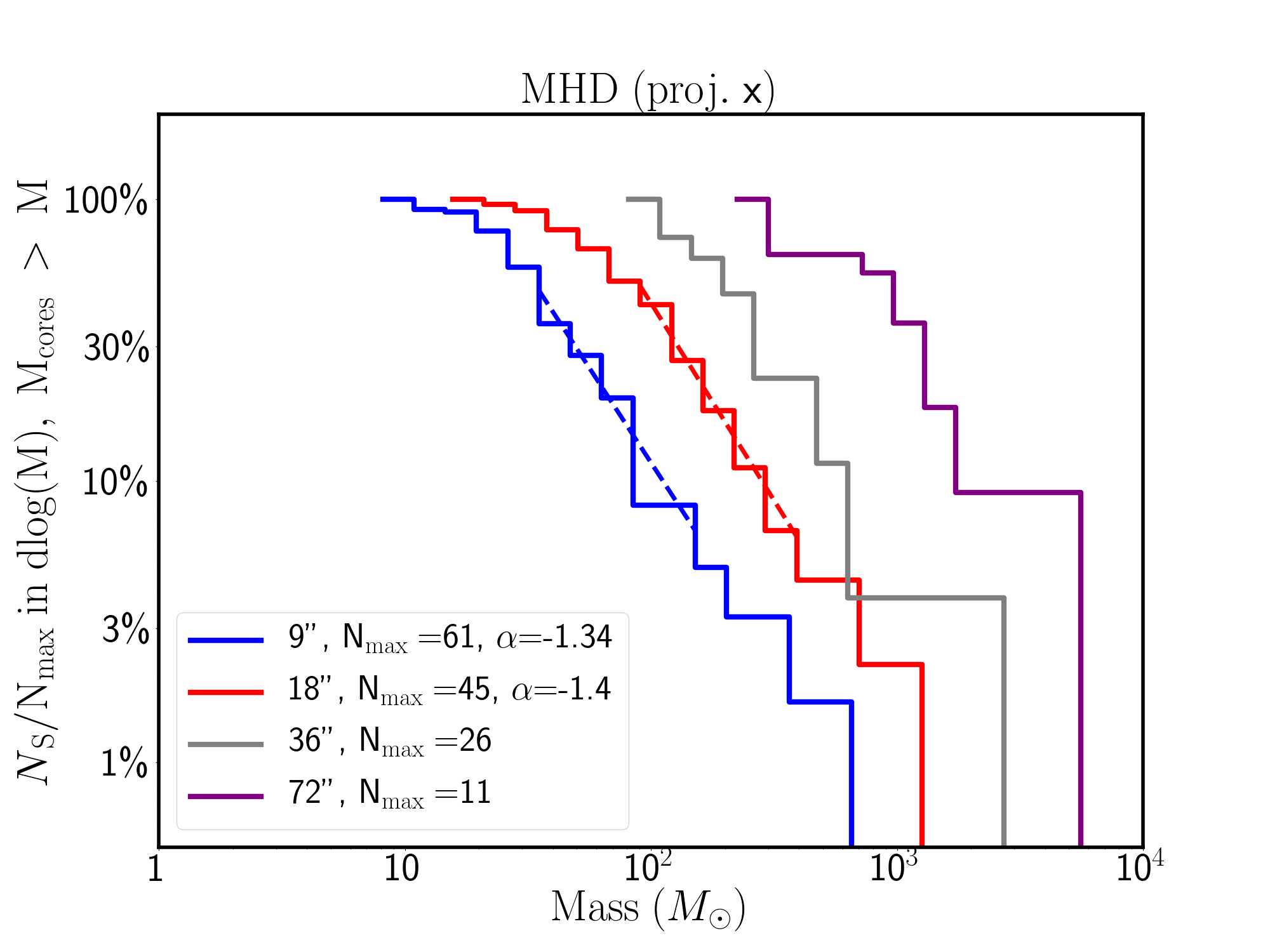}}\\
\subfloat{\includegraphics[trim=3cm 1.2cm 1cm 4.5cm, width=0.34\linewidth]{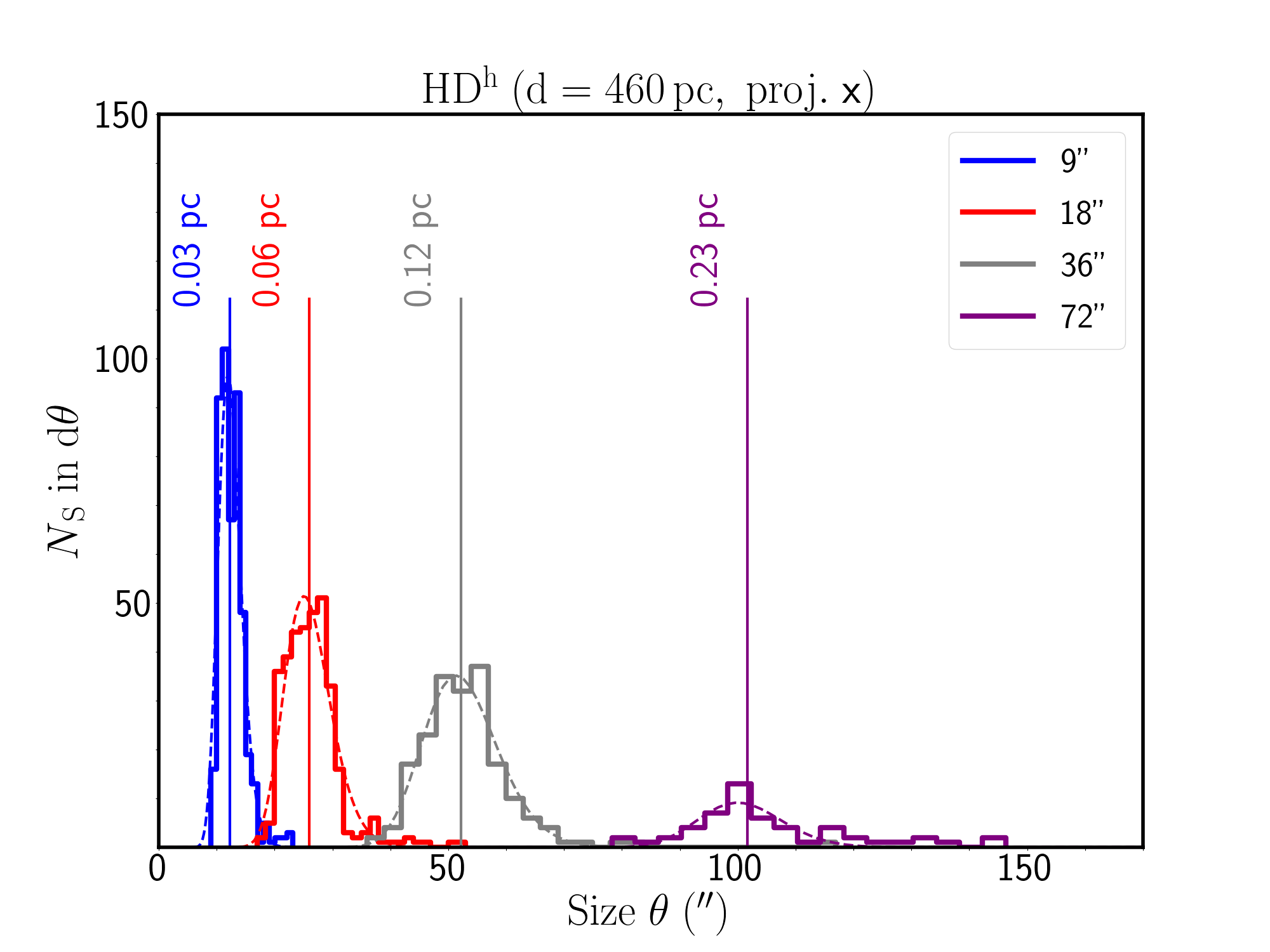}}
\subfloat{\includegraphics[trim=3cm 1.2cm 1cm 4.5cm, width=0.34\linewidth]{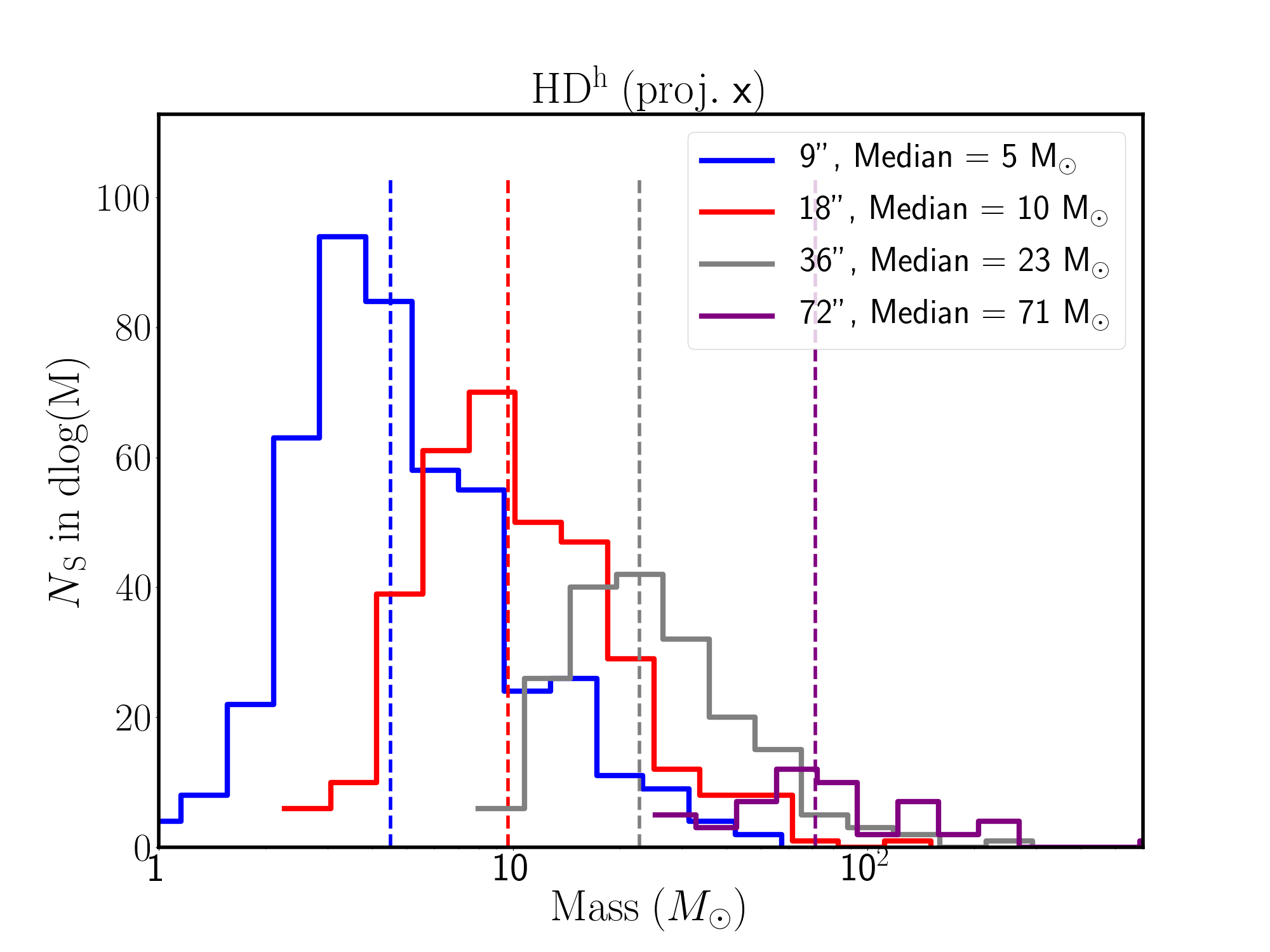}}
\subfloat{\includegraphics[trim=3cm 1.2cm 1cm 4.5cm, width=0.34\linewidth]{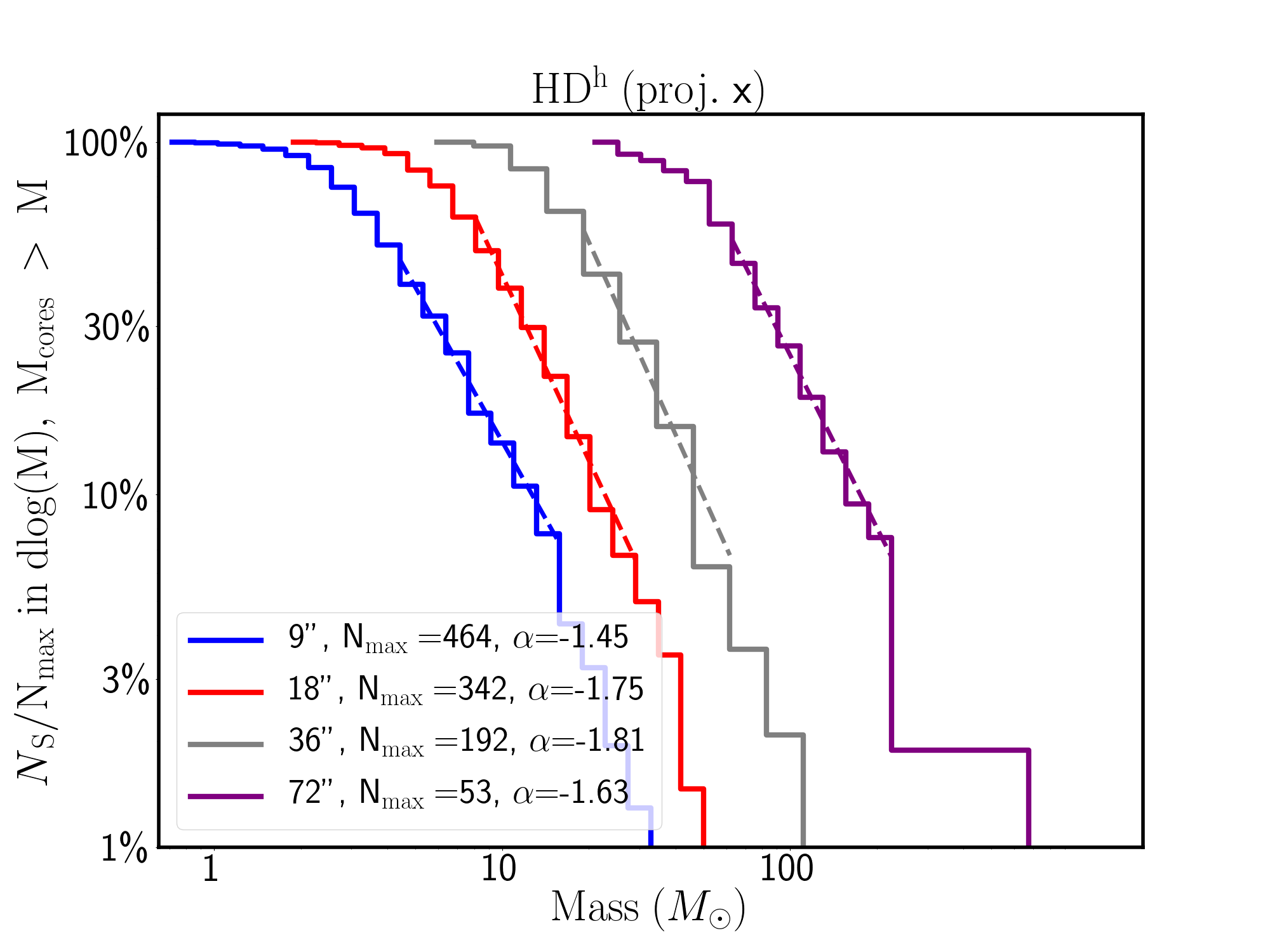}}
\caption
{ 
Effects of different angular resolutions on the sizes and masses of the bound sources in the simulated star-forming regions HD (\emph{top}), MHD (\emph{middle row}), and HD$^\mathrm{h}$ (\emph{bottom}). We show the source size function (\emph{left}), the source mass function (\emph{middle}), and the cumulative mass function (\emph{right}), obtained for the \emph{x} projection of the respective surface density maps. The other projections on the \emph{y} and \emph{z} directions are displayed in Figs.~\ref{f:reseffect-simu} and \ref{f:reseffect-simu-z}. 
} 
\label{f:reseffect-simu-x}
\end{figure*}

\begin{figure*} 
    \centering
\subfloat{\includegraphics[trim=3cm 1.2cm 1cm 5cm, width=0.34\linewidth]{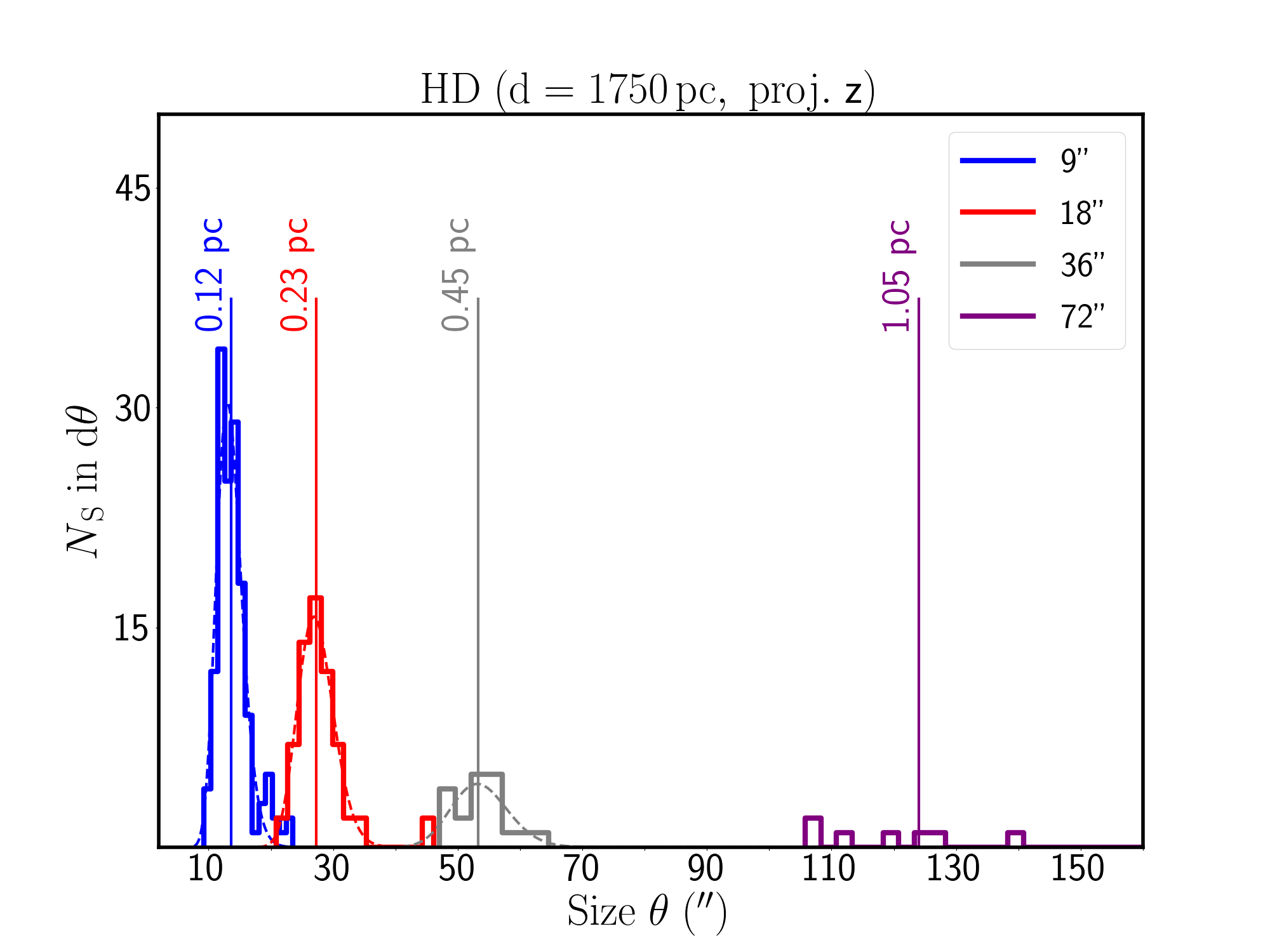}}
\subfloat{\includegraphics[trim=3cm 1.2cm 1cm 5cm, width=0.34\linewidth]{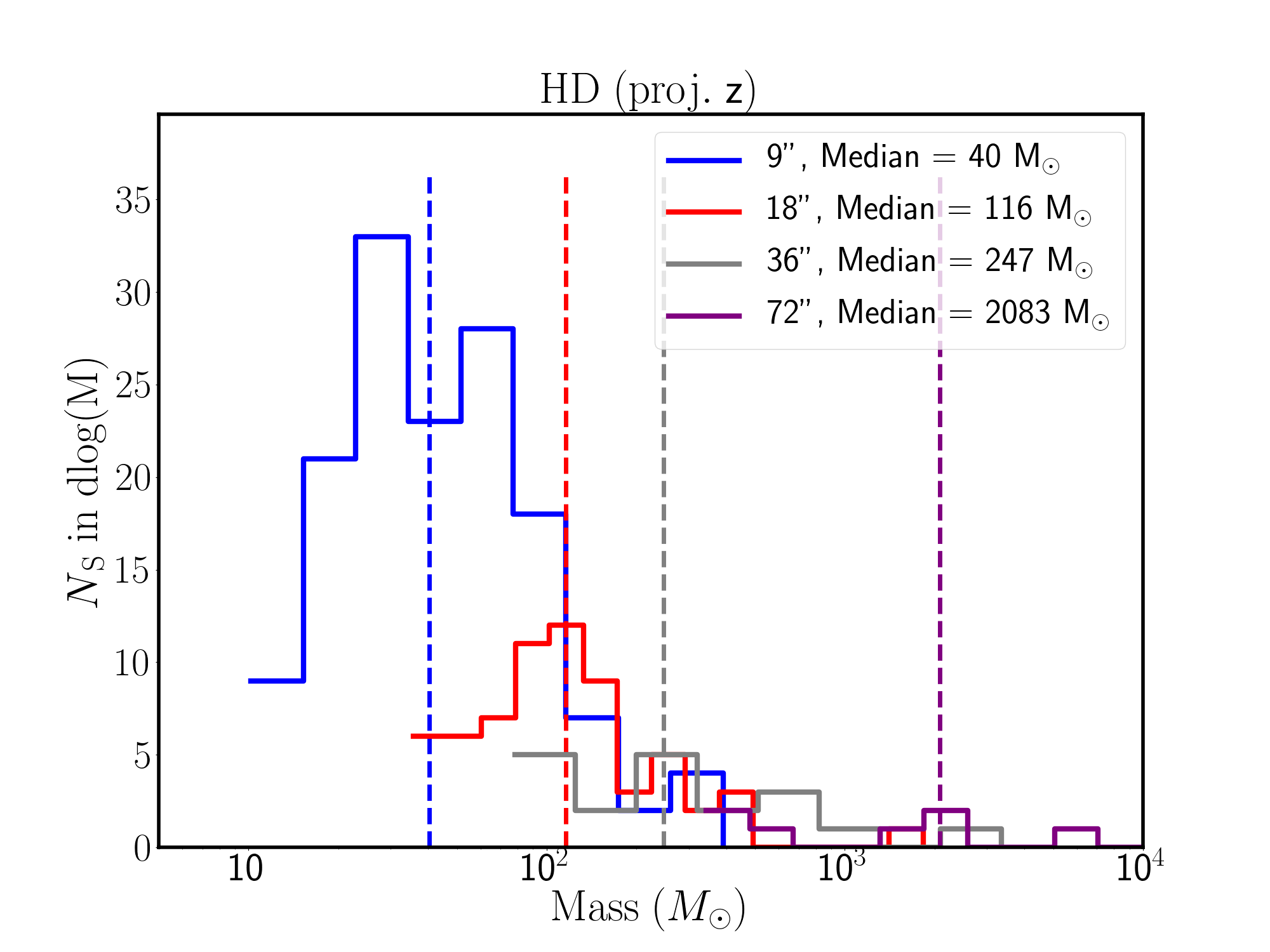}}
\subfloat{\includegraphics[trim=3cm 1.2cm 1cm 5cm, width=0.34\linewidth]{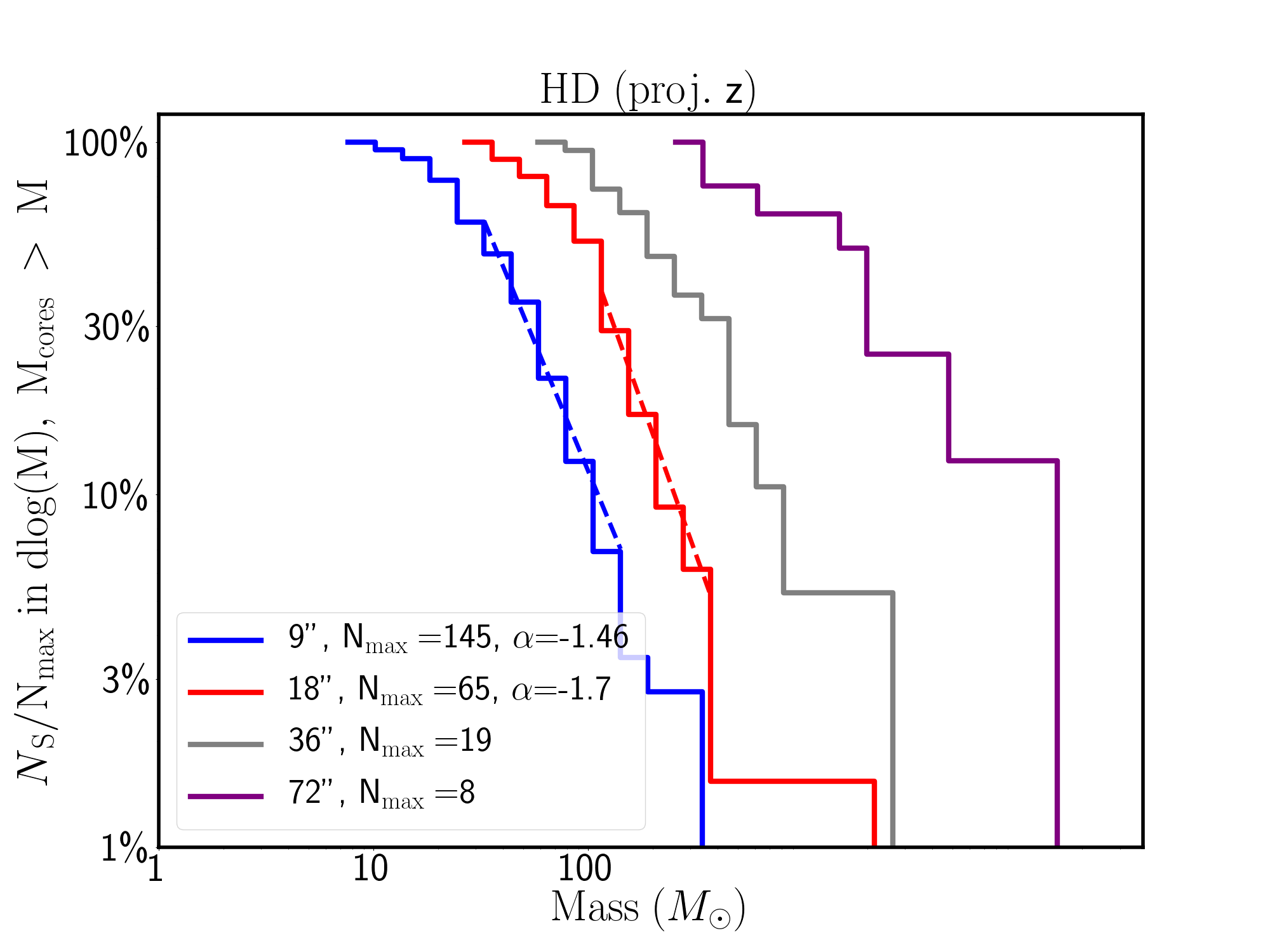}}\\
\subfloat{\includegraphics[trim=3cm 1.2cm 1cm 4.5cm, width=0.34\linewidth]{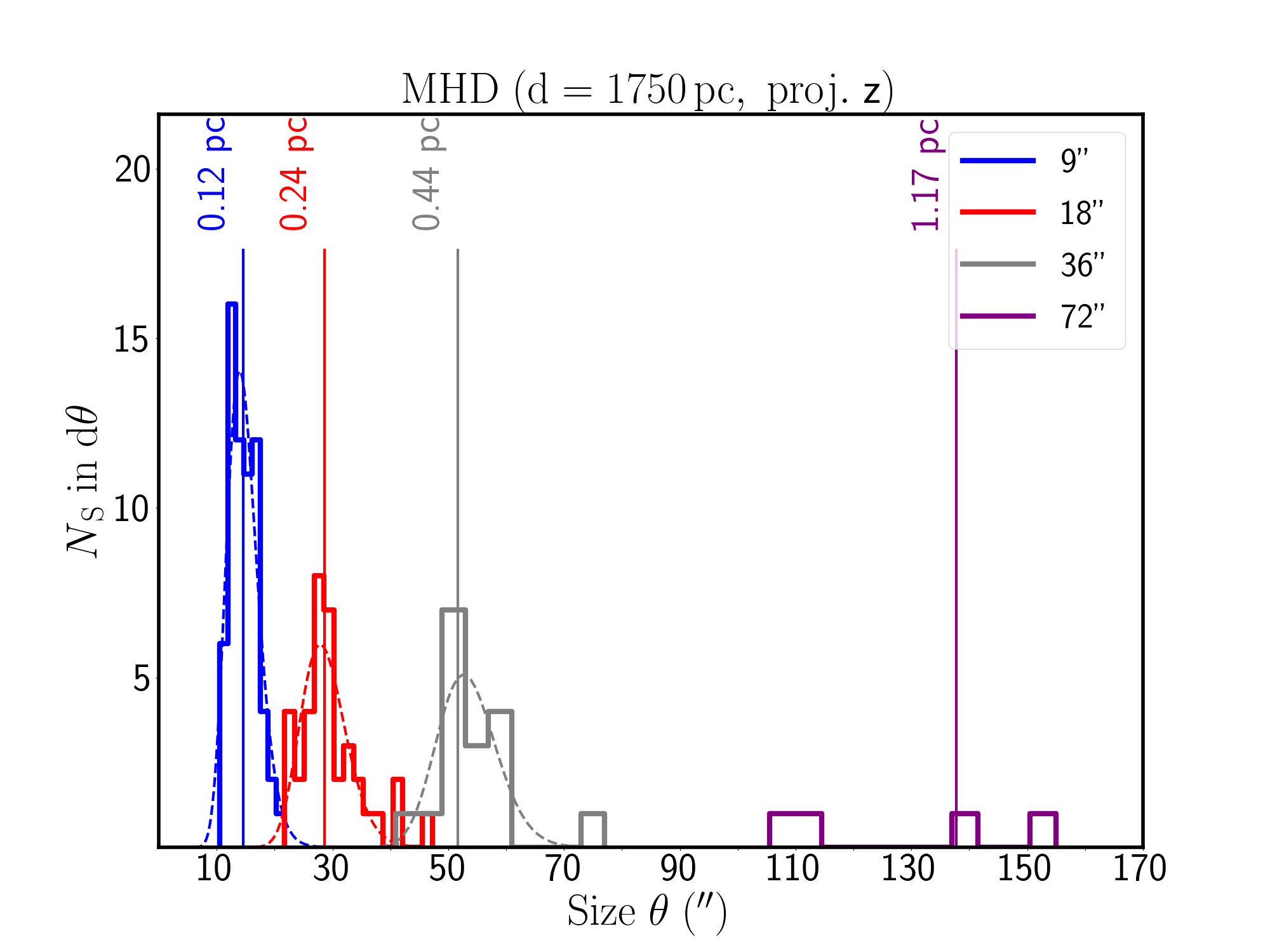}}
\subfloat{\includegraphics[trim=3cm 1.2cm 1cm 4.5cm, width=0.34\linewidth]{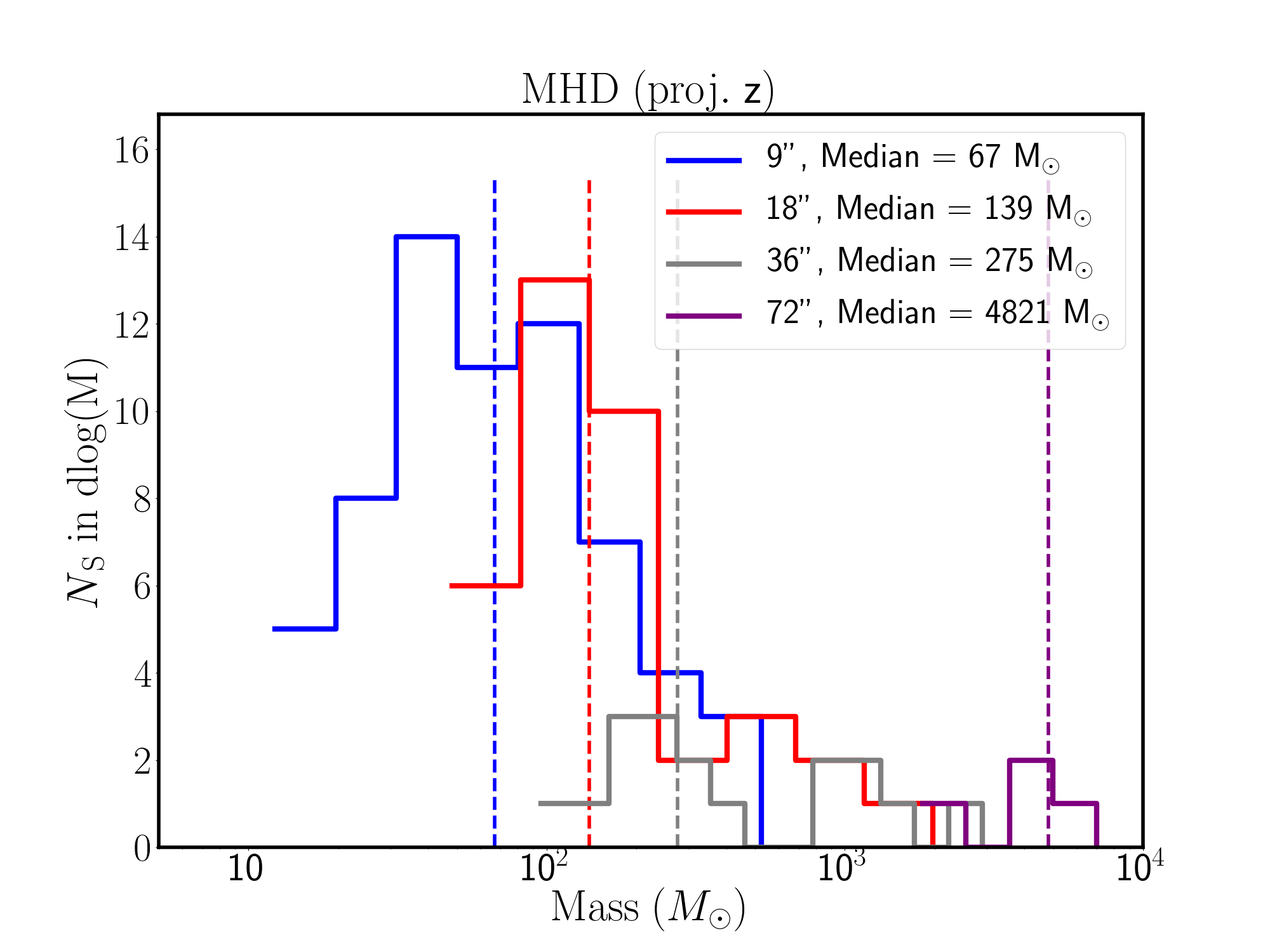}}
\subfloat{\includegraphics[trim=3cm 1.2cm 1cm 4.5cm, width=0.34\linewidth]{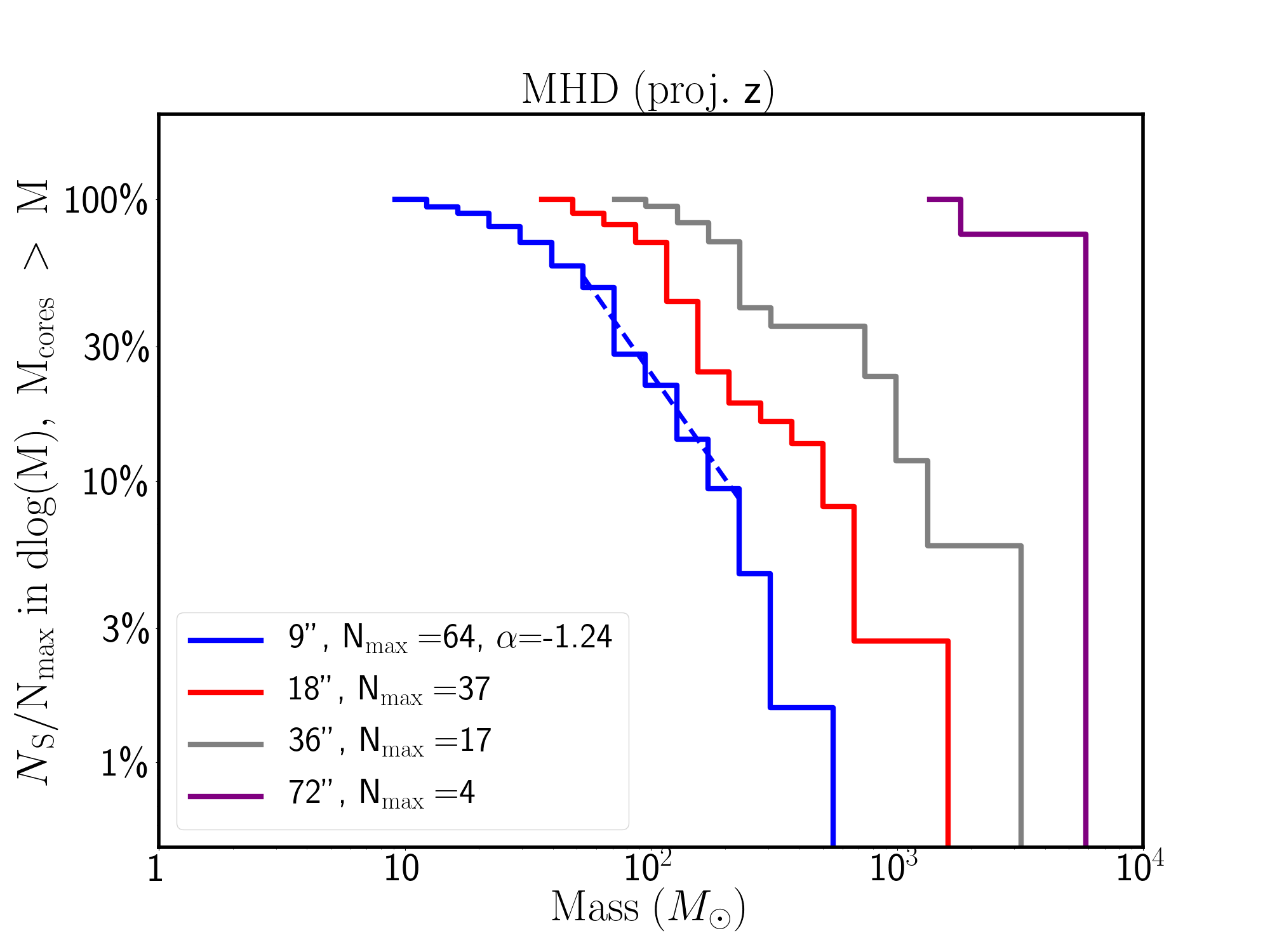}}\\
\subfloat{\includegraphics[trim=3cm 1.2cm 1cm 4.5cm, width=0.34\linewidth]{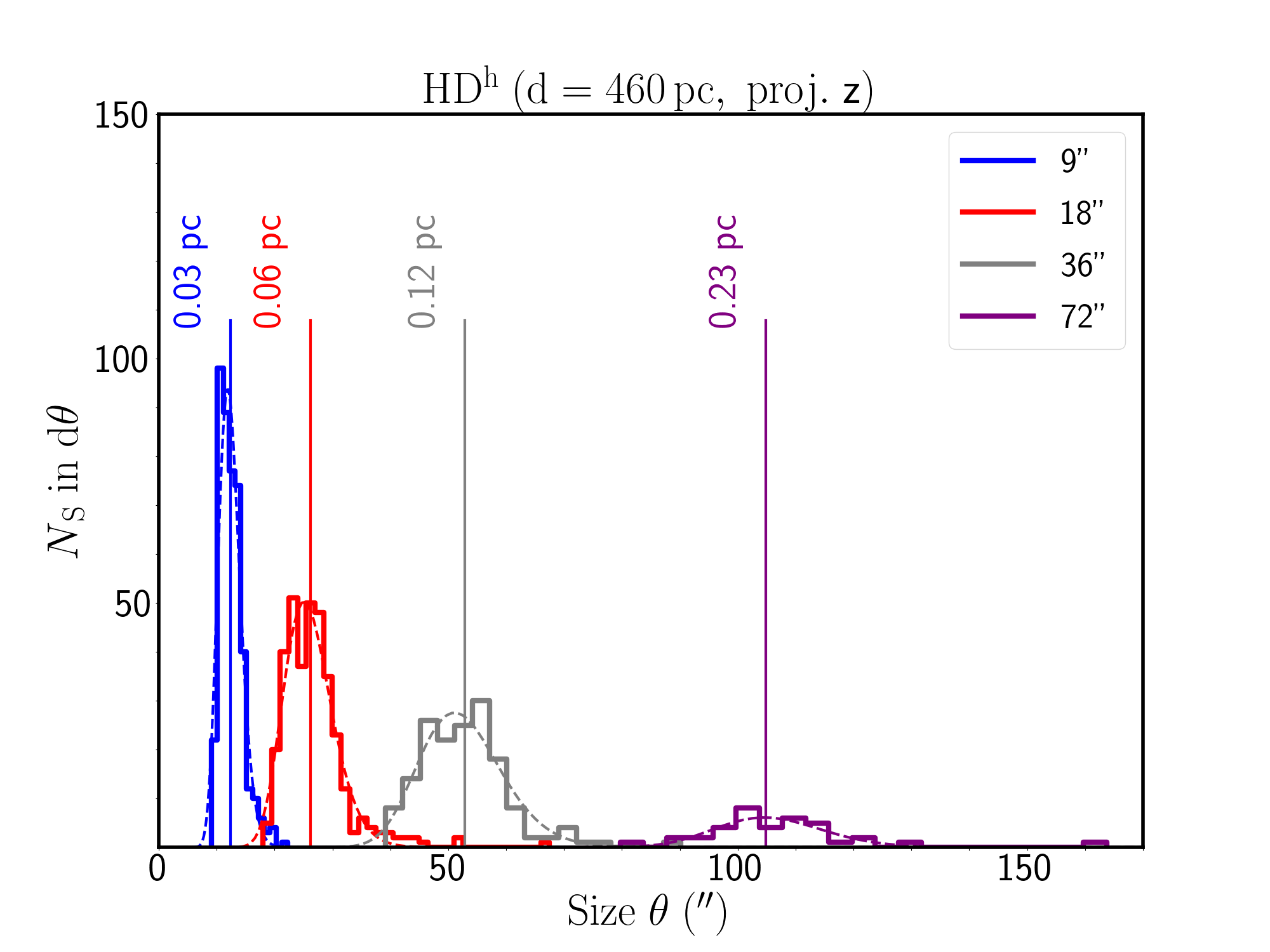}}
\subfloat{\includegraphics[trim=3cm 1.2cm 1cm 4.5cm, width=0.34\linewidth]{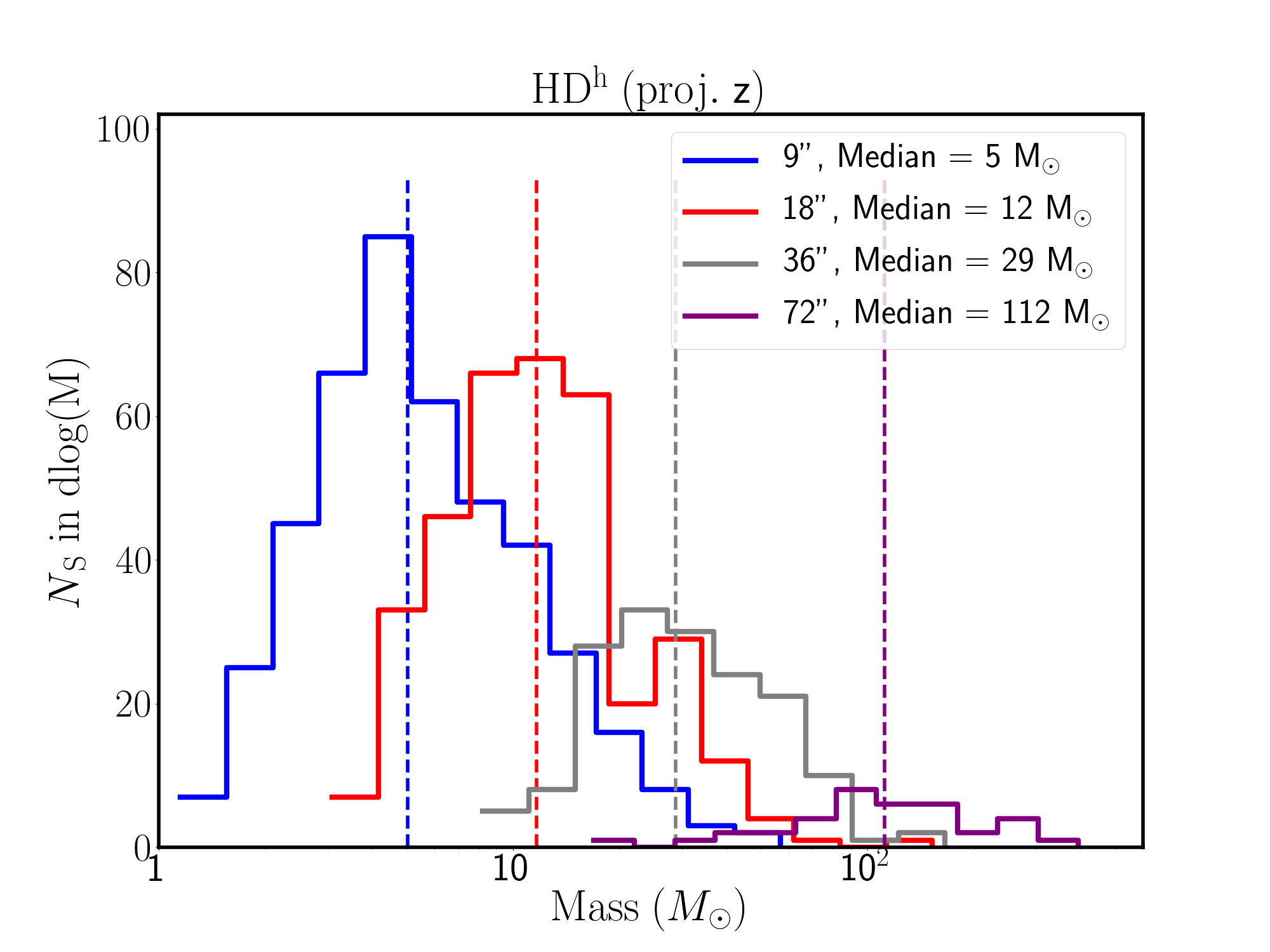}}
\subfloat{\includegraphics[trim=3cm 1.2cm 1cm 4.5cm, width=0.34\linewidth]{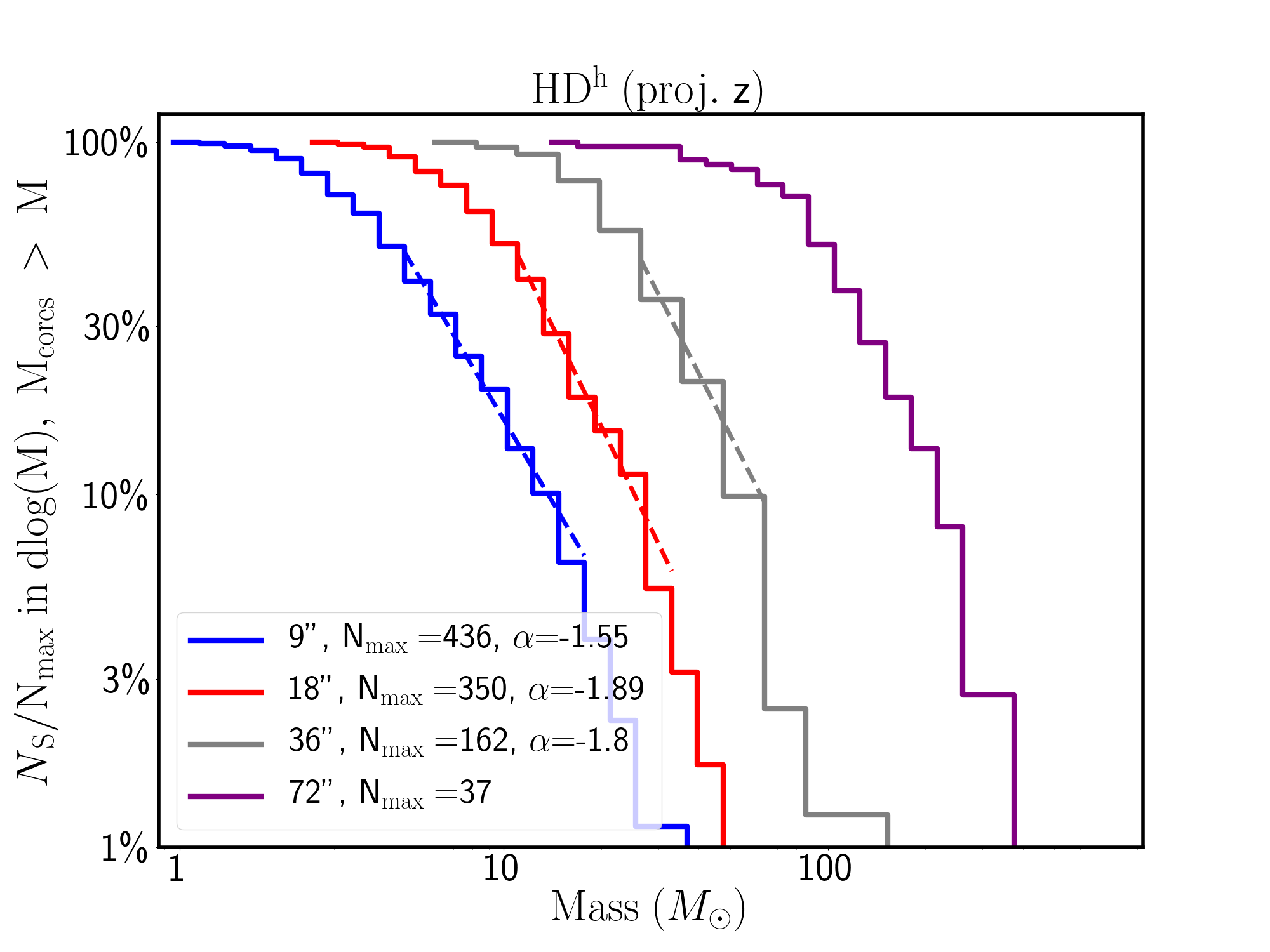}}
\caption{
Effects of different angular resolutions on the sizes and masses of the bound sources in the simulated star-forming regions HD (\emph{top}), MHD (\emph{middle row}), and HD$^\mathrm{h}$ (\emph{bottom}). We show the source size function (\emph{left}), the source mass function (\emph{middle}), and the cumulative mass function (\emph{right}), obtained for the \emph{z} projection of the respective surface density maps. The other projections on the \emph{x} and \emph{y} directions are displayed in Figs.~\ref{f:reseffect-simu} and~\ref{f:reseffect-simu-x}. 
} 
\label{f:reseffect-simu-z}
\end{figure*}

\end{appendix}
\end{document}